\documentclass[prb,nofootinbib,citeautoscript,10pt,longbibliography,notitlepage]{revtex4-2}

\usepackage{graphicx}
\usepackage{dcolumn}
\usepackage{bm}
\usepackage{amsmath}
\usepackage{tabularx,graphicx}
\usepackage{epstopdf}
\usepackage{latexsym}
\usepackage{amssymb}
\usepackage{amsmath}
\usepackage{colortbl}
\usepackage{psfrag}
\usepackage{bbm}
\usepackage{titlesec}
\usepackage{dsfont}
\usepackage{feynmp}
\usepackage{slashed}
\usepackage{multirow}
\usepackage[tight]{subfigure}

\usepackage[papersize={8.5in,11in}]{geometry}

\usepackage{color}
\definecolor{darkblue}{rgb}{0.,0.,0.4}
\definecolor{darkred}{rgb}{0.5,0.,0.}
\definecolor{BlueViolet}{RGB}{138,43,226}
\definecolor{SkyBlue}{RGB}{30,144,255}
\definecolor{DarkGreen}{RGB}{0,100,0}
\usepackage[pdftex,colorlinks=true,linkcolor=darkblue,citecolor=blue,urlcolor=darkred]{hyperref}

\def \nn{\nonumber \\}

\begin{document}


\title{Anatomy of plasmons in generic Luttinger semimetals}

\author{Jing Wang}
\altaffiliation{E-mail: jing$\textunderscore$wang@tju.edu.cn}
\affiliation{Department of Physics $\&$ Tianjin Key Laboratory of Low Dimensional Materials Physics and Preparing Technology, Tianjin University, Tianjin 300072, People's Republic of China}

\author{Ipsita Mandal}
\altaffiliation{E-mail: ipsita.mandal@gmail.com}
\affiliation{Department of Physics, Shiv Nadar Institution of Eminence (SNIoE), Gautam Buddha Nagar, Uttar Pradesh 201314, India
}

\begin{abstract}
We investigate the parameter regimes favourable for the emergence of plasmons in isotropic, anisotropic, and band-mass symmetric and asymmetric Luttinger semimetals (LSMs). An LSM harbours a quadratic band-crossing point (QBCP) in its bandstructure, where the upper and lower branches of dispersion are doubly degenerate.
While a nonzero temperature ($T$) can excite particle-hole pairs about the Fermi level due to thermal effects (even at zero doping), a finite doping ($\mu$) sets the Fermi level away from the QBCP at any $T$, leading to a finite Fermi surface (rather than a Fermi point). Both these conditions naturally give rise to a finite density of states. A nonzero value of $T$ or $\mu$ is thus a necessary condition for a plasmon to exist, as otherwise the zero density of states at the QBCP can never lead to the appearance of this collective mode. In addition to $T$ and $\mu$, we consider the effects of all possible parameters like cubic anisotropy, band-mass asymmetry, and a material-dependent variable $X$ that is proportional to the mass (of the quasiparticle) and the number of fermion flavours. We implement a random-phase-approximation to compute the quasiparticle decay rate $ \tau^{-1} $ (also known as the inelastic scattering rate) resulting from screened Coulomb interactions. A well-defined sharp peak in the profile of $\tau^{-1}$ signals the appearance of a plasmon. From our results, we conclude that $X$ turns out to be a crucial tuning parameter, as higher values of $X$ assist in the emergence of plasmons. On the other hand, the features are broadly insensitive to cubic anisotropy and band-mass asymmetry.
\end{abstract}


\maketitle

\tableofcontents


\section{Introduction}

In contemporary research, three-dimensional (3d) semimetals
with a quadratic band-crossing point (QBCP) are being extensively studied~\cite{Moon2013PRL,kondo2015NC,
Herbut2017PRB,Herbut2017PRB_2,Nandkishore2017PRB,Herbut2016PRB,ips-qbt-sc,Mandal2018PRB,
Wang2019EPJB,Mandal2019AP,
ips-herm1,ips-herm2,ips-herm3,ips-tunnel-qbcp,Bera2021}.
Distinct from Dirac/Weyl semimetals possessing
band-crossings with linear energy dispersions~\cite{Neto2009RMP},
such a bandstructure can be realized in materials like pyrochlore iridates $\mathrm{A_2Ir_2O_7}$
(where $\mathrm{A}$ is a lanthanide element~\cite{Yanagishima2001,Matsuhira2007}),
gray tin ($\alpha$-Sn) \cite{paul,gray-tin}, and HgTe~\cite{Tsidilkovski1997Book}. Such systems are also known as Luttinger semimetals (LSMs), since the Luttinger Hamiltonian of inverted band-gap semiconductors describes the low-energy physics~\cite{Luttinger1956PR,abrikosov1996,Shuichi2004PRB}.

Long-ranged Coulomb interactions in LSMs cause a quantum phase transition to a
non-Fermi liquid state, dubbed as the Luttinger-Abrikosov-Beneslavski\u{i} (LAB) phase~\cite{Abrikosov,Moon2013PRL}, when the chemical potential is tuned to cut right at the band-crossing point
\footnote{In the presence of electron-electron interactions, when the Fermi level is away from the band-crossing point, it has been predicted that two-dimensional (2d) counterparts of the LSMs are unstable to interaction-driven topological insulating phases~\cite{Fradkin2009PRL,Vafek2014PRB,Venderbos2016,Wu2016,Wang2017PRB,Wang2020PRB}.}.
The possibility of the emergence of plasmons at a finite temperature \cite{Mandal2019AP} or chemical potential has also been explored \cite{Tchoumakov-Krempa2019PRB,polini} in isotropic LSMs. One might wonder whether the fate of these plasmons are robust in the presence of anisotropies, which is a more natural possibility. To this end, in this paper, we compute the behaviour of plasmons for both isotropic and anisotropic LSMs, under the influence of both finite temperature ($T$) and doping (away from the QBCP). We also include the effects of bass-mass asymmetries between the upper and lower bands.

We follow the strategy of treating screened Coulomb interactions in the 3d QBCP within a random-phase-approximation (RPA), as is done in Ref.~\cite{Mandal2019AP}. In the first step, the real and imaginary parts of the bare polarization function (i.e. without any interaction line in the loop) are explicitly derived, and the results tell us if a plasmon mode can exist when we add Coulomb interactions to the system [cf. Eq.~(\ref{Eq_diele})].
For the sake of completeness, all external variables, including the frequencies and momenta are considered at the same footing. In addition to keeping $T$ and chemical potential ($\mu$) finite, a suitable material-dependent parameter $X$ is introduced, which can influence the emergence of plasmons \footnote{Plasmons cannot exist at $T=\mu=0$ because of the vanishing of density of states at the QBCP.}. Finally, the results for polarization bubble in various regimes are used to calculate the inelastic scattering rate of the LSM quasiparticles due to Coulomb interactions. We investigate various parameter regimes to figure out when it is possible for plasmons to exist, and this is captured by sharp peaks in the inelastic scattering rates.
We would like to point out that the deviations from isotropic LSM, by breaking rotational symmetry and introducing asymmetry between the band masses of the upper and lower branches of dispersion, only slightly modify the zeros of the real part of effective dielectric function, and cannot lead to the generation of plasmons in the absence of finite $T$ and/or $\mu$. This is to be expected as those asymmetries do not lead to a finite density of states at the band-crossing points (which is essential for plasmons to exist), other than making the dispersions anisotropic.

The paper is organized as follows. In Sec.~\ref{Sec_model}, we present the low-energy effective model for a generic
LSM. Sec.~\ref{Sec_bare-Pi} shows the computations of the bare polarization bubble.
In Sec.~\ref{Sec_Results}, we treat the Coulomb interaction within RPA, and figure out the behaviour of the dielectric function, whose zeros capture the dispersions of the plasmons.
Sec.~\ref{Sec_scattering-rate} is devoted to calculating the inelastic scattering rate, as the presence of sharp peaks in the profile of this function indicates long-lived plasmons.
We also compute the spectral function and the wavefunction renormalization (quasiparticle residue). All our computations encompass the isotropic, anisotropic, and band-mass symmetric and asymmetric scenarios. Finally, we end with a summary and discussion in Sec.~\ref{Sec_Summary}.

\section{Model}

\label{Sec_model}

We consider a paramagnetic bandstructure, arising in a spin-orbit coupled system, for which the states near the center of the Brillouin zone (i.e., the $\Gamma$-point) transform under the symmetry group operations as the four spin states of a
particle with angular momentum $j=3/2$, and four bands cross at that point.
The corresponding angular momentum operators $\boldsymbol{\mathcal J} = \lbrace {\mathcal J}_1, {\mathcal J}_2, {\mathcal J}_3 \rbrace $ transform as the $ \mathrm{T_2}$ representation of the cubic group. The low-energy effective continuum model near the band-touching point is obtained from the $\mathbf k \cdot \mathbf p $ theory, and is captured by the Luttinger Hamiltonian ~\cite{Luttinger1956PR,Shuichi2004PRB,Moon2013PRL}
\begin{align}
\label{eqham}
\mathcal{H}_0 =  \left( \frac{ \alpha \, \mathbf{p}^2} {2\, m}-\mu \right) \Gamma_0
+\sum^5_{a=1}d_a(\mathbf{p})\, \Gamma_a
+\eta\sum^5_{a=1}s_a \, d_a(\mathbf{p}) \,\Gamma_a \,,
\end{align}
at chemical potential $\mu$.
Here, the set of five matrices $\lbrace \Gamma_a \rbrace $ gives the rank-four irreducible representation of the Euclidean Clifford algebra, satisfying anticommutation relation $ \lbrace \Gamma_a, \Gamma_{a'}  \rbrace = 2\,\delta_{a a'} $. In 3d, the basis for a generic $4\times 4$ Hermitian matrix is constituted by the identity matrix (denoted by $\Gamma_0$) and the five $ \lbrace \Gamma_a \rbrace$ matrices (with $a \in[1,5]$).
From the latter, we get ten more distinct matrices defined by $\Gamma_{ab} =\frac{1}{2\,i} [\Gamma_a, \Gamma_b]$.
The three components of the spin operator $\boldsymbol{\mathcal J}$ can be expressed in terms of the $\Gamma_{ab}$-matrices via linear relations \cite{Moon2013PRL}. The functions denoted by $d_a(\mathbf{p})$ are the
$\ell =2$ spherical harmonics given by~\cite{Herbut2012PRB,Herbut2016PRB,Mandal2019AP}
\begin{align}
d_1(\mathbf{p}) & =\frac{\sqrt{3} \, p_y \,p_z} {2 \, m} \,,\quad
d_2(\mathbf{p})=\frac{\sqrt{3} \, p_x \, p_z}{2\, m} \,, \quad
d_3(\mathbf{p})=\frac{\sqrt{3} \, p_x \, p_y} {2 \, m} \,, \nn
d_4(\mathbf{p}) & =\frac{\sqrt{3} \left (p^2_x-p^2_y \right )} {4 \,m} \,,\quad
d_5(\mathbf{p})=\frac{ 2 \,p^2_z-p^2_x-p^2_y} {4 \,m} \,.
\label{Eq_d_a}
\end{align}

\begin{figure}
\centering
\subfigure[]{\includegraphics[width=0.23\textwidth]{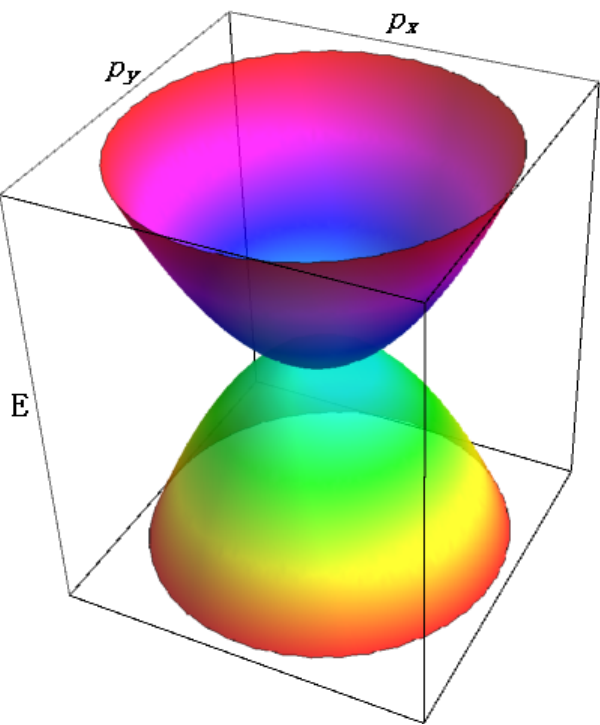}}\hspace{ 0 cm}
\subfigure[]{\includegraphics[width=0.23\textwidth]{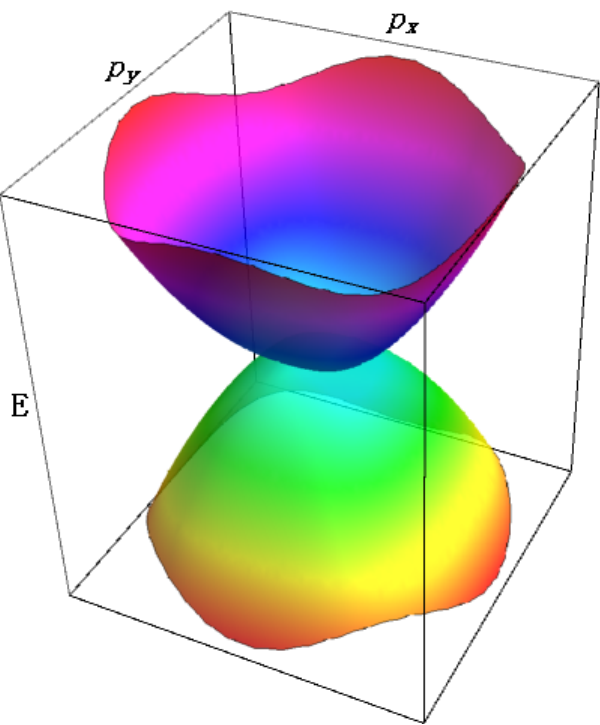}}
\subfigure[]{\includegraphics[width=0.23\textwidth]{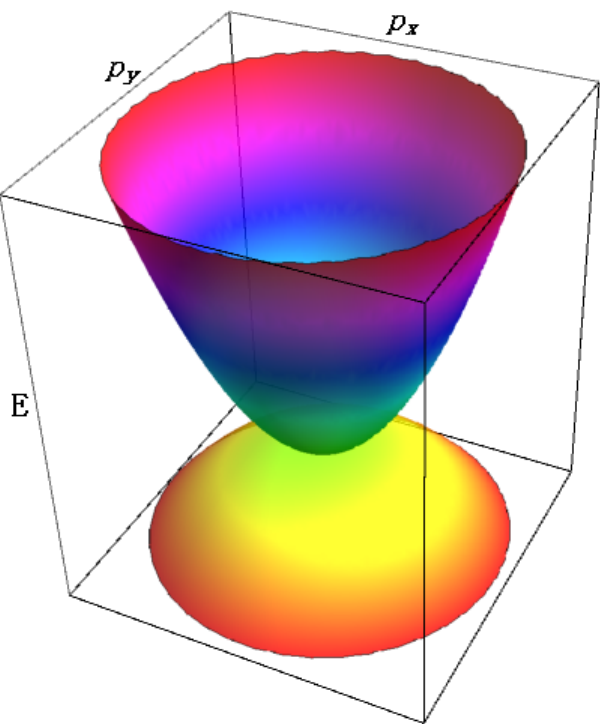}}\hspace{ 0 cm}
\subfigure[]{\includegraphics[width=0.23\textwidth]{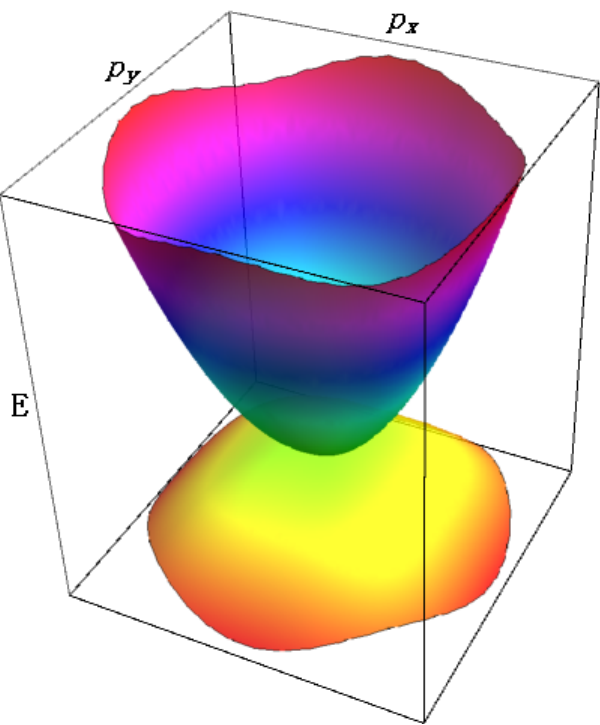}}
\caption{Schematic dispersions of the Luttinger semimetals in the $p_x$-$p_y$ plane
for (a) $\alpha=0$ and $\eta=0$;
(b) $\alpha\neq0$ and $\eta=0$; (c) $\alpha=0$ and $\eta\neq0$; and (d) $\alpha\neq0,\eta\neq0$.
\label{Fig_dispersion_LS}}
\end{figure}

The $\Gamma_a$-matrices can always be chosen such that three are real and two are imaginary \cite{Herbut2012PRB}. We choose a representation in which $\lbrace \Gamma_1, \Gamma_2, \Gamma_3 \rbrace $ are real and $\lbrace \Gamma_4, \Gamma_5 \rbrace $ are imaginary.
Using the same conventions as in Ref.~\cite{Herbut2016PRB}, we set (i) $s_a=1$ for the
off-diagonal (i.e., for $a = \lbrace 2,3,4 \rbrace $);
and (ii) $s_a =-1$ for the diagonal (i.e., for $ a = \lbrace 1,5 \rbrace $) $\Gamma_a$-matrices.
Furthermore, the parameters $\alpha$ and $\eta$ are directly related to
the Luttinger parameters~\cite{Herbut2016PRB}, affecting the various symmetries of an LSM.
More specifically, the band-mass and rotational symmetries are unbroken only when $\alpha = 0$ and $\eta = 0$, respectively. The schematic band dispersions about the QBCP are shown in Fig.~\ref{Fig_dispersion_LS}. Since the two branches of dispersions must have opposite curvatures, we must have $0\leq  \alpha < 1$.

\section{Bare polarization bubble}
\label{Sec_bare-Pi}

In this section, we will calculate the one-loop bare polarization bubble, which means that we will not include any interaction line in the one-loop diagram. While evaluating the results, we will express the energies/frequencies in units of $\Lambda_0$, which is the inverse of the lattice cut-off. Hence, we use the variables $T = \tilde T \, \Lambda_0$, $ m = \tilde m \, \Lambda_0$, and $\mu = \tilde \mu \,\Lambda_0$. According to the same logic, we use the momentum and frequency scalings, such as $q = \tilde q\,\Lambda_0$ (where $q \equiv |\mathbf q|$) and $\omega = \tilde \omega \,\Lambda_0$. In short, we will use tilde to denote the dimensionless quantities obtained from their dimensionful counterparts scaled by appropriate powers of $\Lambda_0 $.

In the Matsubara space, the bare polarization function can be written as
\begin{align}
\label{eqmatpol}
\Pi(i  \, \omega_n  ,\mathbf{q})
=-
\int\frac{d^3\mathbf{p}}{(2 \, \pi)^3}
\int^\infty_{-\infty}\frac{d\epsilon_1}{2 \, \pi}\int^\infty_{-\infty}\frac{d\epsilon_2}{2 \, \pi}\,
\mathrm{Tr}\left [
A^{(0)}(\epsilon_1,\mathbf{p}) \, A^{(0)}(\epsilon_2,\mathbf{p}+\mathbf{q})
\right ]
\left(\frac{1}{\beta}
\sum_{ \Omega_m}
\frac{1}{i  \, \Omega_m-\epsilon_1}
\frac{1}{i  \, \Omega_m+i  \, \omega_n  -\epsilon_2}\right),
\end{align}
where $A^{(0)}(\epsilon,\mathbf{p})$ denotes the bare fermion spectral function, $\omega_n$ represents the fermionic Matsubara frequency, $ \Omega_m$ is the bosonic Matsubara frequency, and $\beta=1/T$ (setting the Boltzmann constant $k_B =1 $).

For the isotropic case, the bare fermion propagator in the Matsubara space takes the simple form
\begin{align}
G_0(i\,\omega_n,\mathbf{p})
 = \frac{1}{ \left( -i \, \omega_n  -\mu  \right) \Gamma_0
 +\mathbf{d}(\mathbf{p})\cdot {\mathbf \Gamma}} \,,
\label{Eq_G_0-mu}
\end{align}
leading to
\begin{align}
A^{(0)}(\epsilon,\mathbf{p})
=-\frac{ \pi \, \mathrm{sgn}(\epsilon +\mu)
\left [\epsilon +\mu+\mathbf{d}(\mathbf{p})\cdot {\mathbf \Gamma} \right ]
}
{|\mathbf{d}(\mathbf{p})|}
\sum \limits_{\zeta=\pm}\delta(\epsilon +\mu+\zeta  \, |\mathbf{d}(\mathbf{p})|)\,.\label{Eq_spectral_function_iso_free}
\end{align}

\begin{figure}
\centering
\subfigure[]{\includegraphics[width=0.45\textwidth]{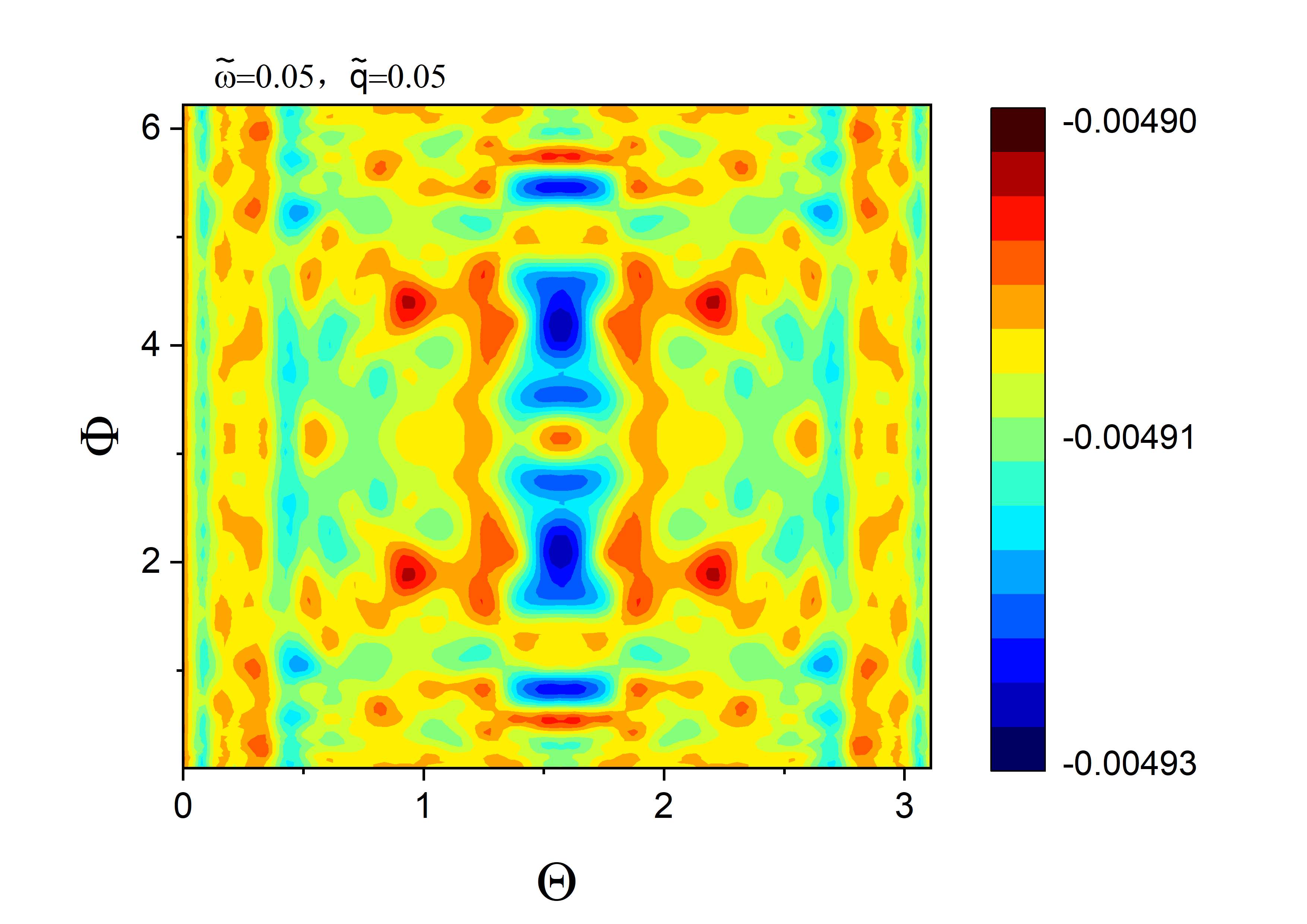}}\quad
\subfigure[]{\includegraphics[width=0.45\textwidth]{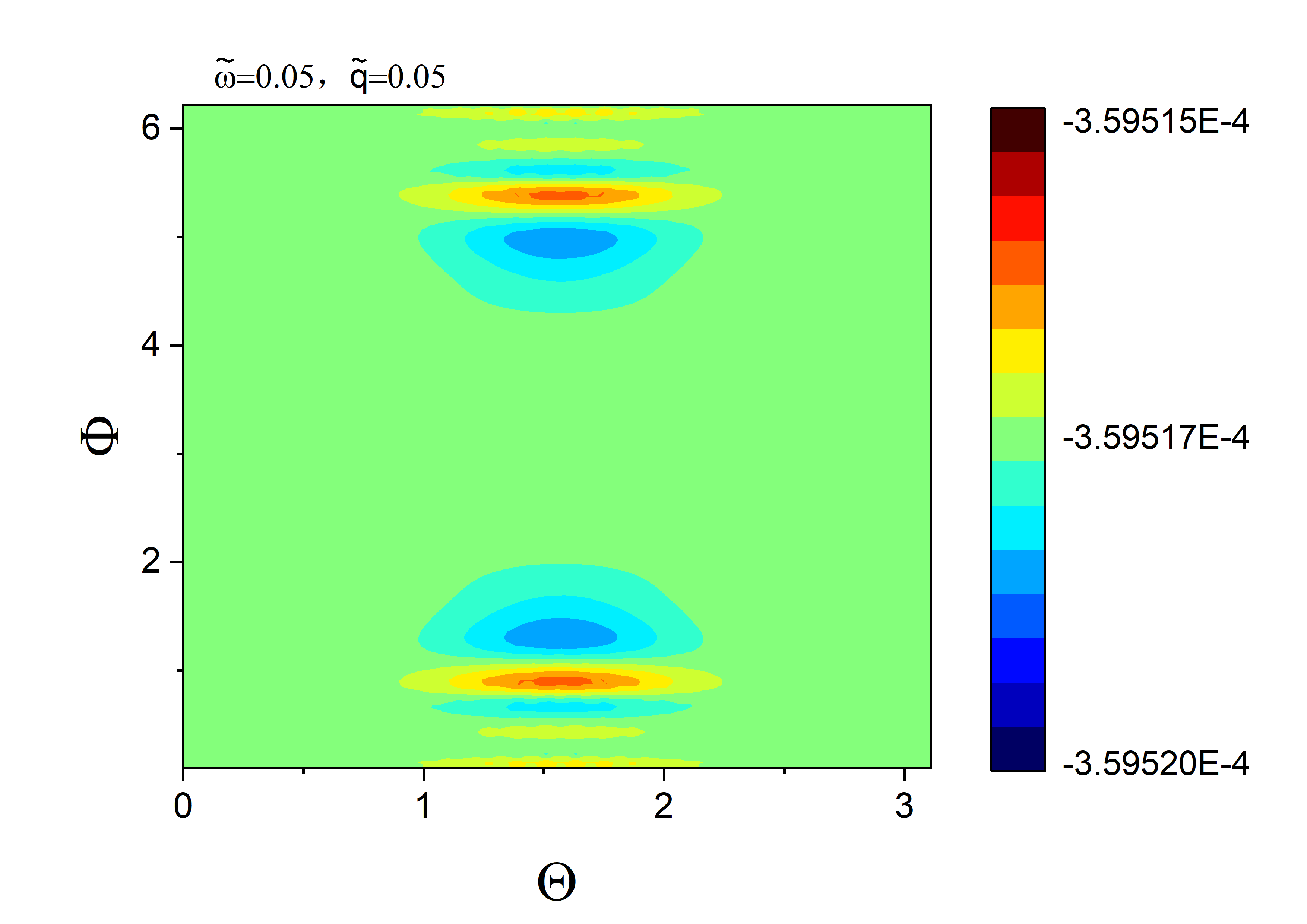}}\\
\subfigure[]{\includegraphics[width=0.45\textwidth]{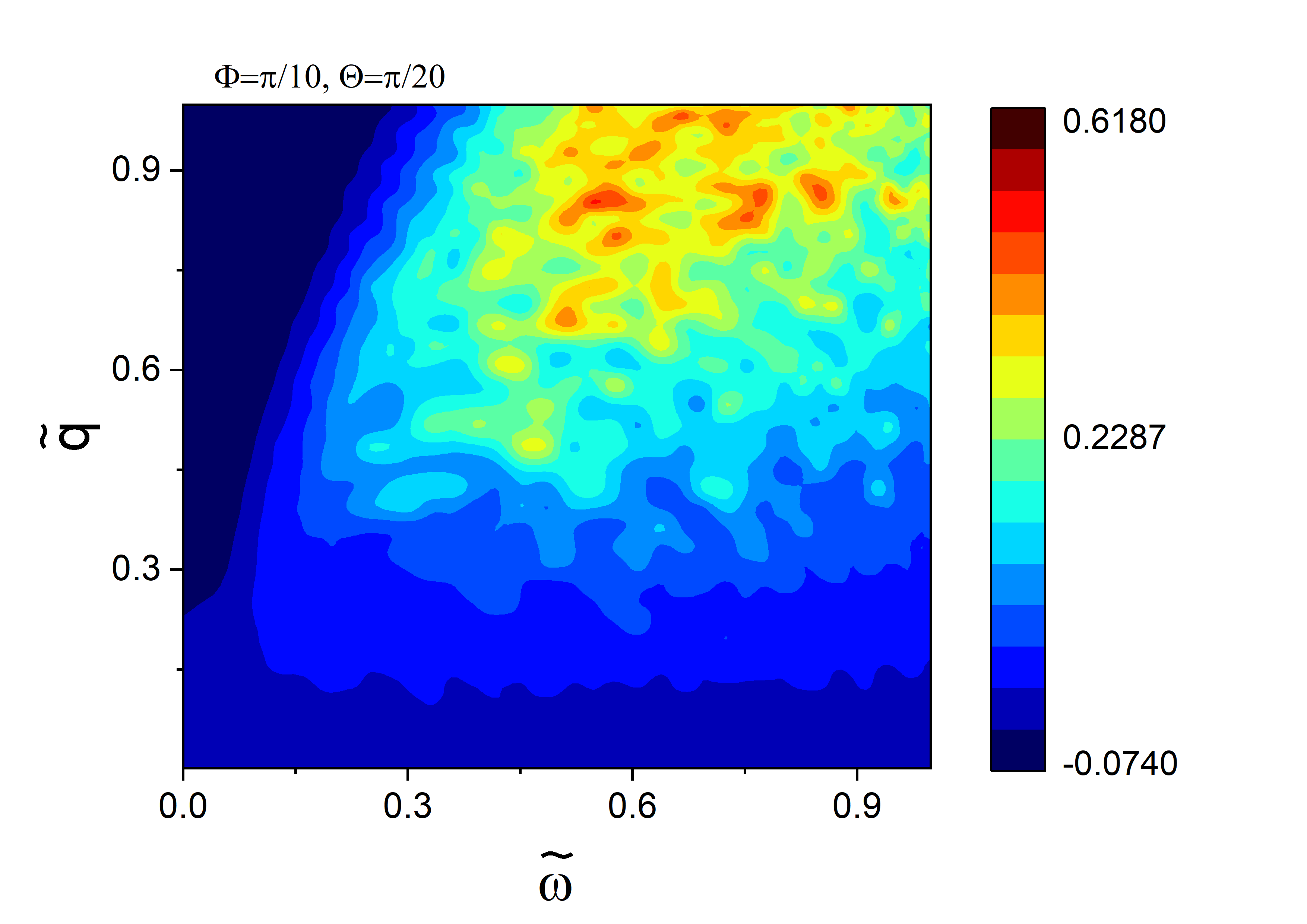}}\quad
\subfigure[]{\includegraphics[width=0.45\textwidth]{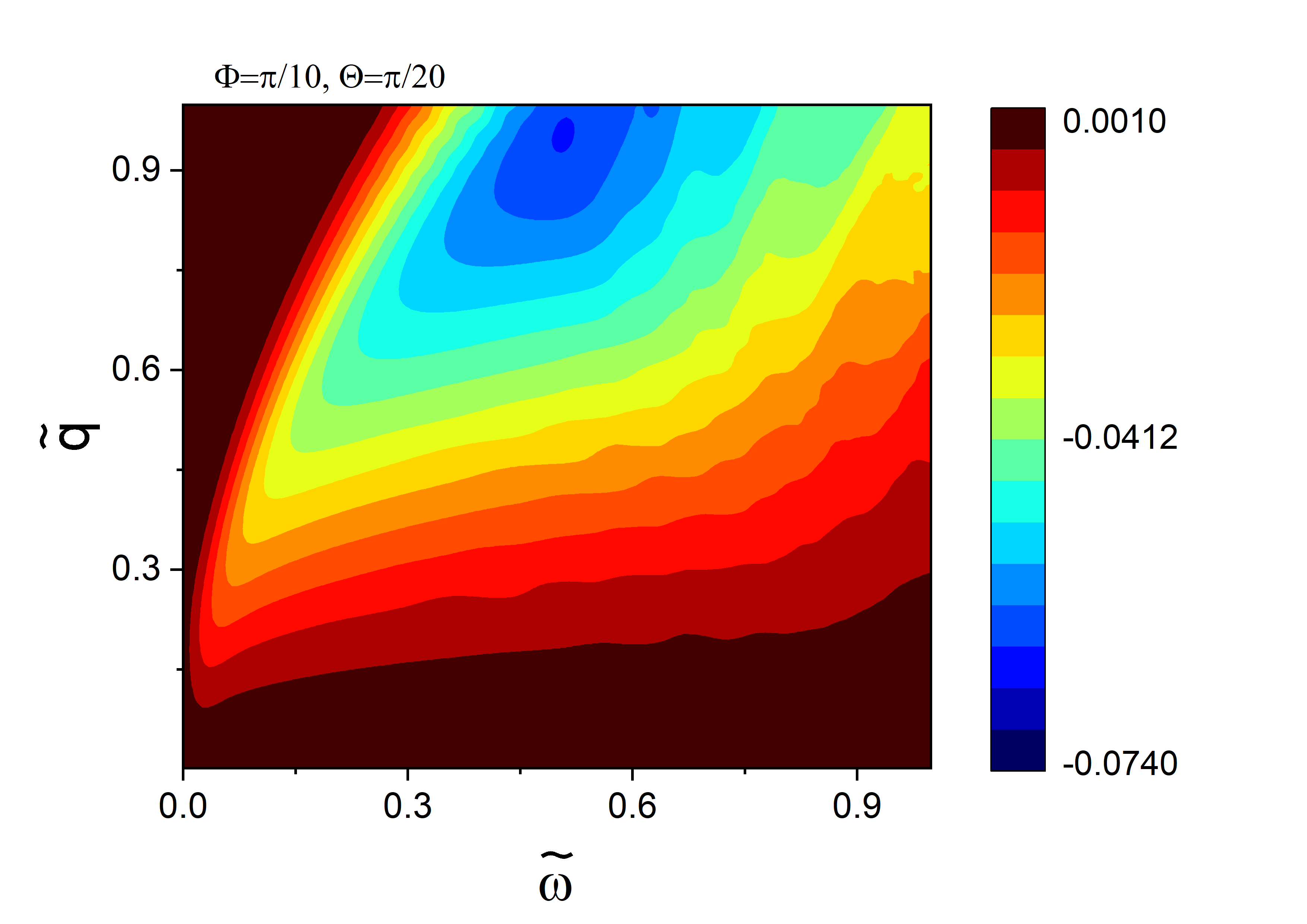}}\\
\caption{
Behaviour of the bare polarization function for an isotropic LSM with $\tilde T=0.1$ and $\tilde \mu=0.1$, and $\tilde \omega$ restricted to positive values:
Subfigures (a) and (b) show the contourplots of $\tilde m\,\Lambda_0^2 \,\mathrm{Re} \, \Pi^R$ and $\tilde m\,\Lambda_0^2 \, \mathrm{Im} \,\Pi^R$, respectively, as functions of the angular variables $\Theta$ and $\Phi$, for $\tilde \omega=0.05$ and $ \tilde q=0.05$.
Subfigures (c) and (d) show the contourplots of $\tilde m\,\Lambda_0^2 \, \mathrm{Re} \, \Pi^R $ and $ \tilde m\,\Lambda_0^2 \, \mathrm{Im}\,\Pi^R $, respectively, as functions of $\tilde \omega$ and $\tilde q$, with $\Phi=\pi/10$ and $ \Theta=\pi/ 20$.
\label{Fig_iso_Pi}
}
\end{figure}

Analytic continuation to real frequencies, after some tedious algebra (see Appendix~\ref{Subsec_A_Pi}), gives
\begin{align}
\label{eqpo1}
\frac{\mathrm{Re} \, \Pi^R(\tilde \omega, \tilde q,\Theta,\Phi)}
{\tilde m  \, \Lambda_0^2 }  =\sum \limits_{\zeta_1, \,\zeta_2 =\pm1 }
\mathcal{K}_{\zeta_1\, \zeta_2 }  (\tilde \omega, \tilde q,\Theta,\Phi)
\,, \quad
\frac{\mathrm{Im} \,\Pi^R(\tilde \omega, \tilde q,\Theta,\Phi)}
{\tilde m  \, \Lambda_0^2}
& = \sum \limits_{\zeta_1, \,\zeta_2 =\pm1 }
\mathcal{F}_{\zeta_1\, \zeta_2 }  (\tilde \omega, \tilde q,\Theta,\Phi) \,,
\end{align}
where $\Lambda_0$ is the inverse of the lattice cut-off, and the explicit expressions for the functions $\mathcal{K}_{\zeta_1\, \zeta_2 } $ and $ \mathcal{F}_{\zeta_1\, \zeta_2 } $ have been shown in
Eq.~\eqref{Eq_iso-KF}. Here we have used the variables $\lbrace \tilde q,\Theta,\Phi \rbrace $ to denote the spherical polar coordinates, which are related to the Cartesian coordinates as
\begin{align}
q_x=   q\sin\Theta\cos\Phi\,, \quad
q_y=   q\sin\Theta\sin\Phi \,, \quad
q_z=   q \cos\Theta \,.
\label{Eq_momenta-transformation}
\end{align}

For the anisotropic case, the bare fermionic propagator (in the Matsubara space) takes a bit more complicated form, captured by
\begin{align}
G_0(i  \, \omega_n  ,\mathbf{p})
&=\frac{1}{
\left( -i  \, \omega_n  + \frac{\alpha \, \mathbf{p}^2}  {2\,m}
-\mu \right) \Gamma_0
+
{\mathbf d}({\mathbf{p}}) \cdot {\mathbf \Gamma} +
\eta \,\sum \limits_{a=1}^5 s_a  \,d_a (\mathbf{p}) \,\Gamma^a }\,,
\label{Eq_aniso-G_0}
\end{align}
such that the fermion spectral function is now given by
\begin{align}
 A^{(0)}(\epsilon )
 &=-\pi  \, \mathrm{sgn} \Bigl(\epsilon +\mu- \frac{\alpha \,
 \mathbf{p}^2} {2\, m}\Bigr)
\left [\epsilon +\mu- \frac{\alpha\, \mathbf{p}^2} {2 \, m}
+ {\mathbf d} ({\mathbf{p}}) \cdot {\mathbf \Gamma} +\eta \,\sum \limits_{a=1}^5
s_a \,d_a(\mathbf{p}) \,\Gamma^a \right ]
\nn & \hspace{0.5 cm} \times
\frac{\sum \limits _{\zeta=\pm}
\delta\bigg (  \epsilon +\mu- \frac{ \alpha \, \mathbf{p}^2} {2\,m}
+\zeta \,\sqrt{(1+\eta) \, {\mathbf d}^2({\mathbf{p}})
+2 \, \eta \, \sum \limits_{a' =1}^5 s_{a'}\, d^2_{a'} (\mathbf{p})}
\, \bigg )}
{\sqrt{(1+\eta) \,{\mathbf d}^2 (\mathbf{p})+2\,\eta \,
\sum \limits_{b =1}^5 s_b \,d^2_b (\mathbf{p})}}.\label{Eq_spectral_function_aniso_free}
\end{align}
The real and imaginary parts of the retarded polarization in this case are found to be
\begin{align}
\label{eqpo2}
\frac{\mathrm{Re} \, \Pi^R(\tilde \omega, \tilde q,\Theta,\Phi)}
{\tilde m  \, \Lambda_0^2 }  =\sum \limits_{\zeta_1, \,\zeta_2 =\pm1 }
\mathcal{M}_{\zeta_1\, \zeta_2 }  (\tilde \omega, \tilde q,\Theta,\Phi)
\,, \quad
\frac{\mathrm{Im} \,\Pi^R(\tilde \omega, \tilde q,\Theta,\Phi)}
{\tilde m  \, \Lambda_0^2}
& = \sum \limits_{\zeta_1, \,\zeta_2 =\pm1 }
\mathcal{J}_{\zeta_1\, \zeta_2 }  (\tilde \omega, \tilde q,\Theta,\Phi)\,,
\end{align}
where the explicit expressions for the functions $\mathcal{M}_{\zeta_1\, \zeta_2 } $ and $ \mathcal{J}_{\zeta_1\, \zeta_2 } $ have been shown in Eq.~\eqref{eq_MJ} of Appendix\ref{Subsec_B_Pi}. As in the isotropic case, we have used the spherical polar coordinates for these final expressions.

\begin{figure}
\centering
\subfigure[]{\includegraphics[width=0.45\textwidth]{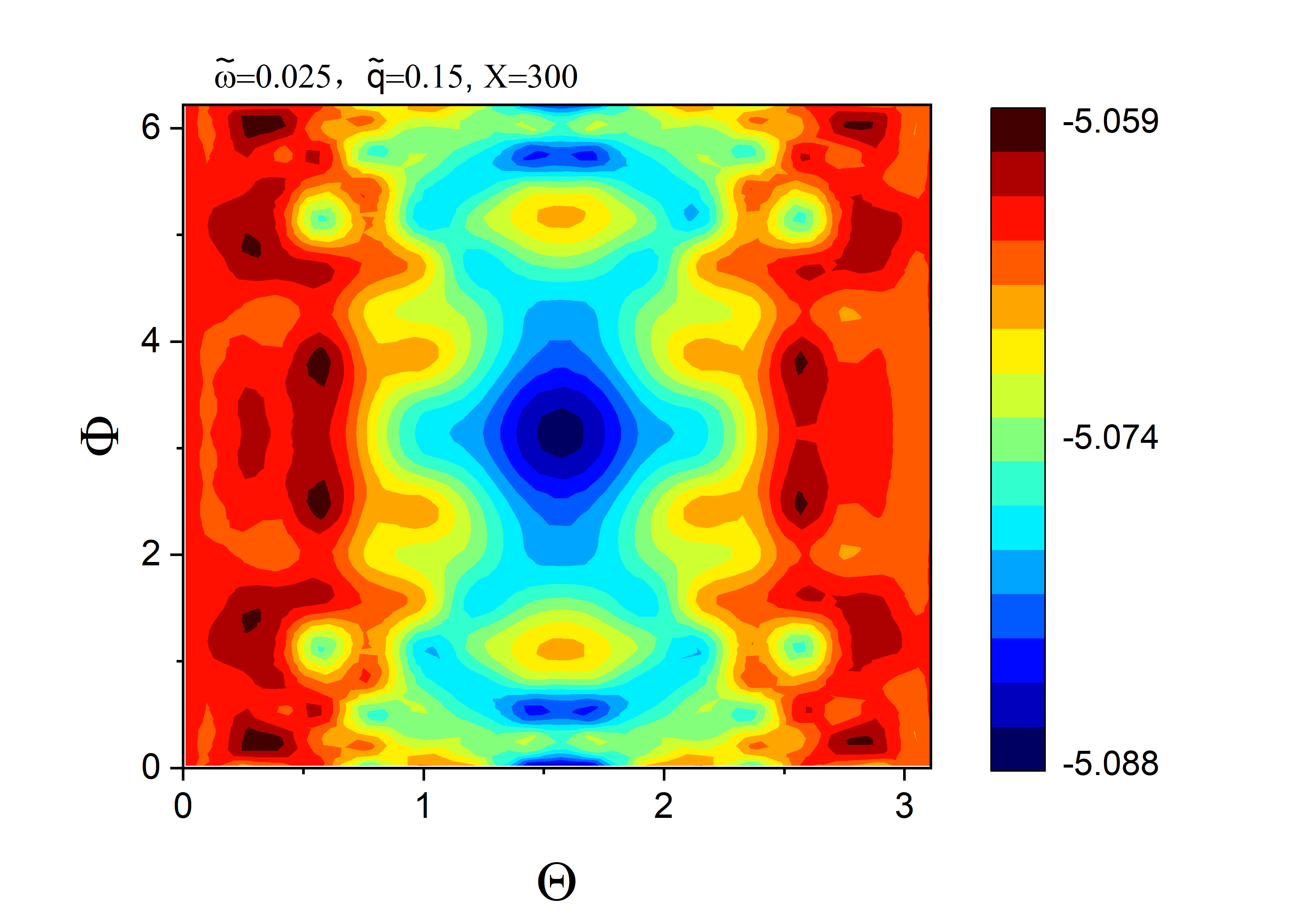}}\quad
\subfigure[]{\includegraphics[width=0.45\textwidth]{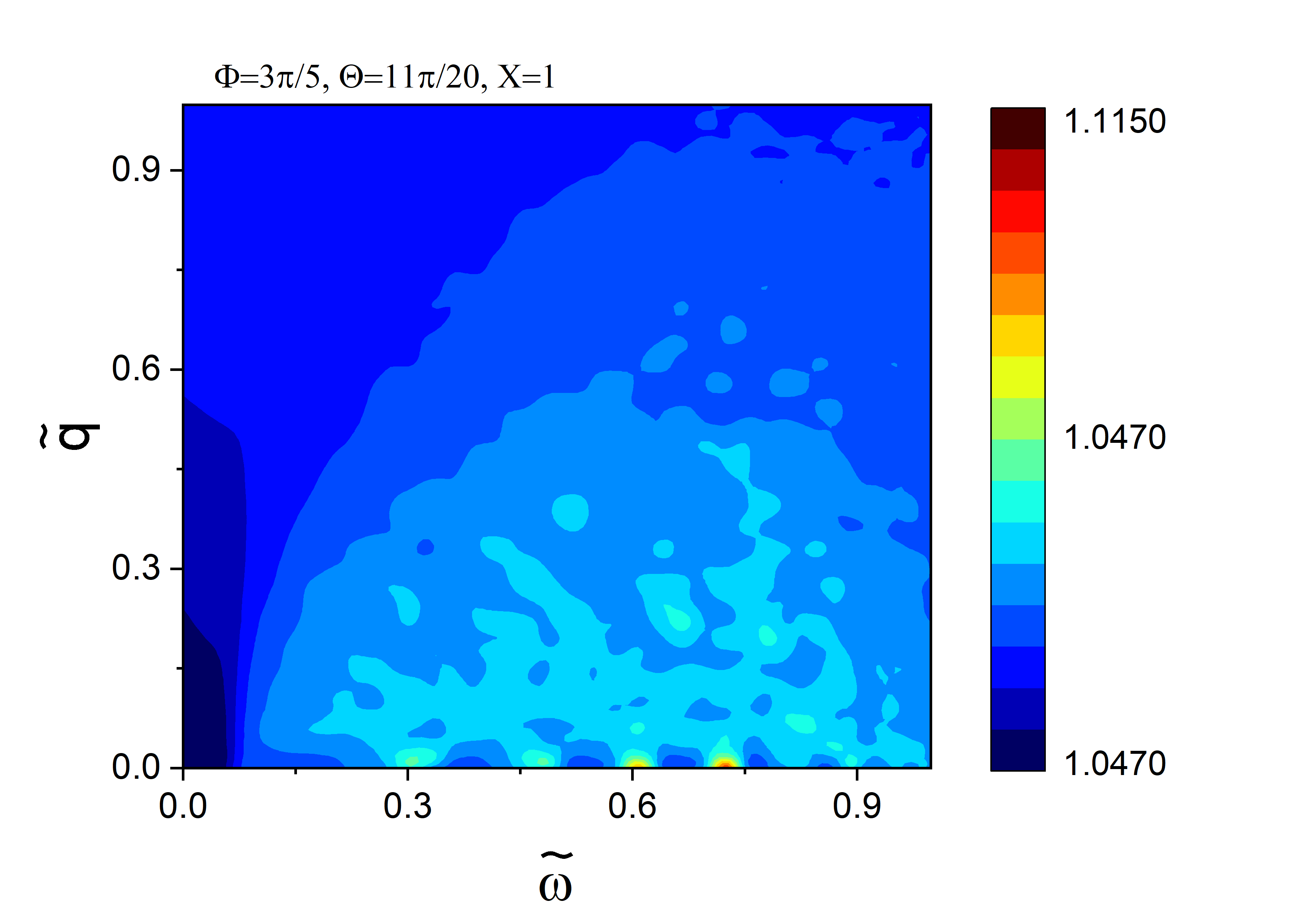}}\\
\subfigure[]{\includegraphics[width=0.45\textwidth]{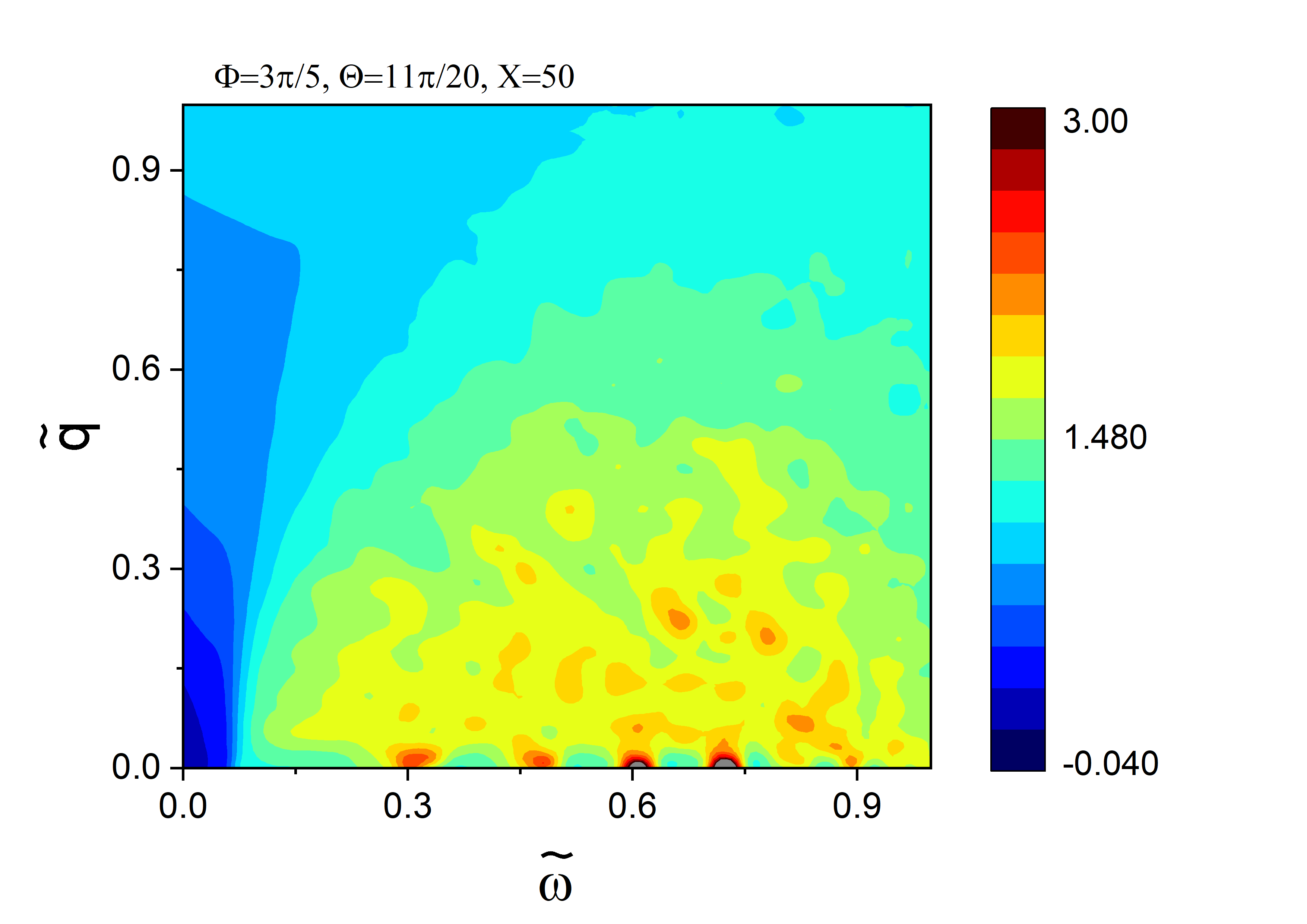}}\quad
\subfigure[]{\includegraphics[width=0.45\textwidth]{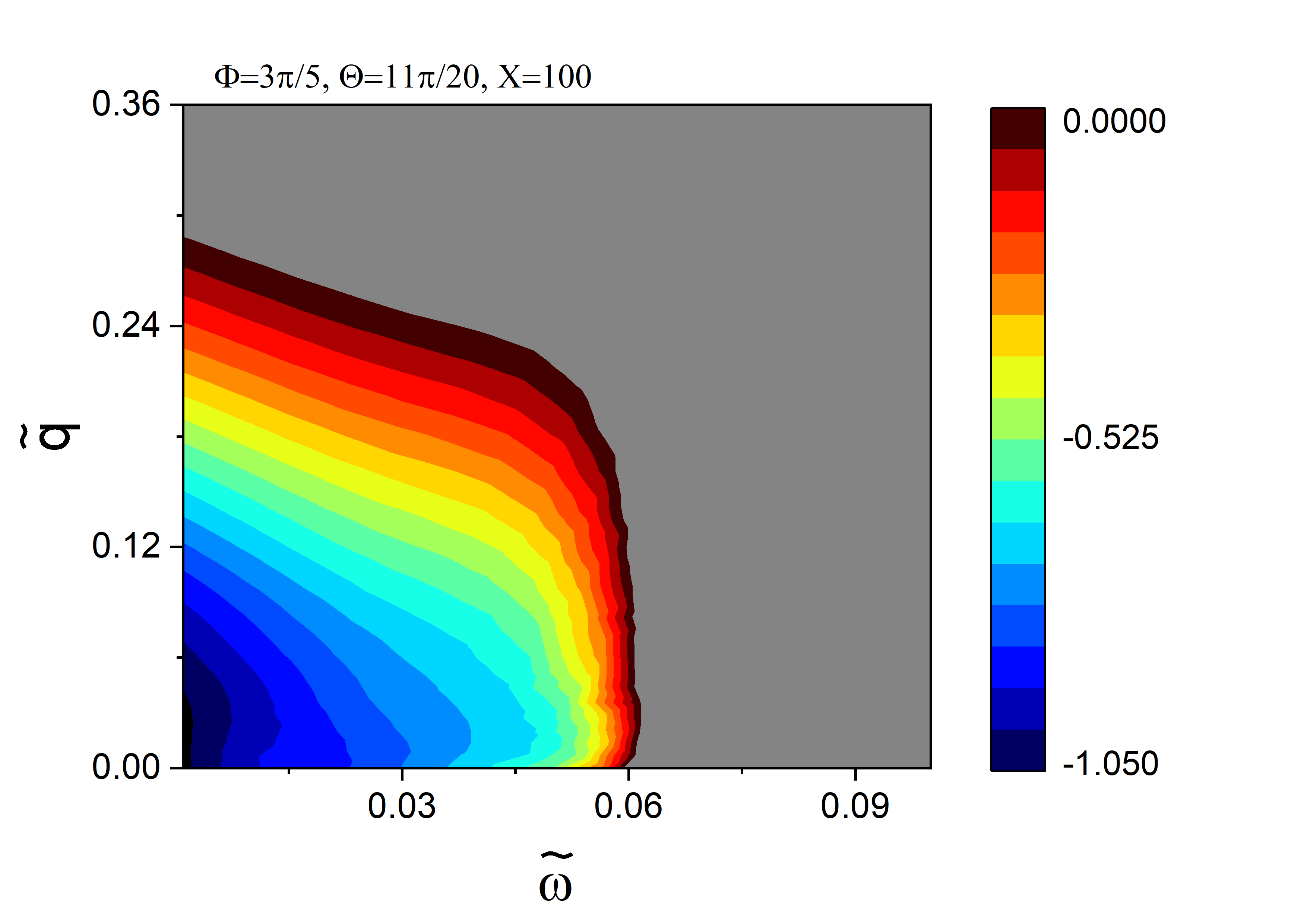}}
\caption{
Behaviour of $\mathrm{Re}\, \mathcal{E}_{\mathrm{eff}}$ for isotropic LSM with $\tilde T=0.1$ and $\tilde \mu=0.1$, and $\tilde \omega$ restricted to positive values: Subfigure (a) captures the dependence on the angular variables $\Theta$ and $\Phi$, for $\tilde \omega=0.025$, $\tilde q=0.15$, and $X =300$.
Subfigures (b), (c), and (d) show the dependence on $\tilde \omega$ and $\tilde q$, with $\Phi= 3\,\pi/ 5 $ and $\Theta= 11 \,\pi /20$, for $X= 1$, $X= 50$, and $X= 100$, respectively.
\label{Fig_iso_Re_E}
}
\end{figure}

\section{Criteria for emergence of plasmons}
\label{Sec_Results}

We now include the effect of screened Coulomb interactions.
Within RPA, this is captured by the effective interaction~\cite{Kozii-Fu2018PRB,Mandal2019AP}
\begin{align}
\label{eqveff}
V^R(\omega,\mathbf q)
=\frac{V_0( \mathbf q)} {1+V_0(\mathbf q) \, N_f \,\Pi^R(\omega, \mathbf q)}\,, \quad
V_0(\mathbf q) =\frac{\alpha_0 } {q^2}\equiv
\frac{4 \, \pi \, e^2}{\varepsilon \,q^2}\,,
\end{align}
where $N_f$ is the number of fermion flavors,
$V_0$ denotes the bare Coulomb interaction in a material with a dielectric constant
$\varepsilon$, and $e$ is the electron charge. In the literature, the coefficient $e^2/\varepsilon$
is usually known as the effective fine structure constant.

The effective interaction $V^R(\omega, \mathbf q)$ can be interpreted as the photon propagator in
the given medium. If the function has any pole, that will indicate the emergence of a plasmon mode, resulting from collective photon-electron excitations.
The dispersions of the plasmons are thus given by
$ \left [
1+V_0(\mathbf q)\,N_f \,\Pi^R(\omega,\mathbf q ) \right ] = 0 $, which are the zeros of the dielectric function
\begin{align}
\mathcal{E}(\omega, \mathbf q)
\equiv \varepsilon
\left[1+V_0( \mathbf q) \, N_f \, \Pi^R(\omega, \mathbf q)\right].
\label{Eq_diele}
\end{align}
For inconvenience, we break up the dielectric constant into real and imaginary parts as follows:
\begin{align}
\label{eqPidielec}
\frac{\mathcal{E}(\omega, \mathbf q)}{\varepsilon}
\equiv\mathrm{Re}\,\mathcal{E}_{\mathrm{eff}}(\omega, \mathbf q)
+i\,{\mathrm{Im}} \,\mathcal{E}_{\mathrm{eff}}(\omega, \mathbf q)\,,\quad
\mathrm{Re}\, \mathcal{E}_{\mathrm{eff}}(\omega, \mathbf q)
=1+\frac{N_f \, \alpha_0 \,\mathrm{Re}\, \Pi^R(\omega, \mathbf q)} {q^2} \,,\quad
{\mathrm {Im} } \,\mathcal{E}_{\mathrm{eff}}(\omega, \mathbf q)
=\frac{N_f\,\alpha_0 \,{\mathrm {Im}} \,\Pi^R(\omega, \mathbf q)}{q^2} \,.
\end{align}
The expression for the real part tells us that in order to have the possibility for the effective
photon propagator $V^R $ to have a pole, we must have regions where $\mathrm{Re}\, \Pi^R < 0 $.
On the other hand, the imaginary part ${\mathrm {Im} } \,\mathcal{E}_{\mathrm{eff}}$ quantifies the decay rate of the plasmon mode.
In all the following discussions, we use the definition
\begin{align}
\label{eqXdef}
X \equiv N_f \, \alpha_0 \,\tilde m\,,
\end{align}
which is a material-dependent parameter, and we vary its value to investigate the regimes favourable for getting plasmons. We would like to clarify that although $\tilde m $ is the scaled mass (and depends on the material we choose to employ in an experimental set-up), the number of fermion flavours $N_f$ is a theoretical parameter, as the examples of materials mentioned in the paper will harbour a single band-touching point (implying $N_f =1$). We keep $N_f$ generic mainly for the reason that for the case of unscreened (or long-ranged) Coulomb interactions, one has to use some controlled approximation (as RPA fails in those cases). Dimensional regularization and large-$N_f$ expansion \cite{Abrikosov,Moon2013PRL} are two alternate methods of controlled approximation. Hence, we have considered here a setting with $N_f$ independent fermionic flavours, although the physical case corresponds to $N_f = 1$. In future, if we plan to investigate the emergence of plasmons for unscreened Coulomb interaction, keeping $N_f$ generic will turn out to be handy.

\begin{figure}
\centering
\subfigure[]{\includegraphics[width=0.45\textwidth]{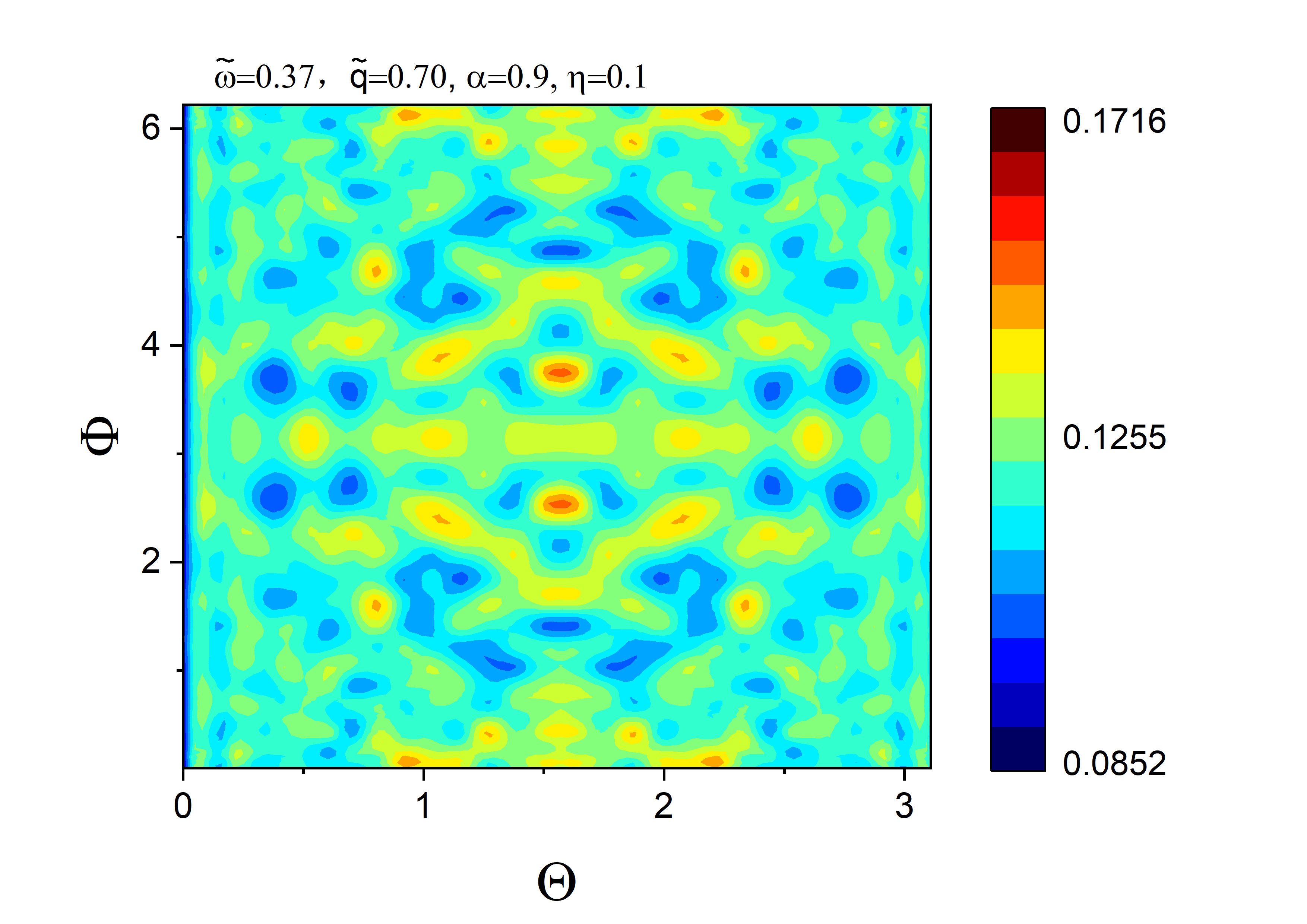}}\quad
\subfigure[]{\includegraphics[width=0.45\textwidth]{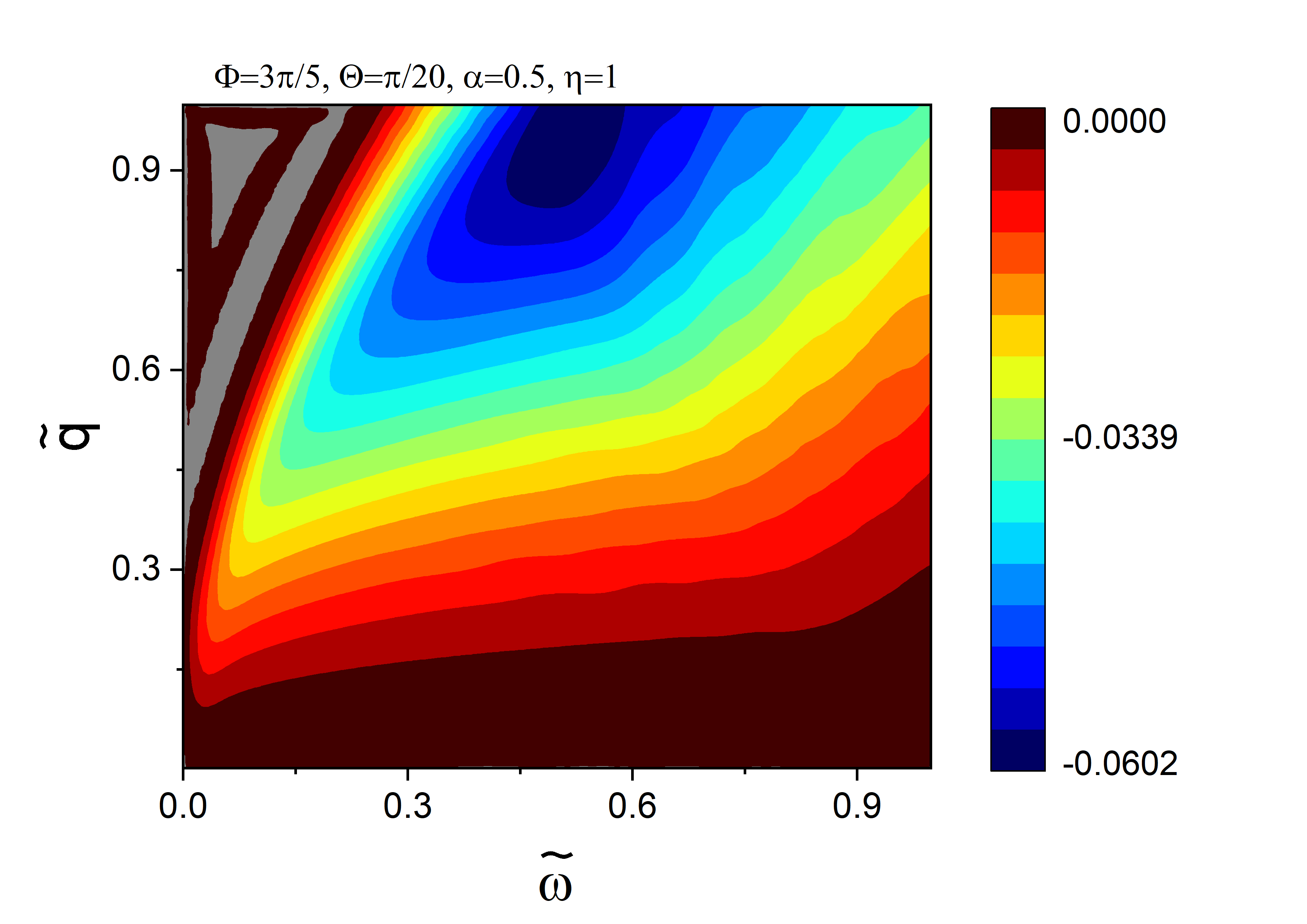}}
\caption{Behaviour of the bare polarization function for a generic LSM with $\tilde T=0.5 $, $\tilde \mu=0.1$, and $\tilde \omega$ restricted to positive values:  Subfigure (a) shows $\tilde m\,\Lambda_0^2 \,\mathrm{Re} \, \Pi^R$ as a function of $\Theta$ and $\Phi$, for $\alpha=0.9$, $\eta=0.1$, $\tilde \omega =  0.37$, and $\tilde q =0.7 $. Subfigure (b) shows $\tilde m\,\Lambda_0^2 \,\mathrm{Im} \, \Pi^R$ as a function of $\tilde \omega $ and $\tilde q$, for $\alpha=0.5$, $\eta=1$, $\Phi= 3\,\pi/5$, and $\Theta= 1\,\pi/20$.
\label{Fig_aniso-Pi}}
\end{figure}

\subsection{Isotropic case}
\label{Subsec_iso}

We show the behaviour of $\Pi^R$ for some chosen parameter regimes in Fig.~\ref{Fig_iso_Pi}, by evaluating the associated integrals numerically. In all the plots, we use the values $\tilde T =0.1 $ and $\tilde \mu=0.1$.
We have checked that the qualitative behaviour for other parameter regimes (varying $\tilde T$ and $\tilde \mu$ in the range $ [0,1]$) is similar to what we have presented here, and the basic conclusions are indeed unchanged
irrespective of the concrete values of $\tilde T$ and $\tilde \mu$ chosen. From the representative plots, we find that the behaviour is quite insensitive to the external angular variables, and is mostly affected by the magnitudes of $\tilde q$ and $\tilde \omega $.
Fig.~\ref{Fig_iso_Pi}(c) shows extended regions where $  \mathrm{Re}\, \Pi^R(\omega,\mathbf q)< 0$, and therefore have the potential to contribute to the emergence of plasmons.

We have checked extensively (covering other potential parameter regimes which are not shown here) that the frequency and momentum dependence of both the real and imaginary parts of the retarded polarization function are insensitive
to the specific values of $T$ and $\mu$, as long as at least one of them is nonzero.
Consequently, the regions with $  \mathrm{Re}\, \Pi^R(\omega,\mathbf q)< 0$ are very robust against the variations of both $T$ and $\mu$.
In the regions where $  \mathrm{Re}\, \Pi^R(\omega,\mathbf q)< 0$, we find that $ {\mathrm { Im}} \,\Pi^R(\omega, \mathbf q)$ has very small values for its magnitude (compared to that of $  \mathrm{Re}\, \Pi^R$), which indicates that the damping in these regions is likely to be very small (almost negligible), contributing to a long lifetime of the emergent plasmons.

Let us now investigate the features of the real part of the dielectric function $\mathcal E$. Since $\mathcal E$ is related to $\Pi^R $ by Eq.~\eqref{Eq_diele}, its susceptibility to the changes in $\Phi$ and $\Theta$ is expected to be the same as that for $\Pi^R $. This is precisely what is seen in Fig.~\ref{Fig_iso_Re_E}(a).
Consequently, the key variable in substantially affecting the behaviour of $\mathcal E$ is the material-dependent parameter $X$. The contours in Fig.~\ref{Fig_iso_Re_E}(b)--(d) show that while for small values of $X$,
$  \mathrm{Re}\, \mathcal{E}_{\mathrm{eff}}(\omega, \mathbf q)$ remain positive (i.e., do not possess any zeros),
we obtain regions with zero and negative values when $X$ is cranked up to higher values of the order of $50$.
Additionally, the regimes where $  \mathrm{Re}\, \mathcal{E}_{\mathrm{eff}}(\omega, \mathbf q)$ vanish
get extended as $X$ is increased.
In analogy with the polarization function, we have also confirmed that the qualitative behaviour
of $  \mathrm{Re}\, \mathcal{E}_{\mathrm{eff}}(\omega, \mathbf q)$ is stable against the
variations of $T$ and $\mu$, as long as at least one of them is nonzero.

To summarize our observations, finite $T$ and / or $\mu$, and a suitable
the material-dependent parameter $X$, create the possibility of obtaining plasmon poles in the isotropic LSMs. The former is a necessary requirement, while the latter provides a crucial tuning parameter.

\begin{figure}
\centering
\subfigure[]{\includegraphics[width=0.45\textwidth]{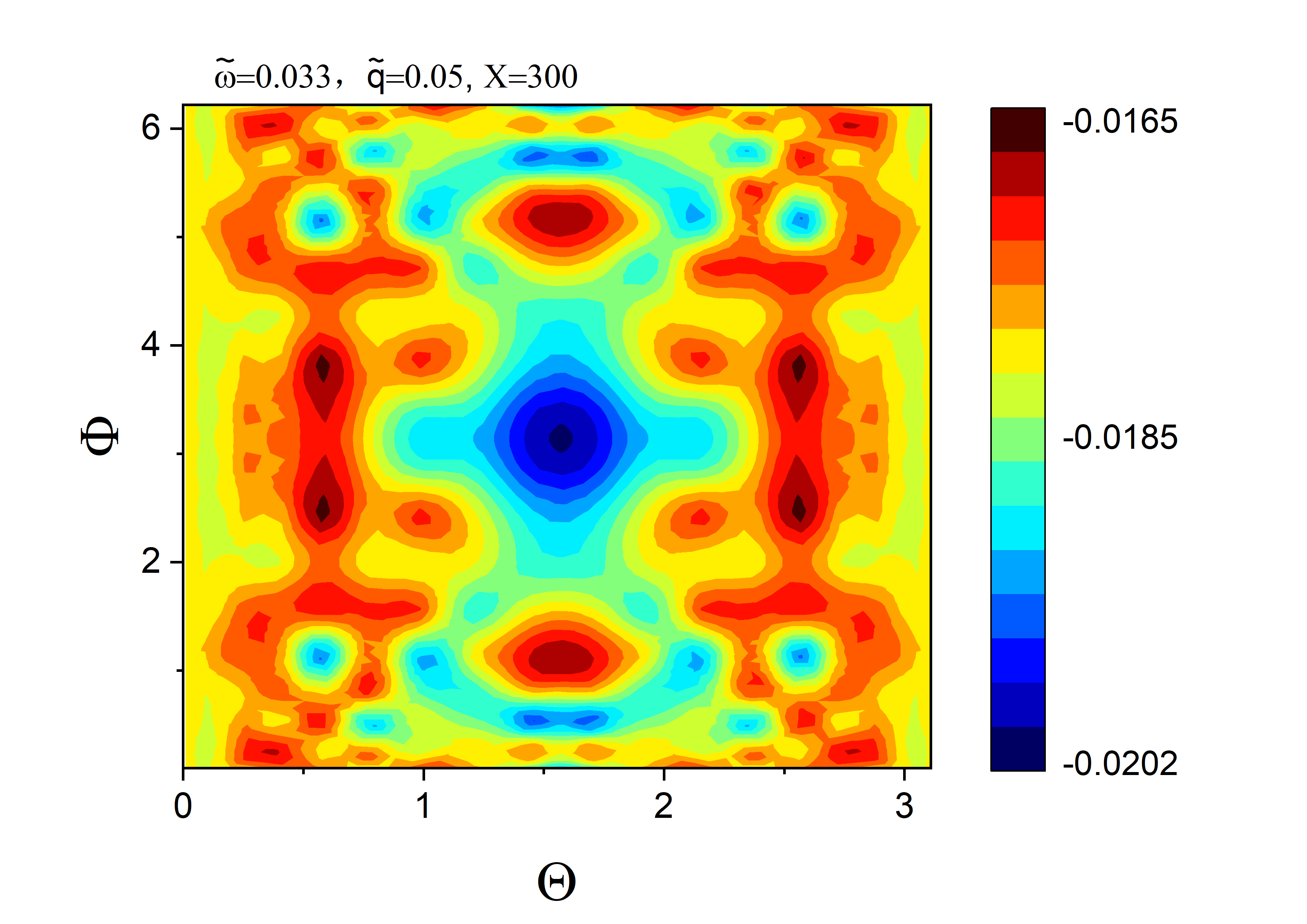}}\quad
\subfigure[]{\includegraphics[width=0.45\textwidth]{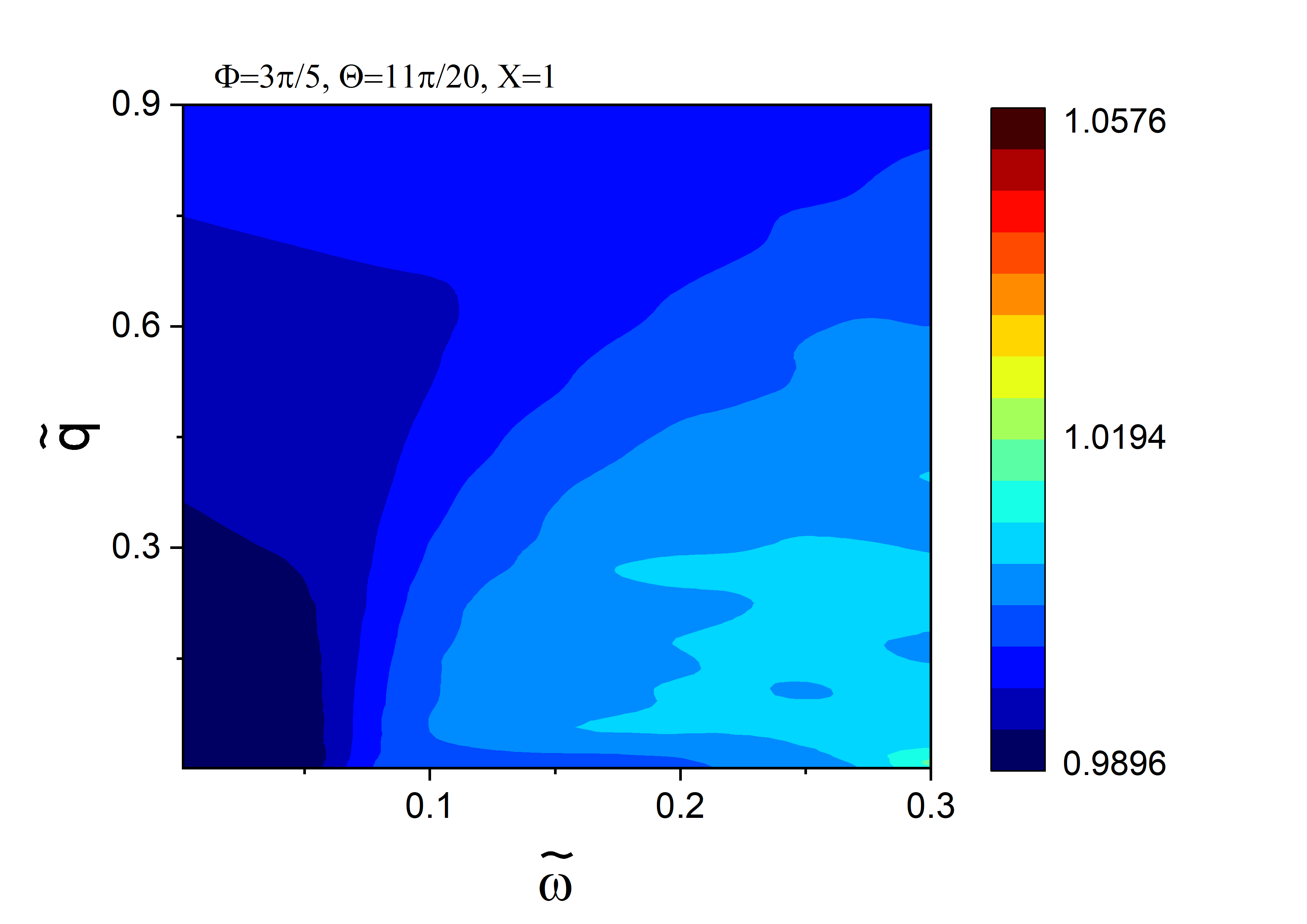}}\\
\subfigure[]{\includegraphics[width=0.45\textwidth]{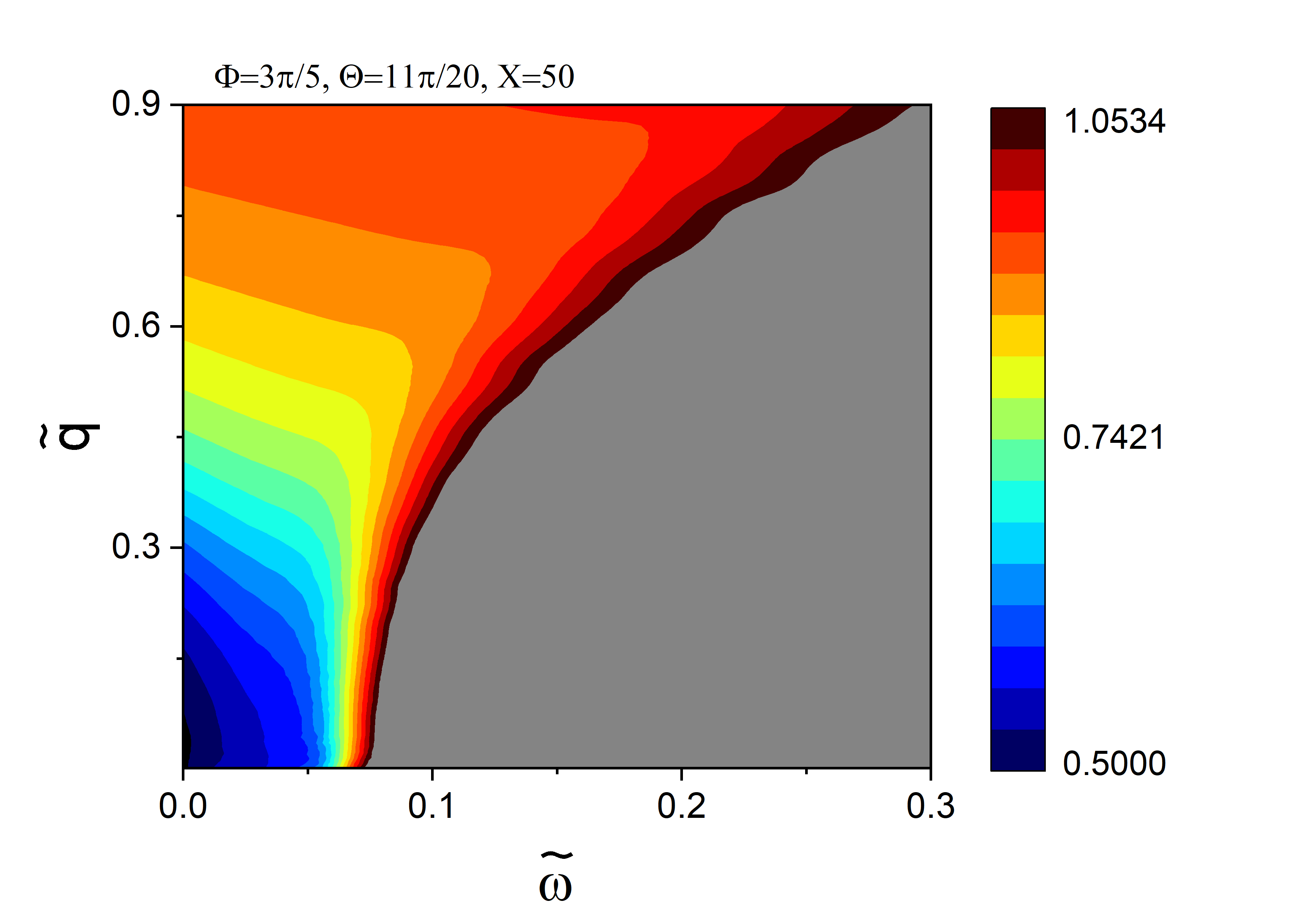}}\quad
\subfigure[]{\includegraphics[width=0.45\textwidth]{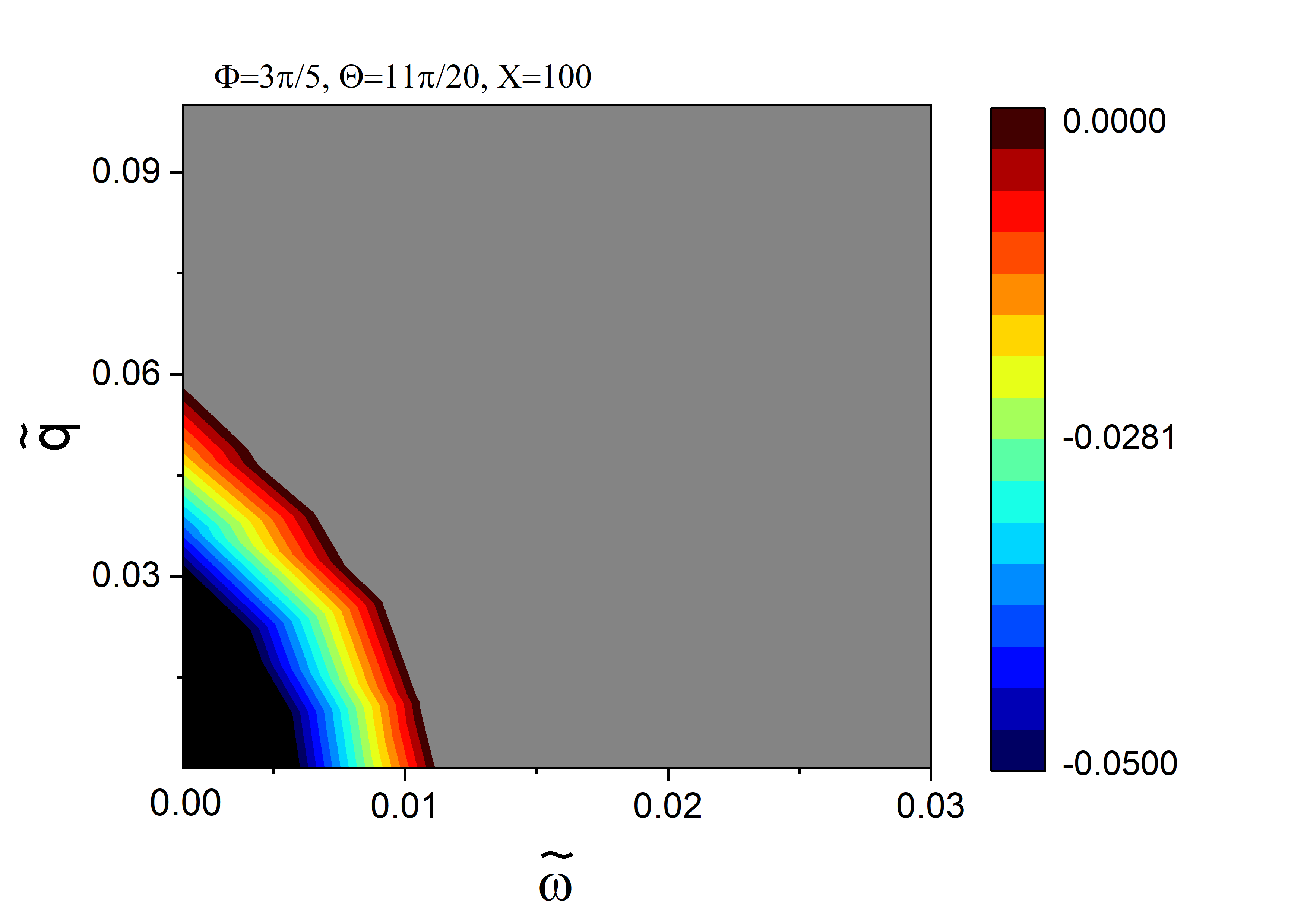}}
\caption{Behaviour of $\mathrm{Re}\, \mathcal{E}_{\mathrm{eff}}$ for a generic LSM
with $\tilde T=0. 5$, $\tilde \mu=0.1$, and $\tilde \omega$ restricted to positive values: Subfigure (a) captures the dependence on the angular variables $\Theta$ and $\Phi$, for $\alpha=0.1$, $\eta=1$, $\tilde \omega= 0.033$, $\tilde q=0.05$, and $X = 300$.
Subfigures (b), (c), and (d) show the dependence on $\tilde \omega$ and $\tilde q$, with $\Phi= 3\,\pi/ 5 $ and $\Theta= 11 \,\pi /20$, for $X= 1$, $X= 50$, and $X= 100$, respectively.
\label{Fig_aniso_Re_E}}
\end{figure}

\subsection{Anisotropic and band-mass asymmetric cases}
\label{Subsection_aniso}

In this subsection, we investigate the effects of nonzero $\alpha$ and/or $\eta$,
and show the results for $\tilde T=0.5$ and $\tilde \mu=0.1$. We have sampled the behaviour at other values of  $\tilde T$ and $\tilde \mu$ in the range $ [0,1]$, which is not shown here, as it does not show significant changes. Fig.~\ref{Fig_aniso-Pi} shows some plots in the regime $\alpha>\eta$. Although not shown here, we have extensively checked the characteristics for $\alpha=\eta$ and $\alpha<\eta$ as well, and have found that they are analogous to the cases presented here. Although the values of both the real and imaginary parts of the polarization function are slightly altered by the presence of finite anisotropic parameters, we have found that they are almost insensitive to the angular variables $\Phi$ and $\Theta$. A representative plot is shown in Fig.~\ref{Fig_aniso-Pi}.
We also observe that the magnitude of $ {\mathrm { Im}} \,\Pi^R(\omega, \mathbf q)$ is much smaller than
that of $  \mathrm{Re}\, \Pi^R(\omega, \mathbf q)$, in the regimes where the latter is negative.

\begin{figure}
\centering
\subfigure[]{\includegraphics[width=0.45\textwidth]{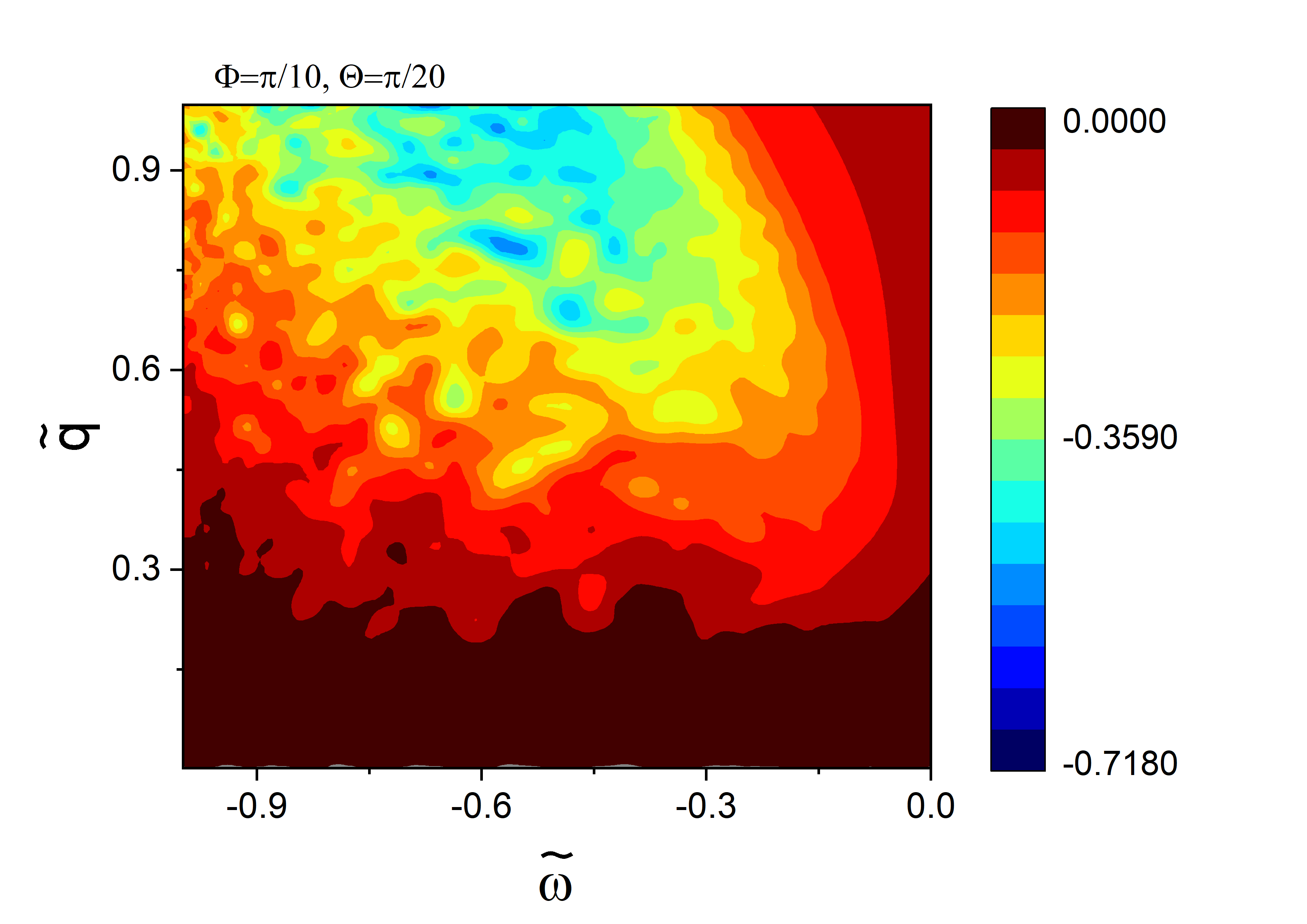}}\quad
\subfigure[]{\includegraphics[width=0.45\textwidth]{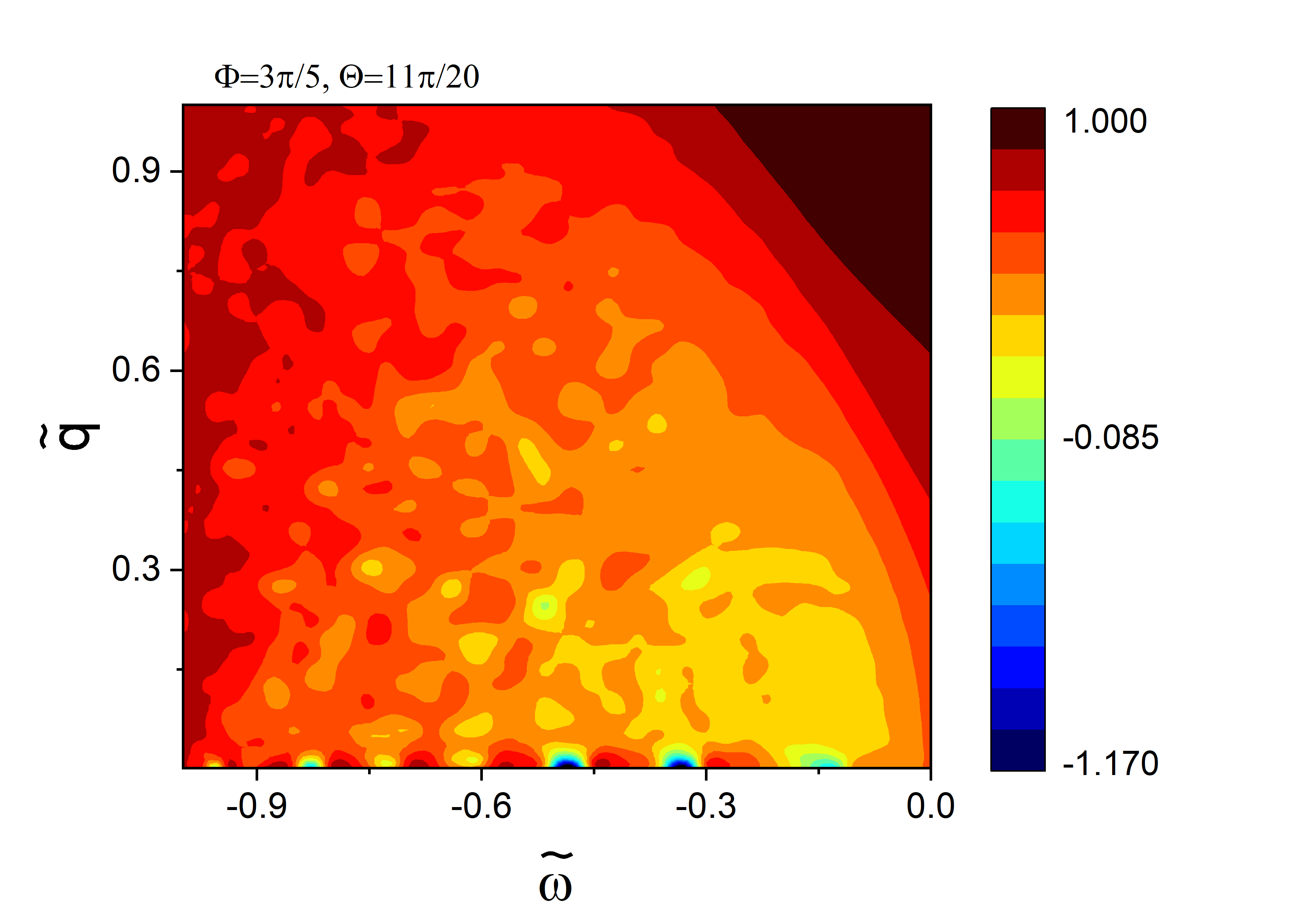}}\\
\caption{\label{Fig_aniso_Pi-E-lower}
Behaviour of (a) $\tilde m\,\Lambda_0^2 \, \mathrm{Re} \, \Pi^R $
and (b) $\mathrm{Re}\, \mathcal{E}_{\mathrm{eff}}$ for a generic LSM with $\tilde T=0.5$, $\tilde \mu=0.1$, $X=100 $, and $\tilde \omega$ restricted to negative values. The values of the other parameters are the same as used in Fig.~\ref{Fig_aniso-Pi} and we have shown only two cases with the aim of comparing the corresponding behaviour shown in those plots (which represent the $\tilde \omega>0$ case). We have not shown the plots for $\tilde \omega<0$ that are the counterparts of the remaining subfigures of Figs.~\ref{Fig_iso_Pi} and~\ref{Fig_iso_Re_E}, because the basic tendency is similar.}
\end{figure}

The real part of the dielectric function $\mathcal E$ follows the trends seen for $\Pi^R$
(cf. Fig.~\ref{Fig_aniso_Re_E}), as far as the dependence on
$\tilde T$, $\tilde \mu$, $\Theta $, and $\phi$ are concerned,
due to the relation in Eq.~\eqref{Eq_diele}.
More specifically, we note that the behaviour of $\mathrm{Re}\, \mathcal E$ is quite insensitive to these variables.
As a result, the key variable which can potentially affect the characteristics of $\mathrm{Re} \,\mathcal E$ drastically is the material-dependent parameter $X$. The plots in Fig.~\ref{Fig_aniso_Re_E}(b)--(d) show that while for small values of $X$, $  \mathrm{Re}\, \mathcal{E}_{\mathrm{eff}}(\omega, \mathbf q)$ remains positive (i.e., does not possess any zeros),
regions with zero and negative values appear when $X$ is tuned to higher values of the order of $50$.
Additionally, the regimes where $  \mathrm{Re}\, \mathcal{E}_{\mathrm{eff}}(\omega, \mathbf q)$ vanish
get extended as $X$ is increased. For the sake of completeness, we also show the behaviour of $  \mathrm{Re} \, \Pi^R $ and $\mathrm{Re}\, \mathcal{E}_{\mathrm{eff}}$ in Fig.~\ref{Fig_aniso_Pi-E-lower}, when $\tilde \omega$ is restricted to negative values.

We would like to point out that we have examined the data obtained for $\eta $-values ranging from 0 to 100, in tandem with various choices for $\alpha \in [0,1)$. The variations due to anisotropy are seen to saturate as $\eta$ is raised to $\sim 100$ and, even then, those variations are not significantly different from the qualitative features outlined above. On the basis of our extensive numerical explorations, we thus conclude that cubic anisotropy and anisotropic band-masses do not substantially change the nature of the poles in the effective interaction, compared to the isotropic case discussed in the previous subsection.

\section{Signature of plasmons from inelastic scattering rates}
\label{Sec_scattering-rate}

In this section, we calculate the inelastic electron scattering rate $\tau^{-1}$, resulting from the screened Coulomb interactions, where $\tau $ parametrizes the average lifetime of the quasiparticles. This is obtained from the imaginary part of the retarded fermion self-energy $\Sigma^R$.

The one-loop correction to the self-energy, caused by the Coulomb interaction, is given by
\begin{align}
\Sigma(i\,\omega_n,\mathbf{q})
=  T  \int\frac{d^3\mathbf{p}}{(2 \, \pi)^3}
\sum_{\Omega_m}
V(i  \, \Omega_m,\mathbf{q}) \,
G_0(i \,\omega_n+i  \, \Omega_m,\mathbf{p+q})
\label{Eq_self_energy}
\end{align}
in the Matsubara space, where $V(i  \, \Omega_m,\mathbf{q}) $ is the Matsubara-frequency-dependent counterpart of the effective retarded interaction shown in Eq.~\eqref{eqveff}.
Analytic continuation to real frequencies leads to
\begin{align}
\label{eqimsigmarealfreq}
 {\mathrm {Im}} \,\Sigma^R(\varepsilon,\mathbf{q})
=\frac{1}{4 }  \int\frac{d^3\mathbf{p}}{(2 \, \pi)^3}\int^\infty_{-\infty}
\frac{d\omega}{2 \, \pi}\,
B(\omega,\mathbf{p}) \, A^{(0)}(\varepsilon+\omega,\mathbf{p+q})
\left[ \coth\bigg (\frac{\omega}{2\,T} \bigg )
+ \tanh\bigg (\frac{\varepsilon+\omega}{2 \, T} \bigg )\right],
\end{align}
where
\begin{align}
B(\varepsilon,\mathbf{q})
\equiv -2 \, {\mathrm { Im}} \, V^R(\varepsilon,\mathbf{q})
=\frac{-2 \, V^2_0(\mathbf{q}) \,
N_f \,{\mathrm { Im}} \,\Pi^R(\varepsilon,\mathbf{q})}
{[1+V_0(\mathbf{q})\, N_f  \,\mathrm{Re}\, \Pi^R(\varepsilon,\mathbf{q})]^2
+[V_0(\mathbf{q})\, N_f \,{\mathrm { Im}} \,\Pi^R(\varepsilon,\mathbf{q})]^2}
\end{align}
is the bosonic spectral function.

Since the self-energy is a $4\times 4$ matrix, it is convenient to parametrize it as
\begin{align}
\Sigma^R(\varepsilon,\mathbf{p})
=\Sigma^R_s \,\Gamma_0 +\Sigma^R_v \, \mathbf{d} (\mathbf{p}) \cdot {\mathbf \Gamma} \,.
\end{align}
One can show that the scattering rate involves
only $\Sigma^R_s$, and hence we only compute ${\mathrm {Im}} \,\Sigma^R_s$, which takes the form:
\begin{align}
 & {\mathrm { Im}} \,\Sigma^R_s(\varepsilon,\mathbf{p})
\nn &= -\frac{\pi}{4}
\int\frac{d^3\mathbf{q}}{(2 \, \pi)^3}
\int^\infty_{-\infty}\frac{d\omega}{2 \, \pi} \,
B(\omega,\mathbf{q})
\left[\coth\bigg (\frac{\omega}{2\,T} \bigg )
+ \tanh \bigg (\frac{\varepsilon+\omega}{2 \, T} \bigg )\right]
\left [
\varepsilon+\omega+\mu-\frac{ \alpha\, (\mathbf{p+q})^2} {2\, m}
\right ]
 \mathrm{sgn} \bigg( \varepsilon+\omega+\mu
-\frac{\alpha\, (\mathbf{p+q})^2} {2\, m}\bigg)
\nn & \hspace{4 cm} \times
\frac{
\sum \limits_{\zeta =\pm}
\delta \bigg (\varepsilon+\omega+\mu-\frac{\alpha\,(\mathbf{p+q})^2} {2\,m}
+\zeta\, \sqrt{(1+\eta) \,d^2_{\mathbf{p+q}}
+2 \, \eta \sum \limits_{a'=1}^5 s_{a'}\,d^2_{a'}( \mathbf{p+q}) } \,\bigg )
}
{\sqrt{(1+\eta) \, |\mathbf d(\mathbf{p+q})|^2
+ 2\,\eta  \sum \limits_{a=1}^5 s_a  \, d^2_a (\mathbf{p+q})
}}
\label{Eq_Im_Sigma_B}
\end{align}
in general.

\begin{figure}
\centering
\subfigure[]{\includegraphics[width=0.45\textwidth]{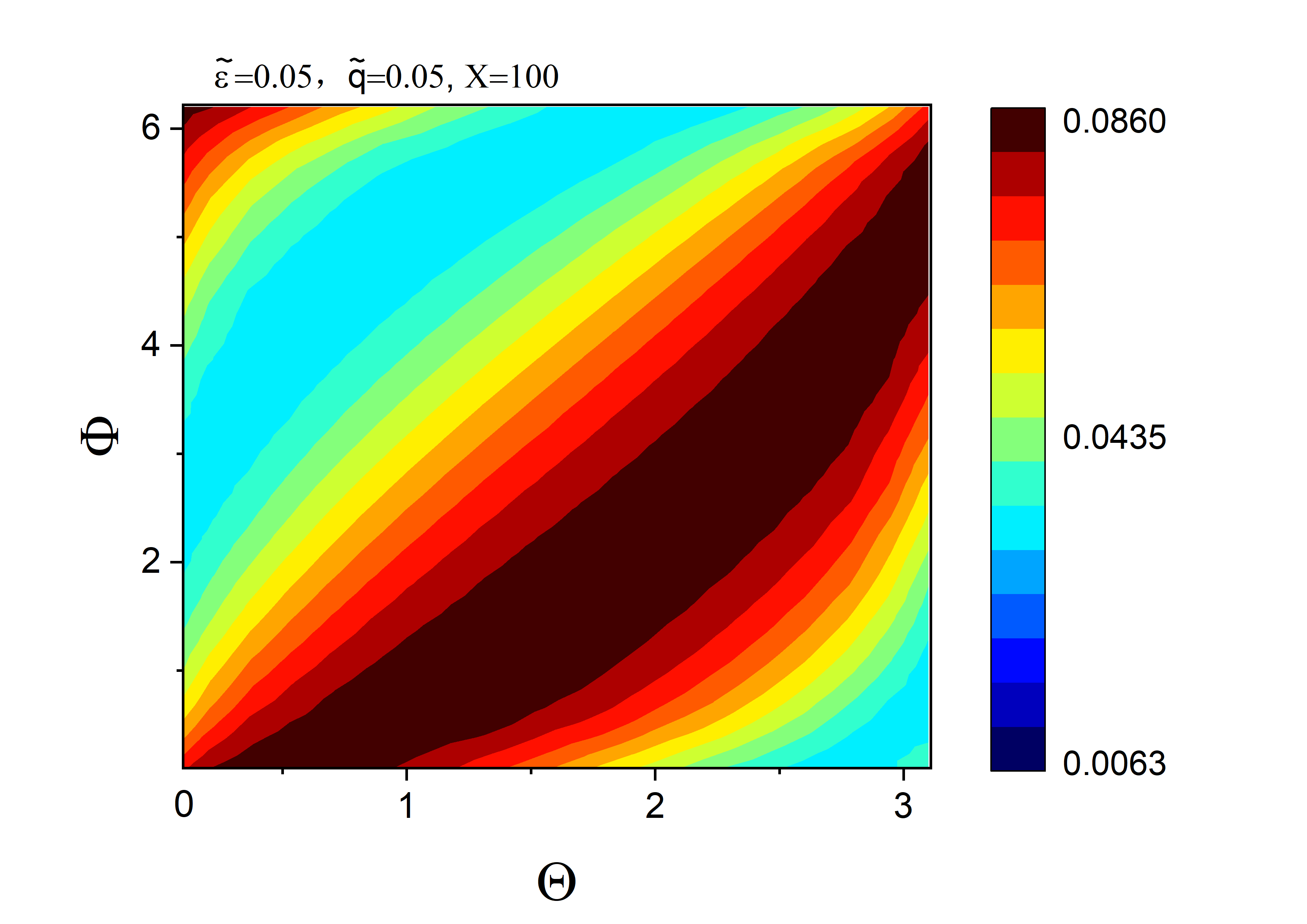}}\quad
\subfigure[]{\includegraphics[width=0.45\textwidth]{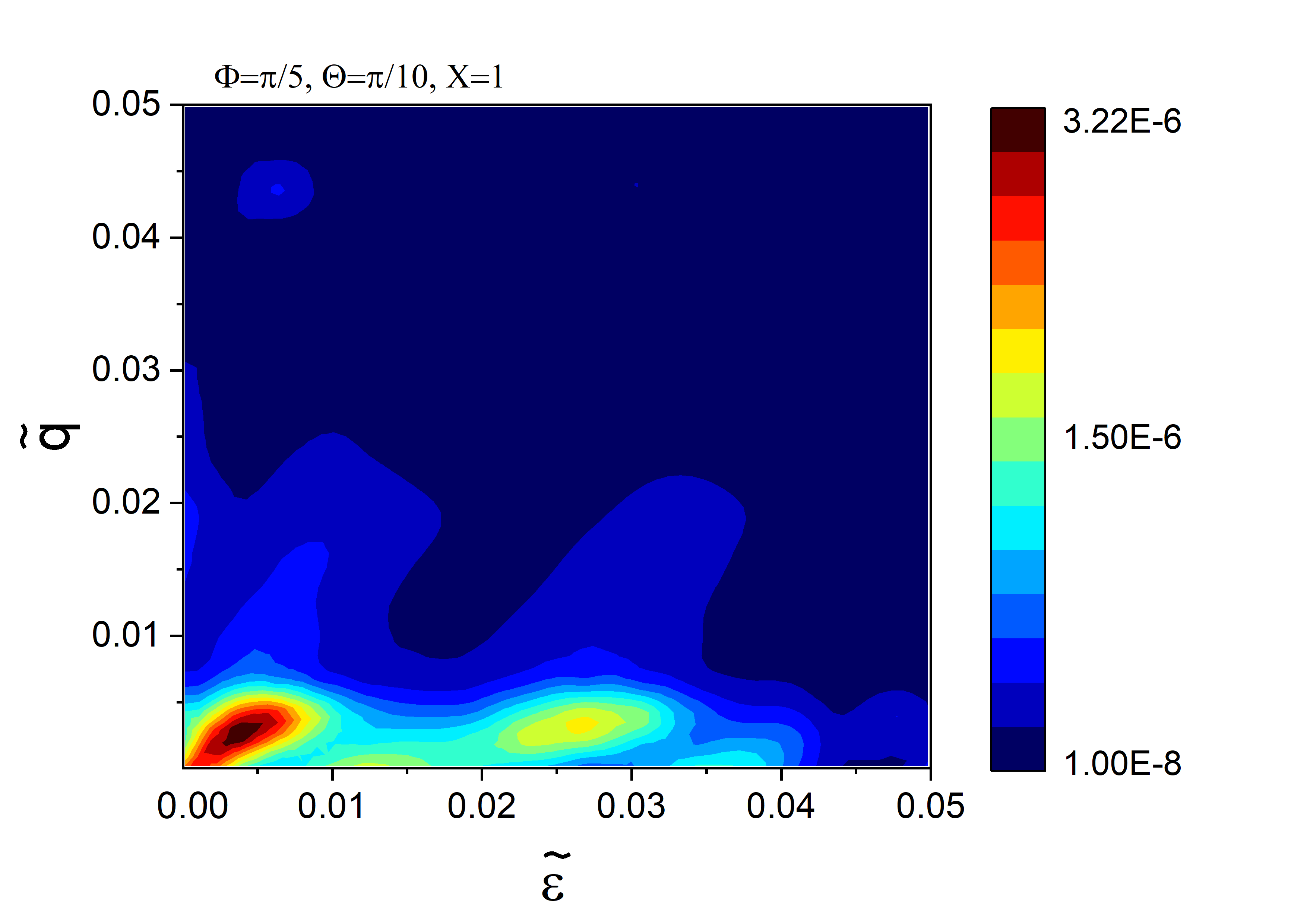}}\\
\subfigure[]{\includegraphics[width=0.45\textwidth]{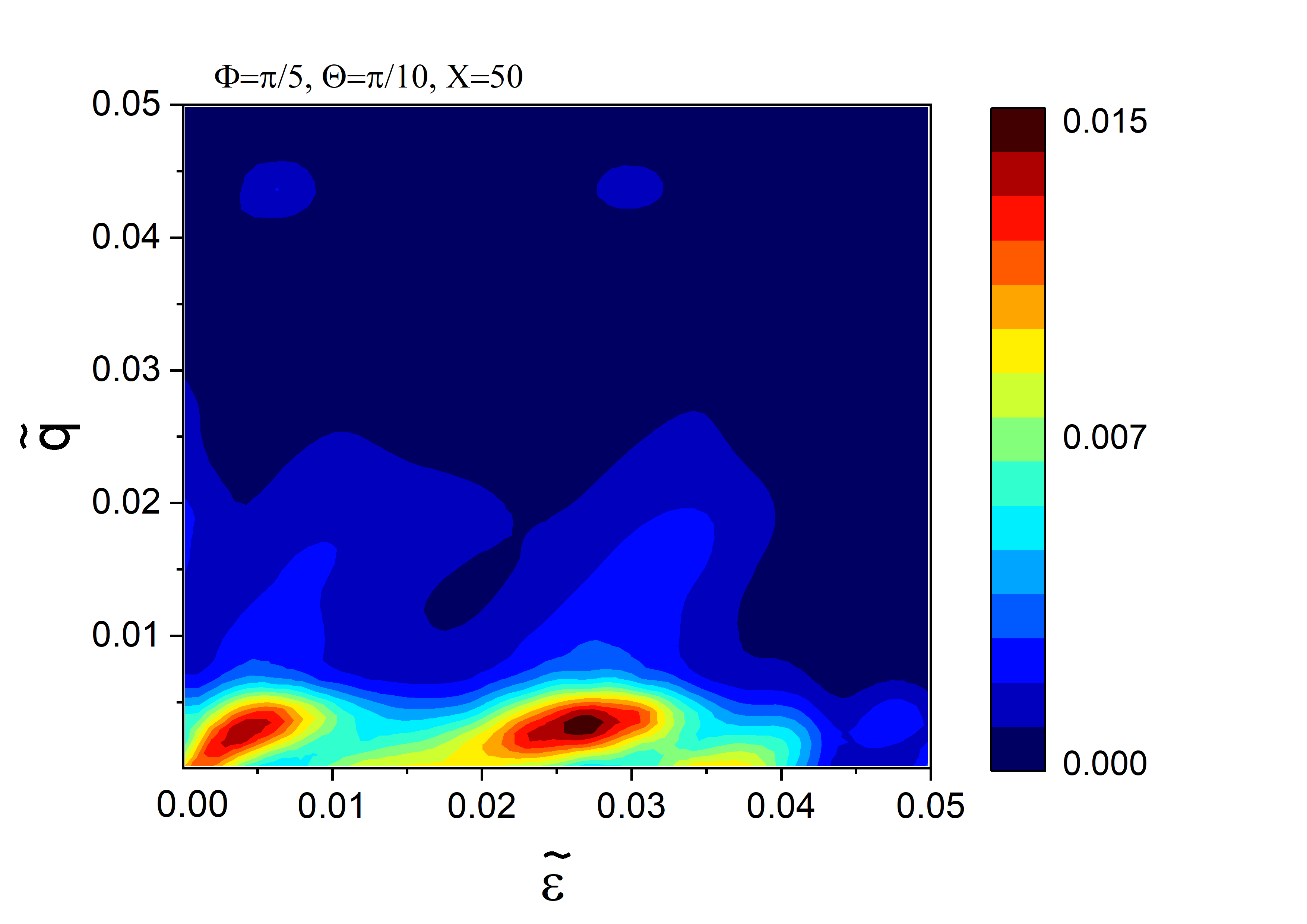}}\quad
\subfigure[]{\includegraphics[width=0.45\textwidth]{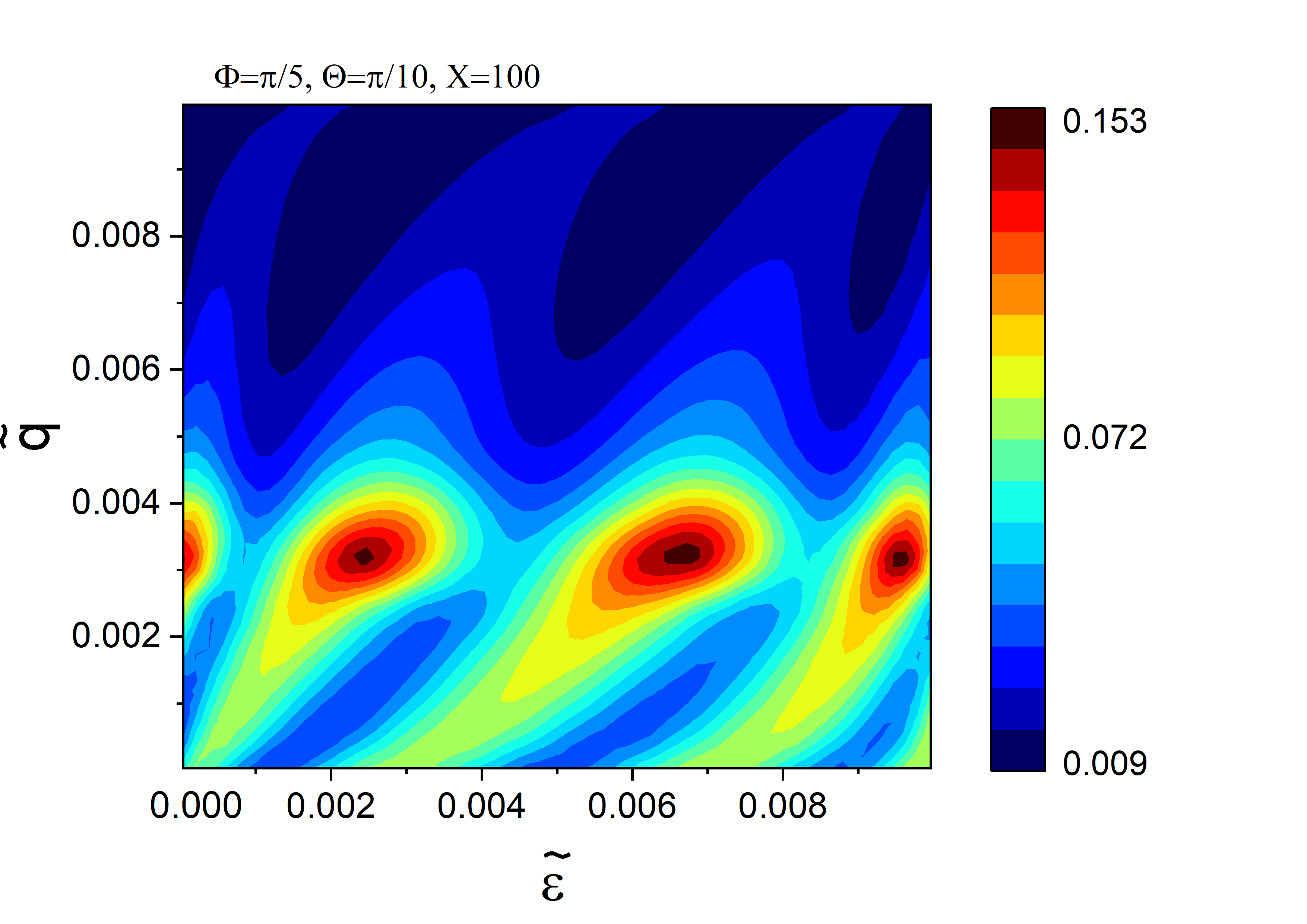}} \\
\caption{
Behaviour of $1/\tau$ for an isotropic LSM with $\tilde T=0. 1$, $\tilde \mu=0.1$, and $\tilde \varepsilon \geq 0$: Subfigure (a) captures the dependence on the angular
variables $\Theta$ and $\Phi$, for $\tilde \varepsilon = \tilde q = 0.05$ and $X = 100 $.
Subfigures (b), (c), and (d) show the dependence on $\tilde \varepsilon $ and $\tilde q$, with
$\Phi= \pi/ 5 $ and $\Theta= \pi / 10$, for $X= 1$, $X= 50$, and $X= 100$, respectively.
\label{Fig_iso_tune-X}
}
\end{figure}

\begin{figure}
\centering
\subfigure[]{\includegraphics[width=0.45\textwidth]{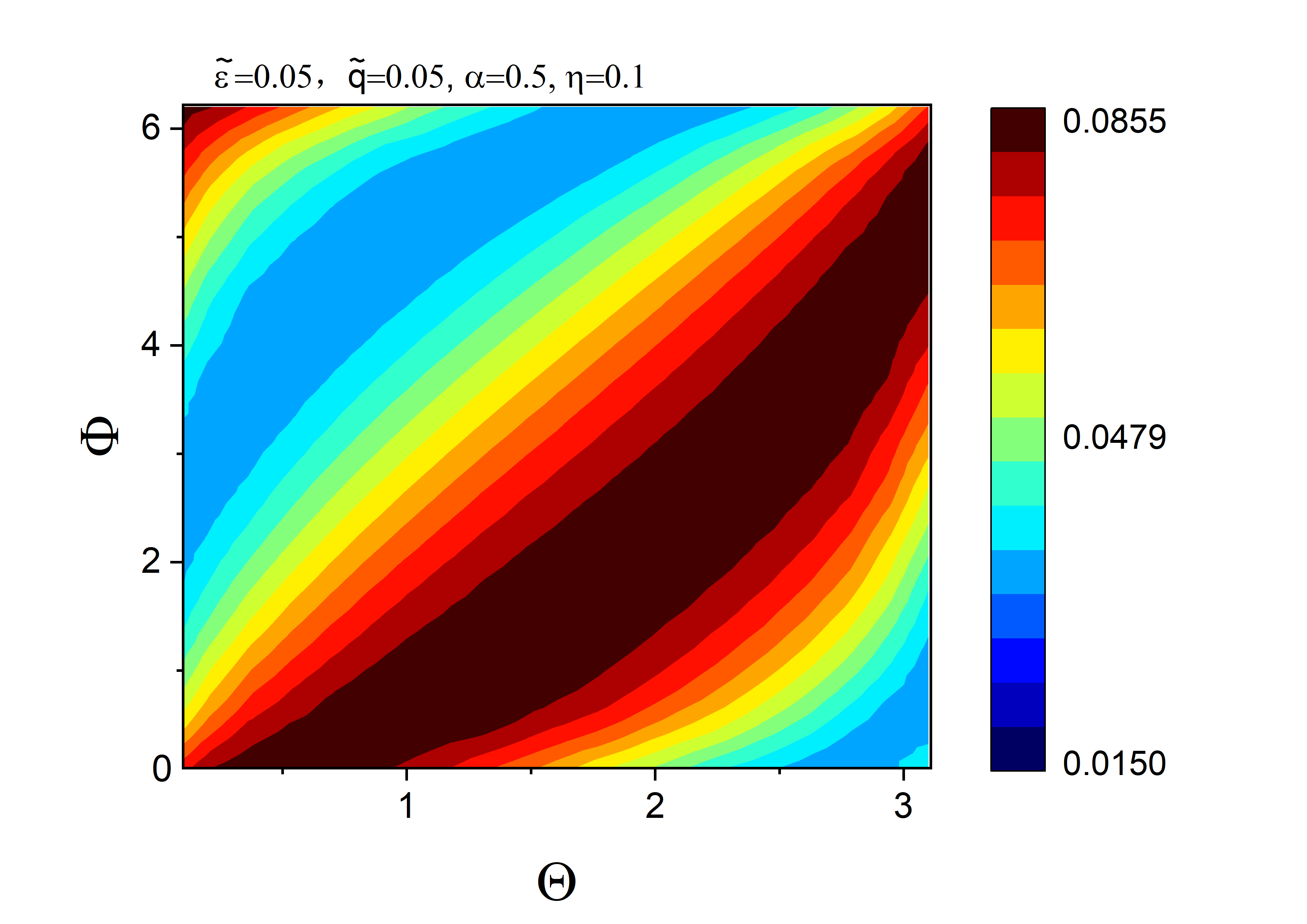}}\quad
\subfigure[]{\includegraphics[width=0.45\textwidth]{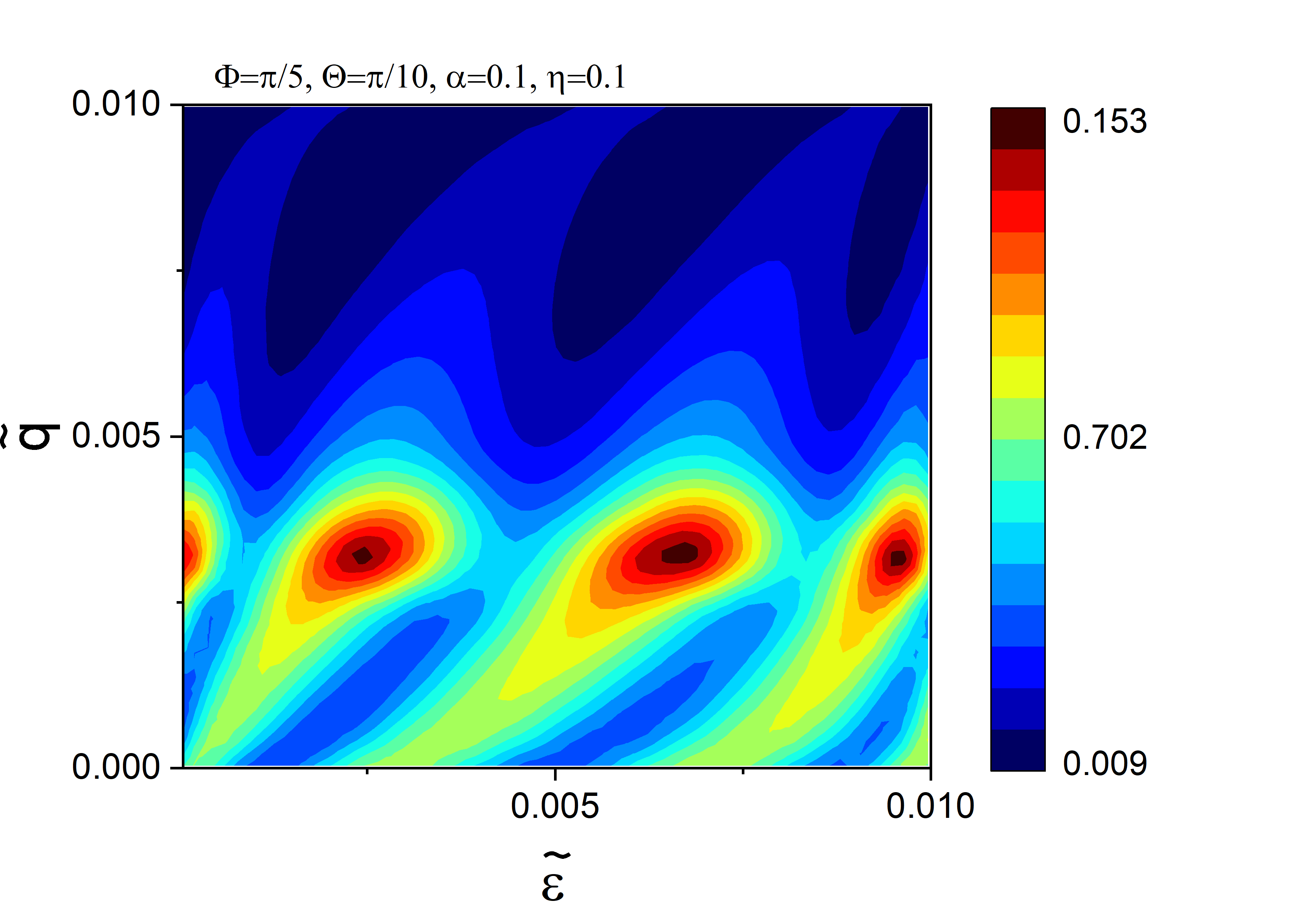}}\\
\subfigure[]{\includegraphics[width=0.45\textwidth]{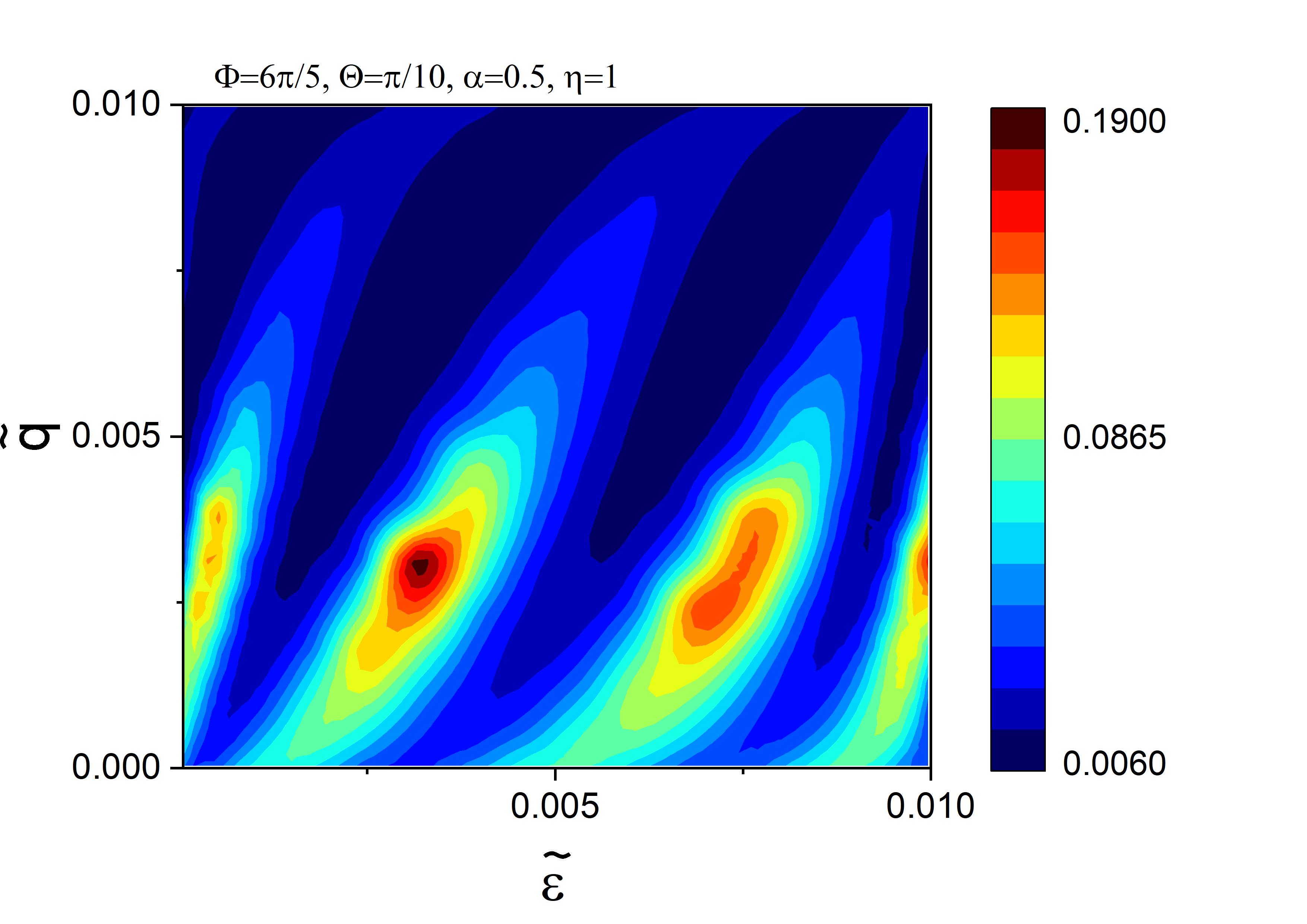}}\quad
\subfigure[]{\includegraphics[width=0.45\textwidth]{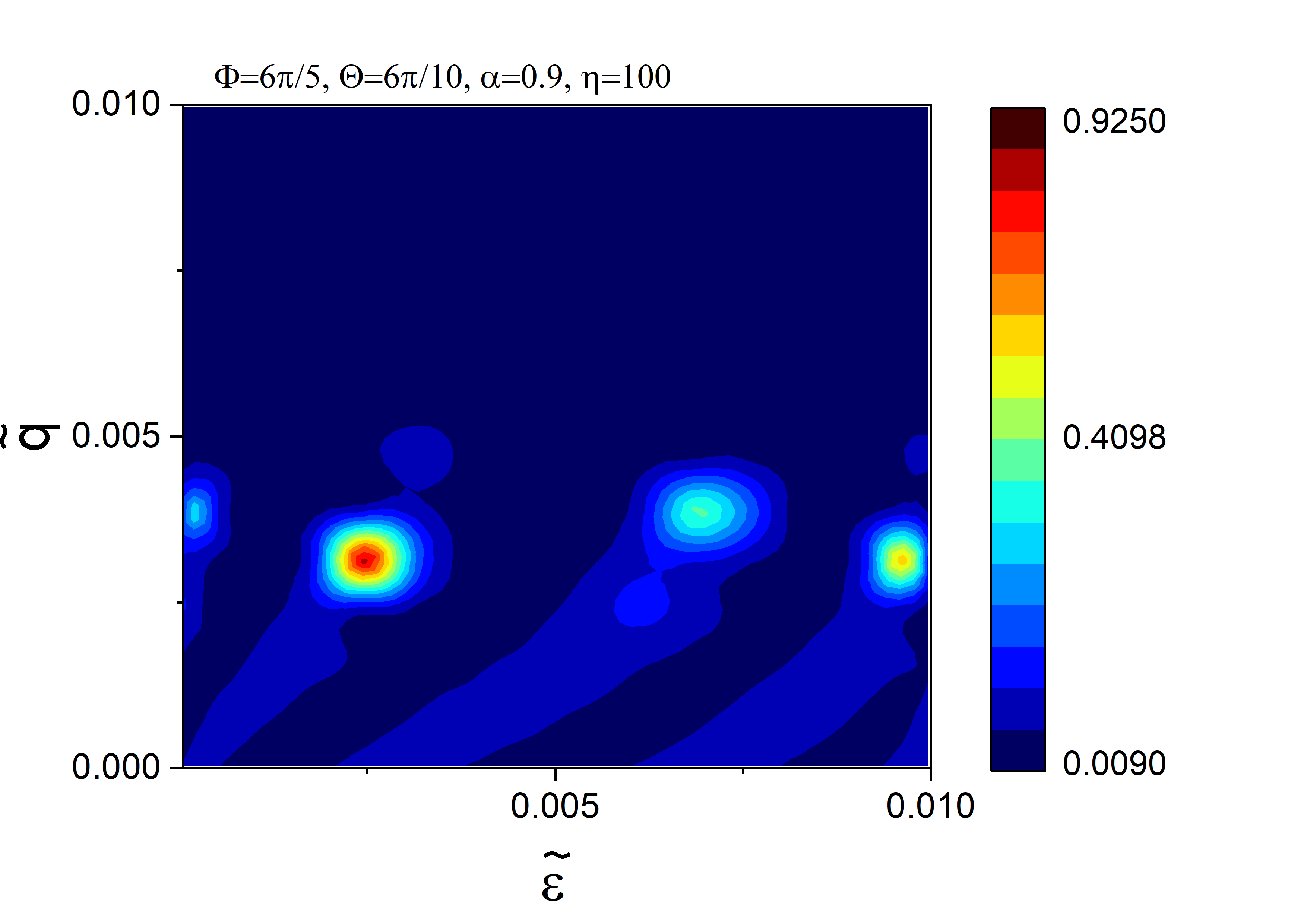}}
\caption{Behaviour of  $1/\tau $ for a generic LSM
with $\tilde T=0.5$, $\tilde \mu=0.1$, $X=100$, and $\tilde \varepsilon \geq 0$: Subfigure (a) captures the dependence on the angular variables $\Theta$ and $\Phi$, for $\tilde  \varepsilon = \tilde q = 0.05$, $\alpha=0.5$, and $\eta=0.1$. Subfigures (b), (c), and (d) show the dependence on $\tilde  \varepsilon $ and $\tilde q$, for $ \lbrace \alpha, \eta, \Phi , \Theta  \rbrace =
\lbrace 0.1, 0.1, \pi/ 5, \pi / 10 \rbrace $, $\lbrace \alpha, \eta, \Phi,\Theta \rbrace =\lbrace 0.5, 1, 6 \, \pi/ 5,\pi / 10 \rbrace$,
and $\lbrace \alpha, \eta , \Phi, \Theta \rbrace = \lbrace 0.9, 100,6 \,\pi/ 5 , 6 \,\pi / 10 \rbrace $, respectively.
\label{Fig_aniso-tau}
}
\end{figure}

\begin{figure}
\centering
\subfigure[]{\includegraphics[width=0.45\textwidth]{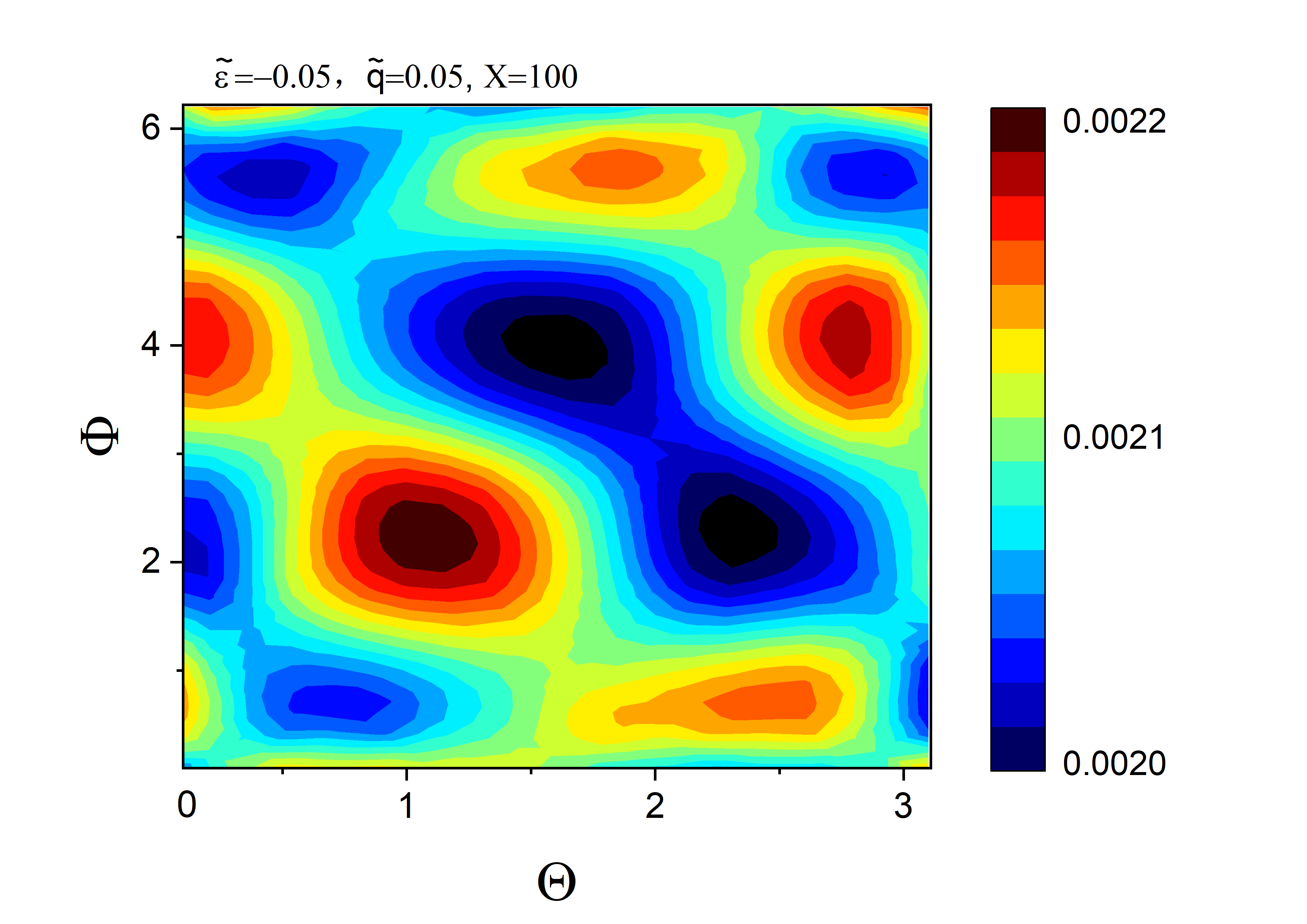}}\quad
\subfigure[]{\includegraphics[width=0.45\textwidth]{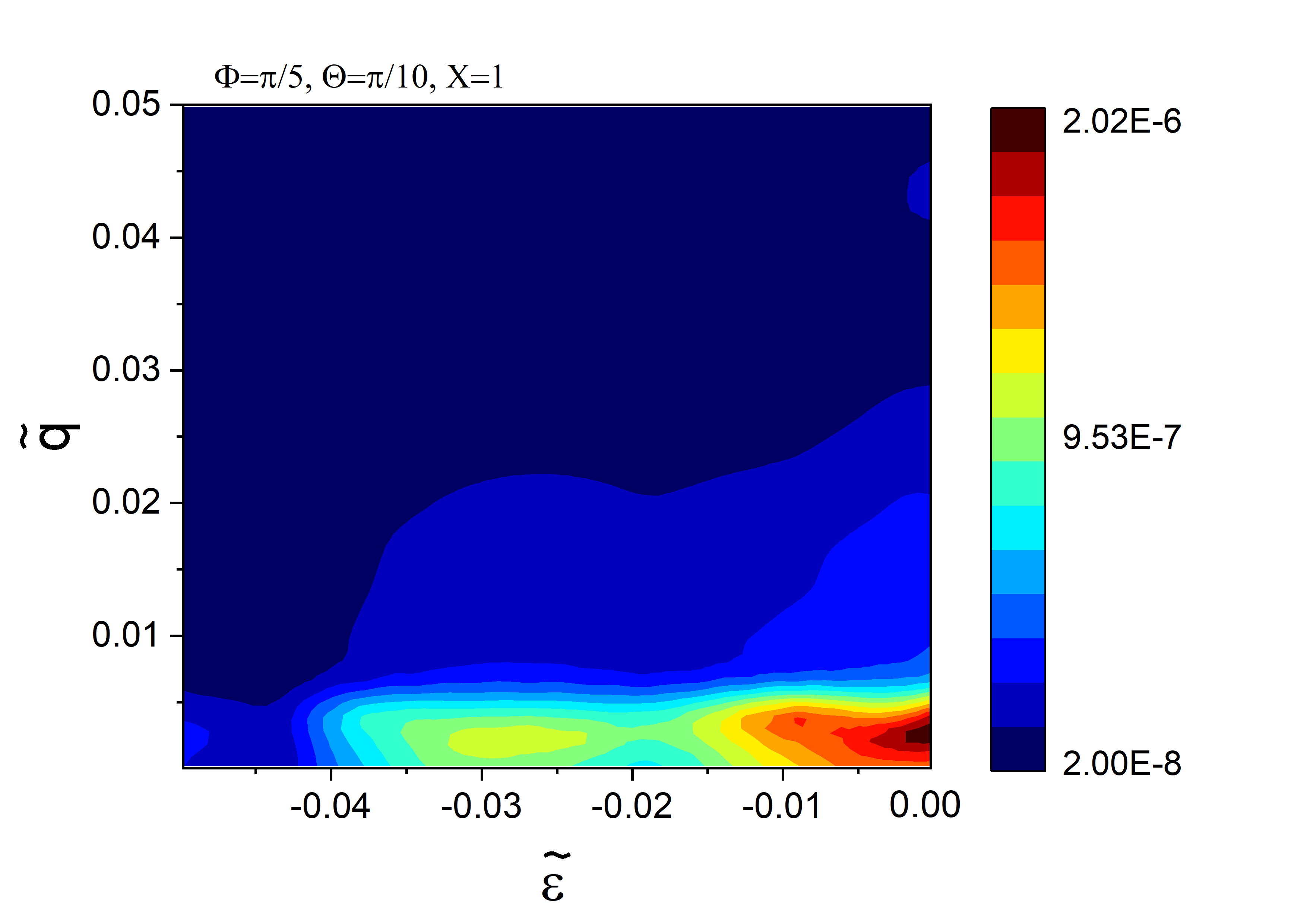}}\\
\subfigure[]{\includegraphics[width=0.45\textwidth]{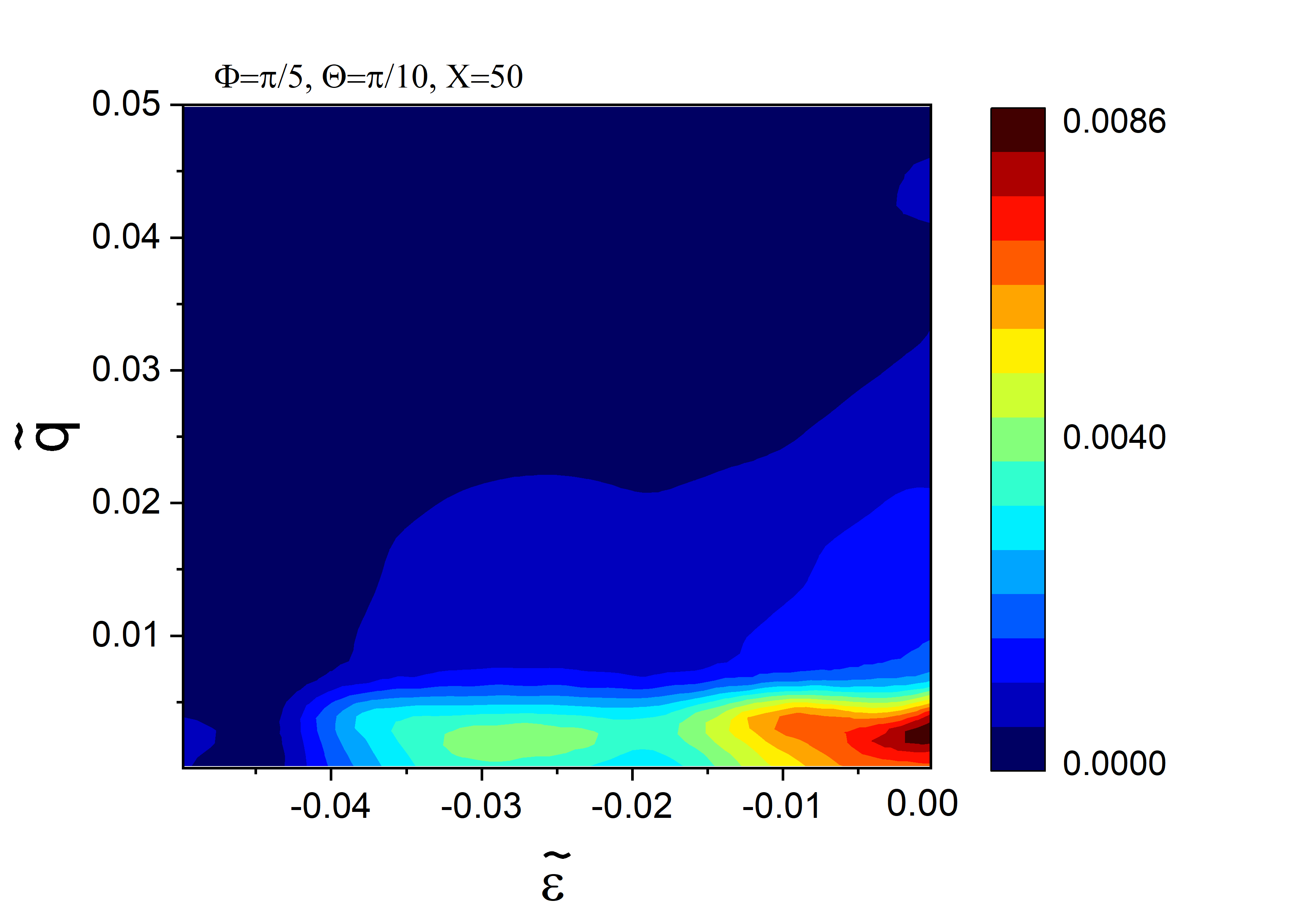}}\quad
\subfigure[]{\includegraphics[width=0.45\textwidth]{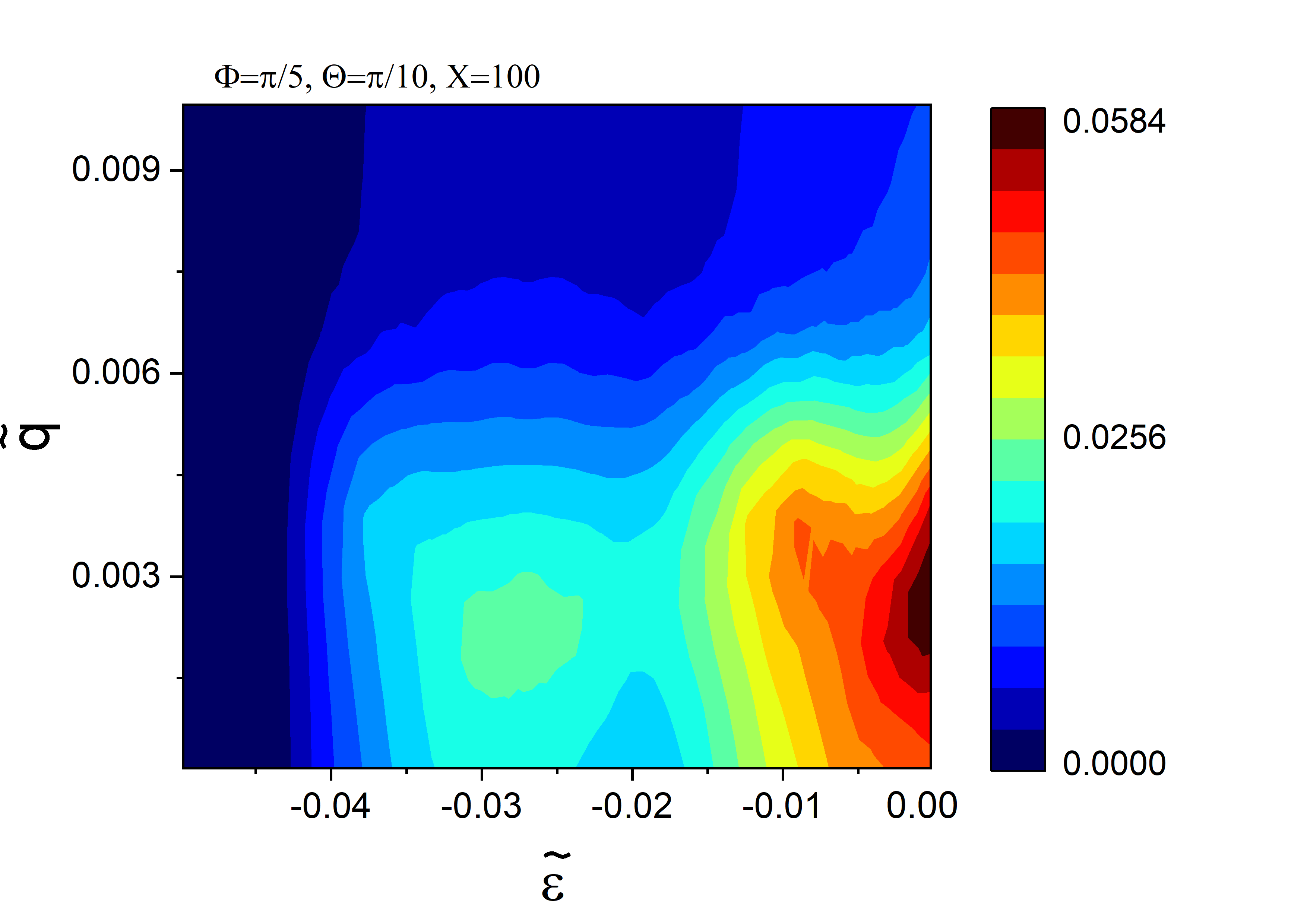}} \\
\caption{Behaviour of $1/\tau$ for an isotropic LSM
with $\tilde T=0. 1$, $\tilde \mu=0.1$, and $\tilde \varepsilon < 0$: Subfigure (a) captures the dependence on the angular variables $\Theta$ and $\Phi$,
for $\tilde \varepsilon = -0.05$, $\tilde q = 0.05$, and $X = 100 $.
Subfigures (b), (c), and (d) show the dependence on $\tilde  \varepsilon $ and $\tilde q$, with
$\Phi= \pi/ 5 $ and $\Theta= \pi / 10$, for $X= 1$, $X= 50$, and $X= 100$, respectively.
\label{Fig_iso_tune-X-omega-minus}
}
\end{figure}

Using the relation~\cite{abrikosov2012methods}
\begin{align}
\frac{1}{\tau(\varepsilon,\mathbf{q})}
\equiv-2 \,{\mathrm { Im}} \,\Sigma^R_s(\varepsilon,\mathbf{q}) \,,
\label{Eq_tau}
\end{align}
we compute the scattering rate $\tau $. Due to band-mass anisotropy, in general,
$\tau(\varepsilon,\mathbf{q})$ is different from $\tau(-\varepsilon,\mathbf{q})$.
First, let us focus on the $\varepsilon\geq 0$ case.
Armed with the results obtained for the polarization bubbles discussed in Sec.~\ref{Sec_Results},
we make a judicious choice of the parameter regimes to numerically compute $\tau^{-1}$.
Figs.~\ref{Fig_iso_tune-X} and~\ref{Fig_aniso-tau} show some representative plots.
Fig.~\ref{Fig_aniso-tau} demonstrates the features for nonzero $\alpha$ and $\eta$, and includes the cases $\alpha>\eta$, $\alpha=\eta$, and $\alpha<\eta$. Corroborating our observations for the results for $\Pi^R $ and $\mathrm{Re}\, {\mathcal{E}}_{\mathrm{eff}}$, we find that nonzero anisotropy parameters
hardly bring about substantial changes to the behaviour of $1/\tau$, compared to an isotropic LSM.
Next, let us consider the $\varepsilon<0$ case. Fig.~\ref{Fig_iso_tune-X-omega-minus} shows some representative plots for the isotropic scenario. We have checked that the basic results turn out to be similar when we consider the anisotropic case --- hence, we have not shown those results here for the sake of brevity.

From the results shown in Figs.~\ref{Fig_iso_tune-X}--\ref{Fig_iso_tune-X-omega-minus}, we notice that the plasmon mode can be triggered more easily for the
$\varepsilon>0$ case than that for the $\varepsilon<0$ case.
We have explored various other parameter regimes (not shown in the plots presented here) similar to  those pointed out in Sec.~\ref{Sec_Results}. Broadly, the features are found to be quite insensitive to variations in $T$, $\mu$, $\alpha$, and $\eta$. Of course, a higher value of $X$ always favours the possibility of the emergence of plasmons.

\begin{figure}
\centering
\subfigure[]{\includegraphics[width=0.45\textwidth]{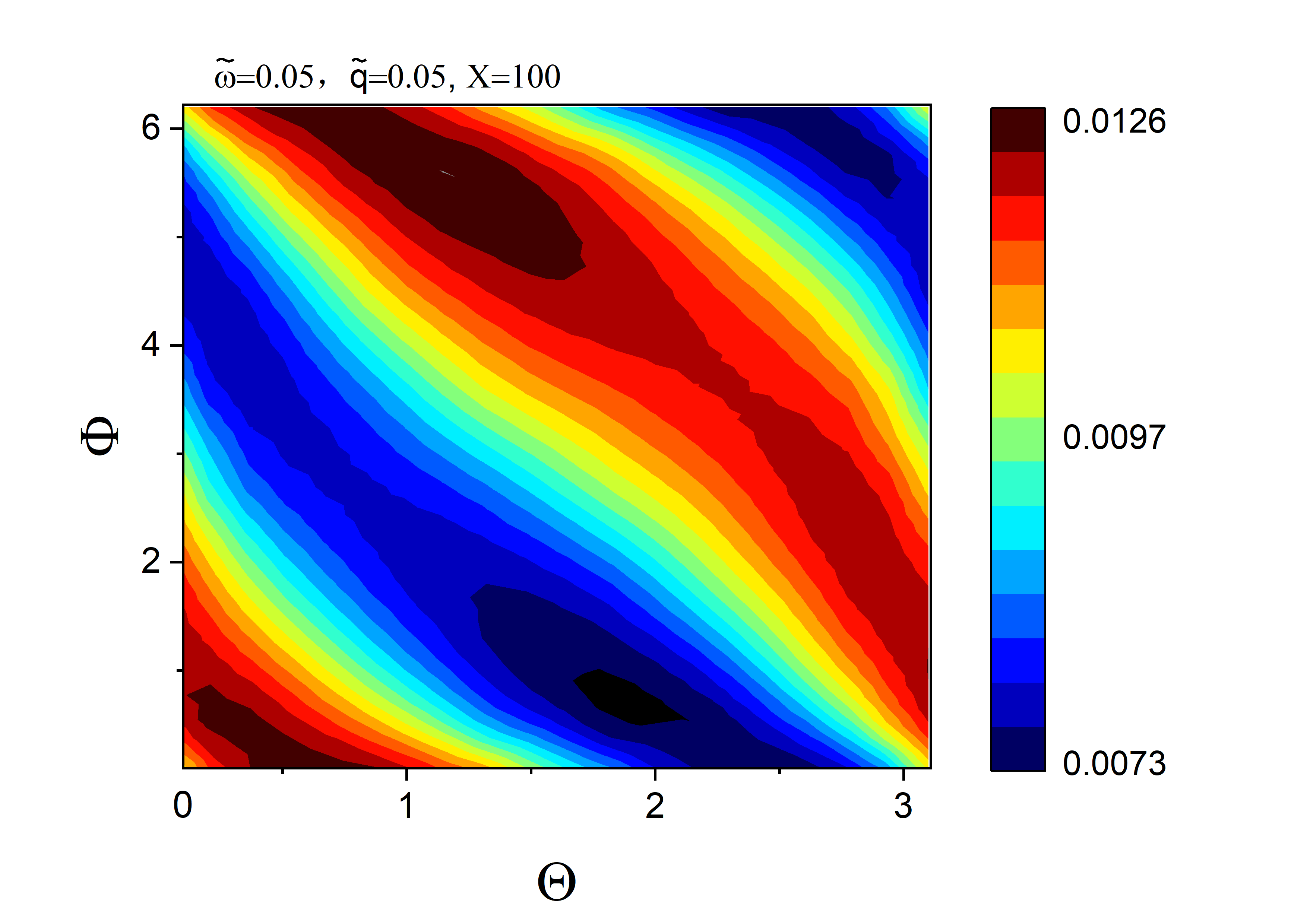}}\quad
\subfigure[]{\includegraphics[width=0.45\textwidth]{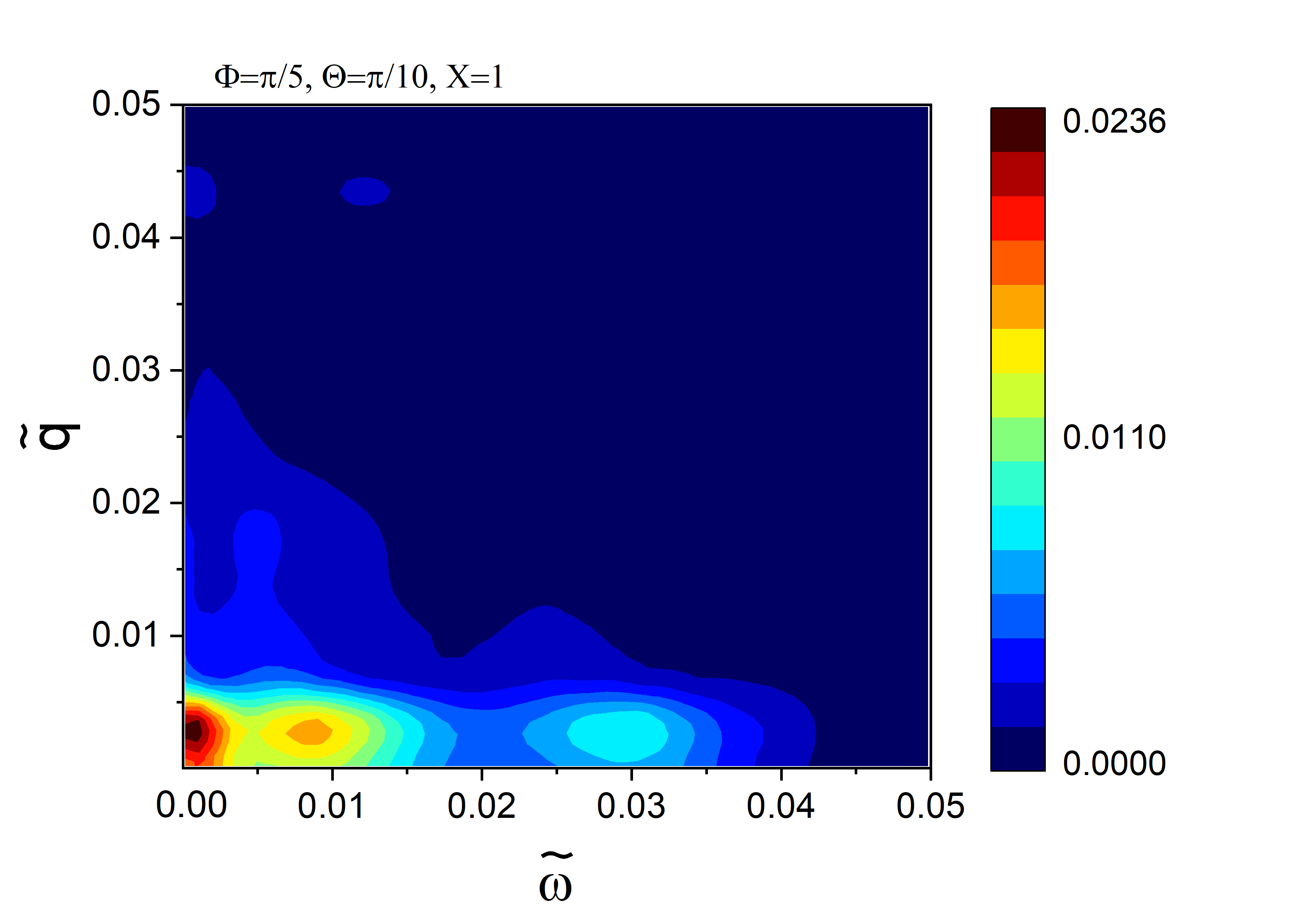}}\\
\subfigure[]{\includegraphics[width=0.45\textwidth]{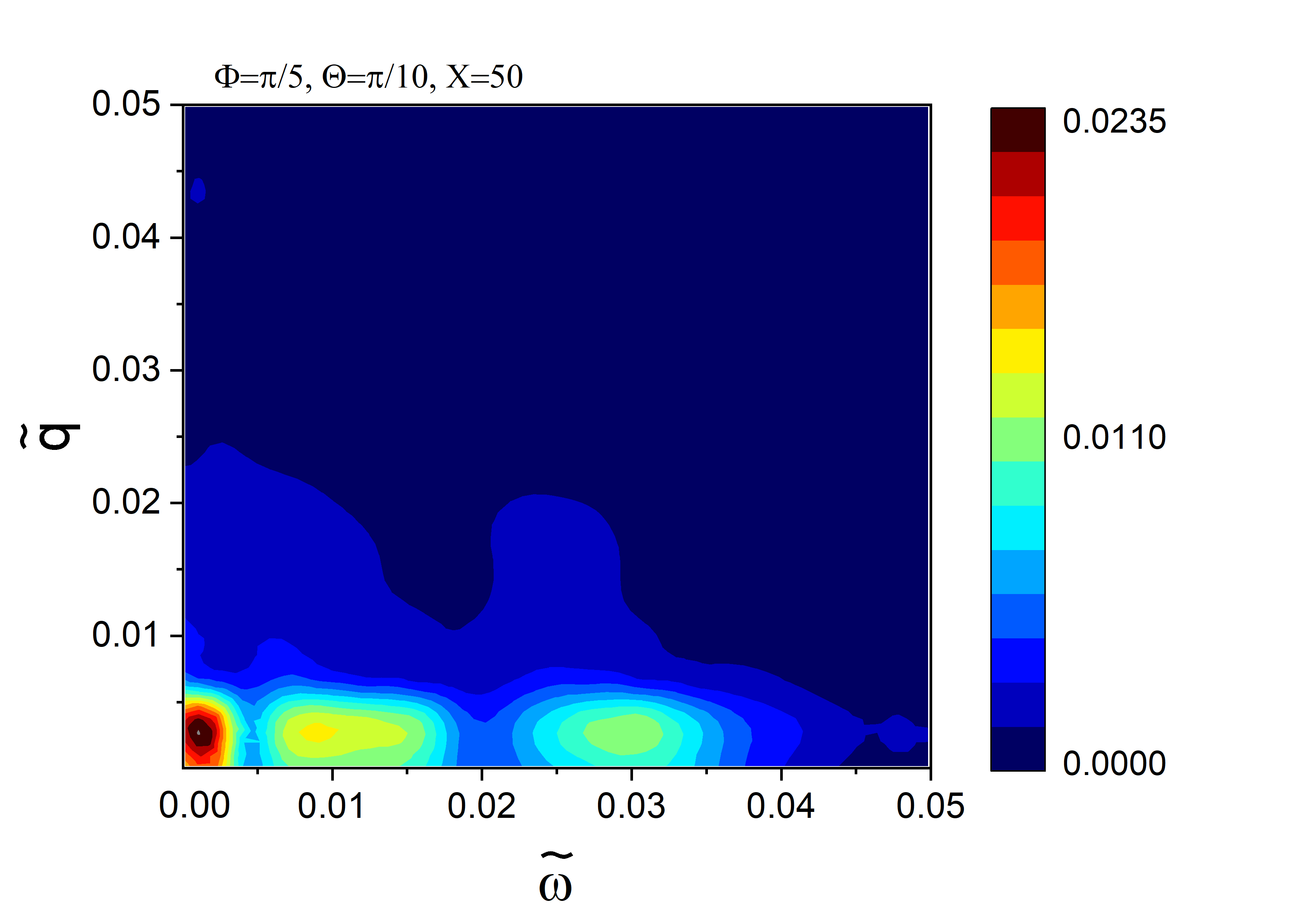}}\quad
\subfigure[]{\includegraphics[width=0.45\textwidth]{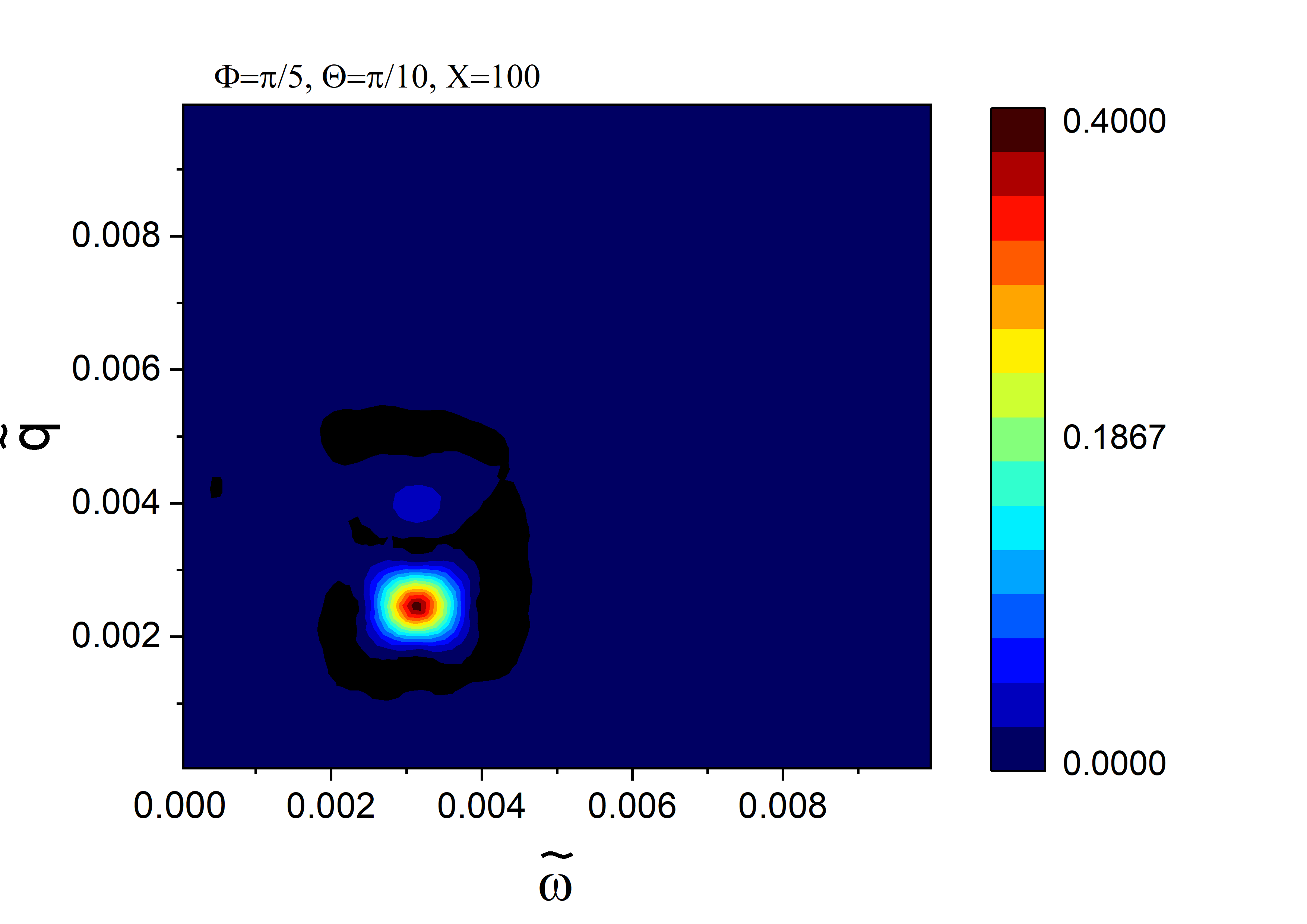}}\\
\caption{Behaviour of the one-loop corrected spectral function $A$ for an isotropic LSM
with $\tilde T=0. 1$ and $\tilde \mu=0.1$:
Subfigure (a) captures the dependence on the angular variables
$\Theta$ and $\Phi$, for $\tilde \omega= \tilde q = 0.05$ and $X = 100 $.
Subfigures (b)--(d) show the dependence on
$\tilde \omega $ and $\tilde q$, with $\Phi= \pi/ 5 $ and
$\Theta= \pi / 10$, for $X= 1$, $X= 50$, and $X= 100$, respectively.
\label{Fig_iso_A_u-p_T01-mu-01-plus}
}
\end{figure}

\begin{figure}
\centering
\subfigure[]{\includegraphics[width=0.45\textwidth]{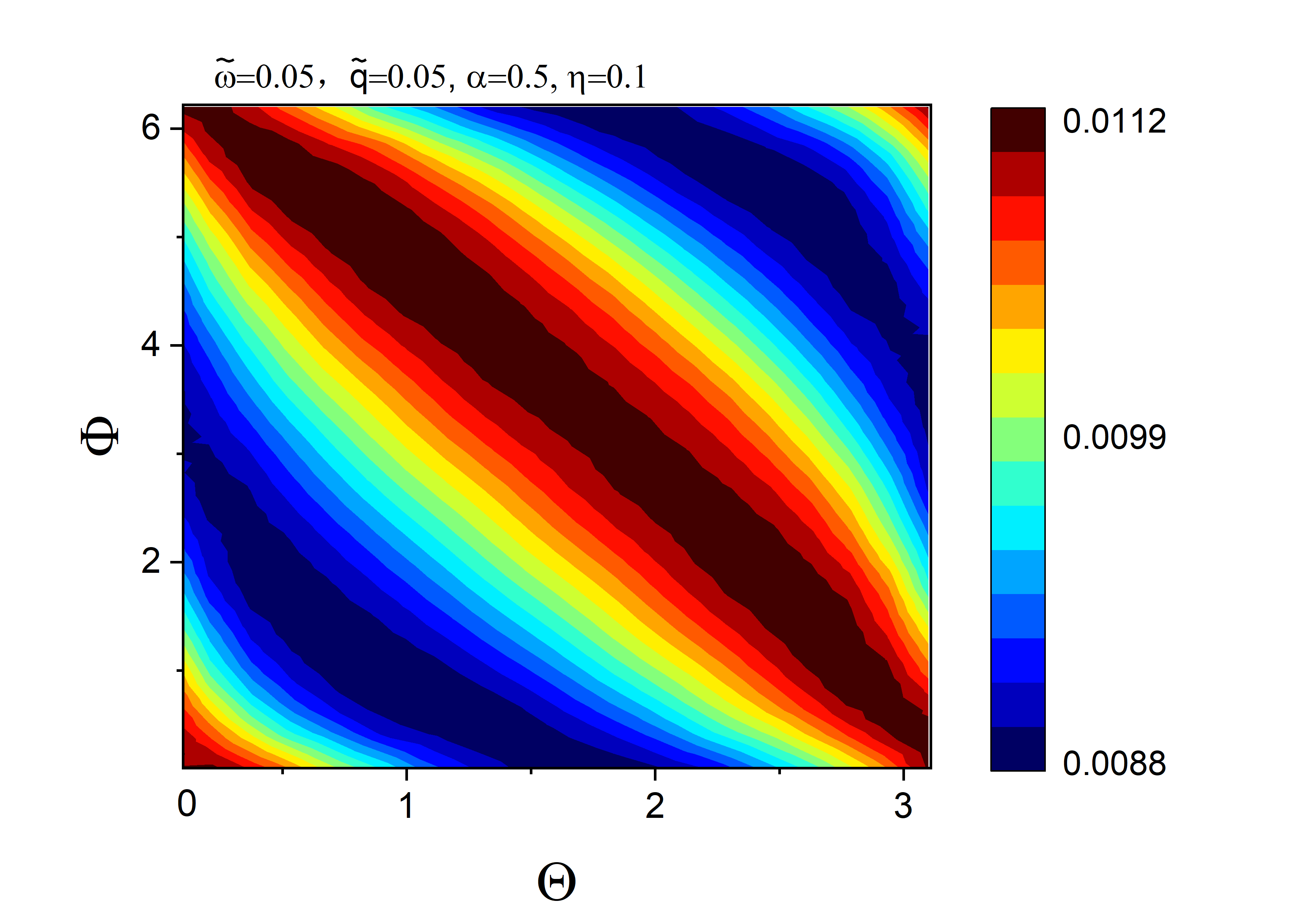}}\quad
\subfigure[]{\includegraphics[width=0.45\textwidth]{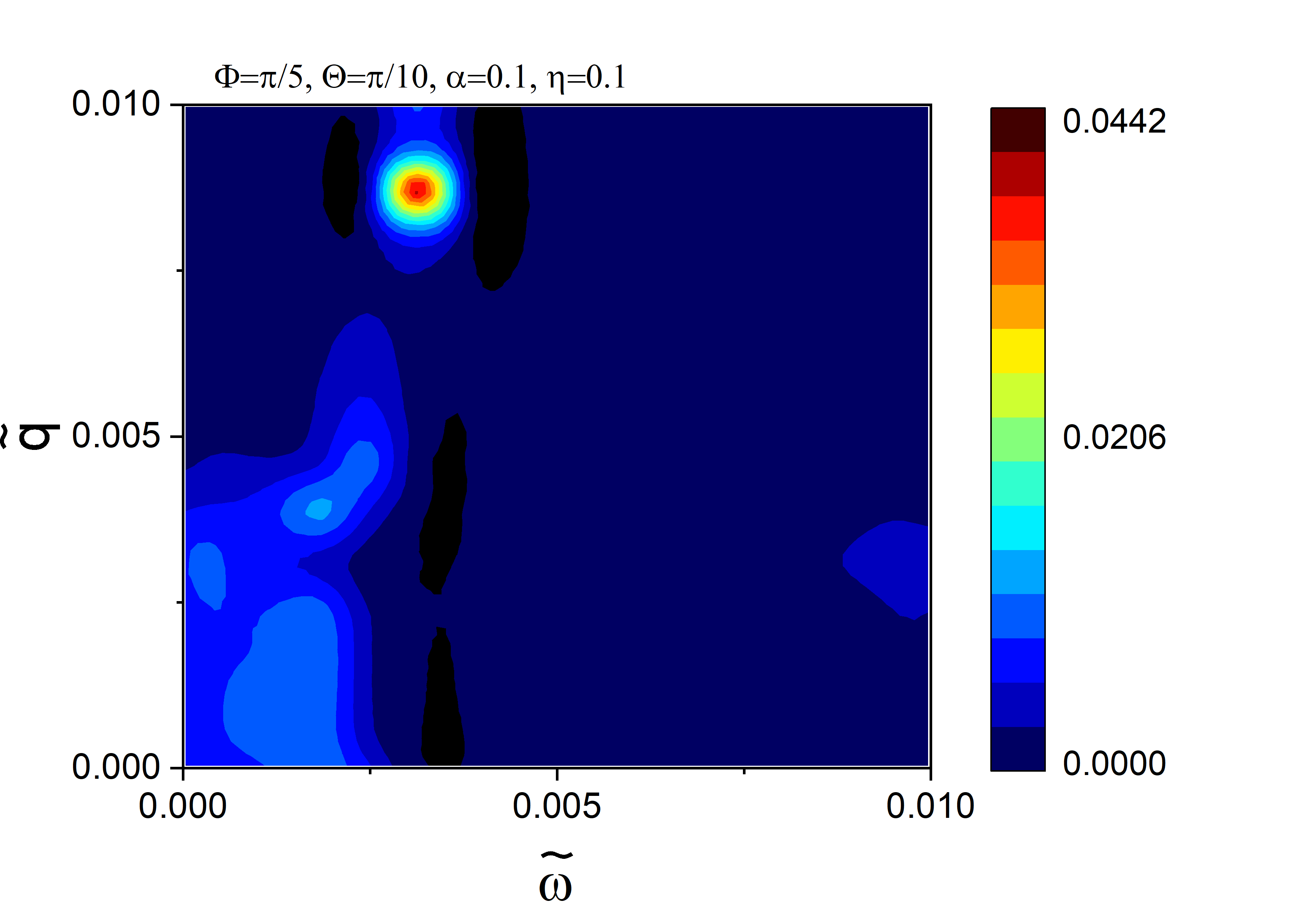}}\\
\subfigure[]{\includegraphics[width=0.45\textwidth]{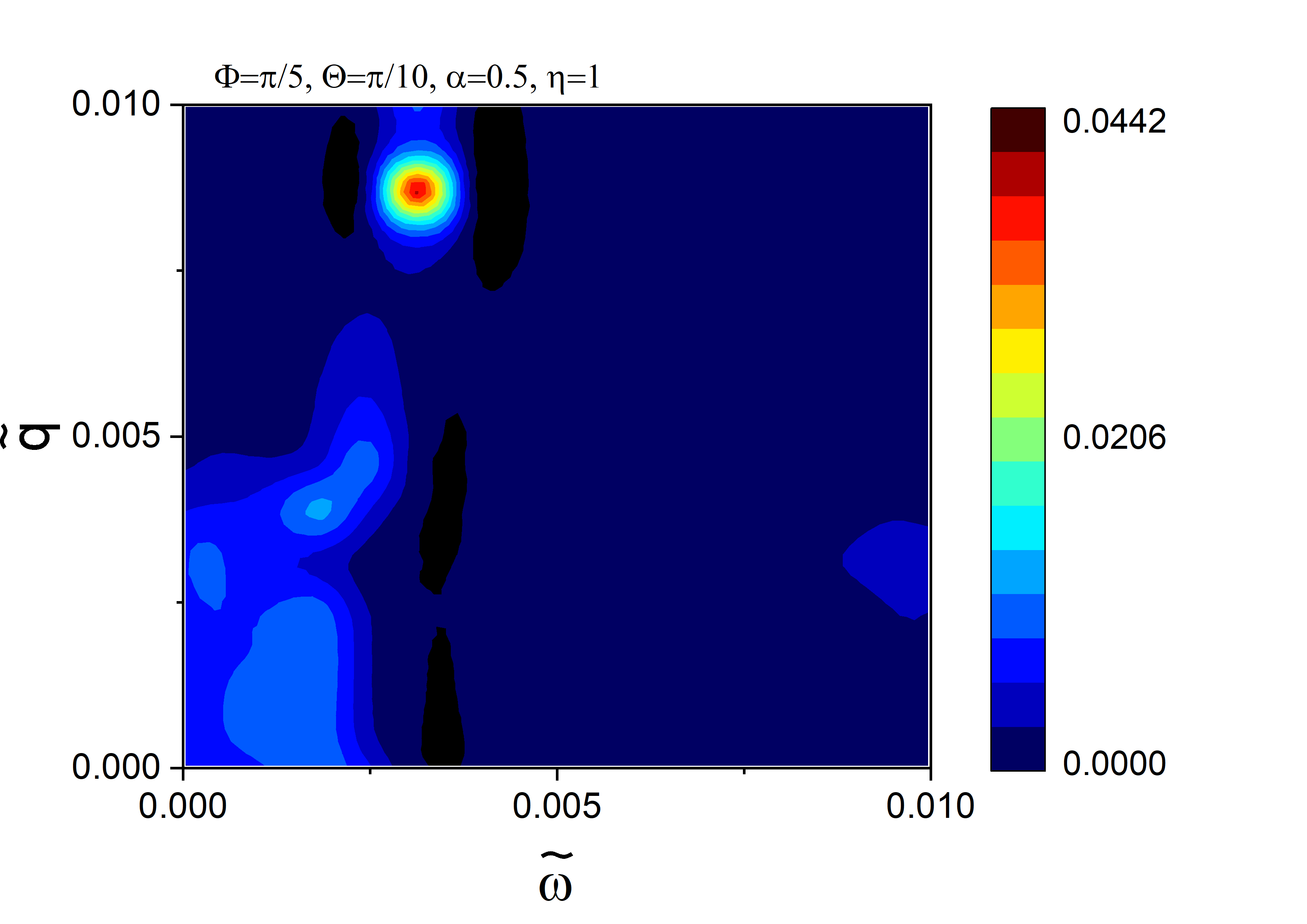}}\quad
\subfigure[]{\includegraphics[width=0.45\textwidth]{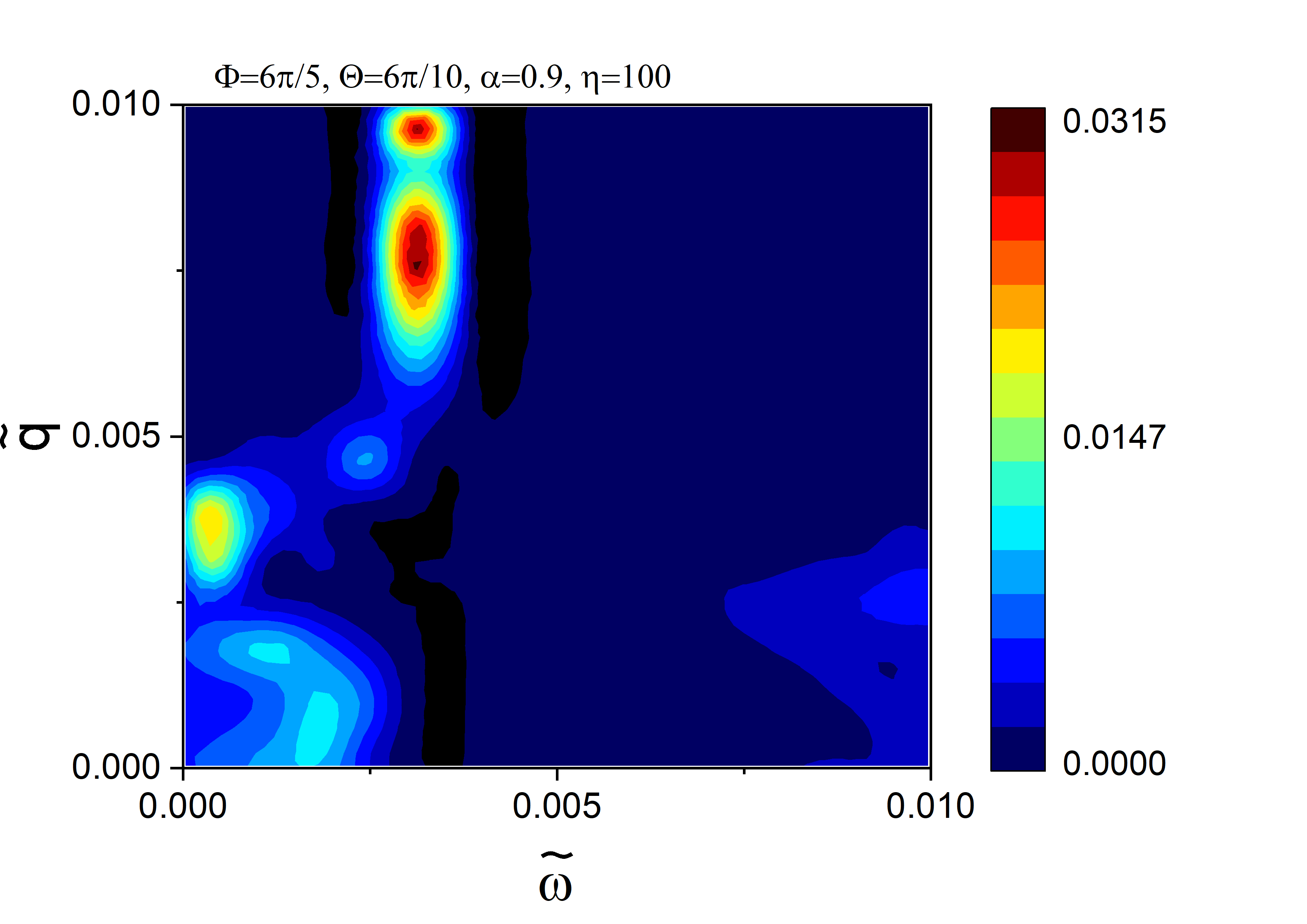}}\\
\caption{Behaviour of the one-loop corrected spectral function $A$ for a generic LSM with $\tilde T=0.5$, $\tilde \mu=0.1$, $X=100$, and $\tilde \omega >0$: Subfigure (a) captures the dependence
on the angular variables $\Theta$ and $\Phi$, for $\tilde \omega= \tilde q = 0.05$,
$\alpha=0.5$, and $\eta=0.1$. Subfigures (b), (c), and (d) show the dependence on $\tilde \omega$ and $\tilde q$, for $ \lbrace \alpha, \eta, \Phi , \Theta  \rbrace =
\lbrace 0.1, 0.1, \pi/ 5, \pi / 10 \rbrace $, $\lbrace \alpha, \eta, \Phi,\Theta \rbrace =\lbrace 0.5, 1,  \pi/ 5,\pi / 10 \rbrace$,
and $\lbrace \alpha, \eta , \Phi, \Theta \rbrace = \lbrace 0.9, 100,6 \,\pi/ 5 , 6 \,\pi / 10 \rbrace $, respectively.
\label{Fig_aniso_A_u-p_T05-mu-01-plus}
}
\end{figure}

\section{Quasiparticle characteristics}
\label{Sec_Z_f-spectral}

In this section, we investigate the behaviour of quasiparticle by computing the quasiparticle residue and spectral function at one-loop order~\cite{Mahan2000Book} in the presence of the Coulomb interactions.

In the Matsubara space, the one-loop corrected fermionic propagator is given by~\cite{Mahan2000Book}
\begin{align}
G^{-1}(i\,\omega_n,\mathbf{p})
&=G^{-1}_0(i\,\omega_n,\mathbf{p})-\Sigma(i \,\omega_n,\mathbf{p}),
\end{align}
where $G_0(i\,\omega_n,\mathbf{p})$ is the bare fermion propagator defined in Eq.~\eqref{Eq_G_0-mu} or Eq.~\eqref{Eq_aniso-G_0}, depending on whether we are dealing with the isotropic or anisotropic case.
Furthermore, $\Sigma(i\,\omega_n,\mathbf{p})$ is the one-loop fermionic self-energy, as defined in Eq.~(\ref{Eq_self_energy}).
Performing the appropriate analytic continuation to real frequencies, the
retarded fermionic propagator can be written as
\begin{align}
G^R(\omega  ,\mathbf{p})
&=\frac{
\left( -\omega + \frac{\alpha \, \mathbf{p}^2}  {2\,m}
-\mu \right) \Gamma_0+{\mathbf d}({\mathbf{p}}) \cdot {\mathbf \Gamma} +
\eta \,\sum \limits_{a=1}^5 s_a  \,d_a (\mathbf{p}) \,\Gamma^a-{\mathrm { Re}} \,\Sigma^R(\omega,\mathbf{p})
+ i  \,{\mathrm { Im}} \,\Sigma^R(\omega,\mathbf{p})
}
{\left( -\omega + \frac{\alpha \, \mathbf{p}^2}  {2\,m}
-\mu -{\mathrm { Re}} \,\Sigma^R(\omega,\mathbf{p})\right)^2
+\left[(1+\eta)  \,  |\mathbf{d}(\mathbf{p})|^2
+2 \,\eta \,\sum \limits_{a'=1}^5 s_{a'}
 \left (d^{a'} (\mathbf{p}) \right )^2\right]
+\left \lbrace {\mathrm { Im}} \,\Sigma^R(\omega,\mathbf{p}) \right \rbrace^2} \,.
\end{align}
This expressions leads to the one-loop corrected spectral function
\begin{align}
A(\omega  ,\mathbf{q})
&=-2 \, \mathrm{Im} \,G^R (\omega  ,\mathbf{q})\,.
\end{align}
The Kramers-Kronig relation connects the real and imaginary parts of any complex function, such that
\begin{align}
\mathrm { Re} \,\Sigma^R(\omega,\mathbf{p})
&=\int^\infty_{-\infty}\frac{d\omega'}{\pi}\frac{\mathrm { Im} \,\Sigma^R(\omega',\mathbf{p})}{\omega'-\omega},
\end{align}
which can be used for deriving $A(\omega  ,\mathbf{q})$, by plugging in the expression for $\mathrm{Im} \,\Sigma^R$ shown in Eq.~\eqref{eqimsigmarealfreq}.
Figs.~\ref{Fig_iso_A_u-p_T01-mu-01-plus} and~\ref{Fig_aniso_A_u-p_T05-mu-01-plus}
show some representative plots for the isotropic and anisotropic cases, respectively, for $\omega \geq0$. The parameter values for these plots are the same as those in Figs.~\ref{Fig_iso_tune-X} and~\ref{Fig_aniso-tau}.
Comparing Figs.~\ref{Fig_iso_tune-X} and~\ref{Fig_iso_A_u-p_T01-mu-01-plus} for an isotropic LSM, we find that the spectral function is strongly suppressed and gets
saturated at the locations where peaks of $1/\tau$ are observed. This implies that quasipartices are not even not well-defined in these regimes where plasmon modes are induced. For the anisotropic LSM, the basic tendencies are analogous, as seen from comparing Figs.~\ref{Fig_aniso-tau} and~\ref{Fig_aniso_A_u-p_T05-mu-01-plus}. The results for the $\omega  < 0$ case are again qualitatively similar, and hence not explicitly shown here.

The quasiparticle weight is given by the residue at the pole of the Green's function, whose nonzero value signals the existence of the quasiparticles. This is captured by
\begin{align}
Z_f(\mathbf{q})
&=\frac{1}{1-\lim \limits_{\omega\rightarrow0}\frac{\partial}{\partial\omega}\mathrm { Re} \,\Sigma^R(\omega,\mathbf{q})}.
\end{align}
Fig.~\ref{Fig_iso_Z_f_u-p_T01-mu-01-q-05}(a) is a representative contourplot of $Z_f$ for an isotropic LSM, using the same parameter values as in Fig.~\ref{Fig_iso_tune-X} (except $\tilde \varepsilon $).
A decrease in $Z_f$ signals a reduction in the weight of the quasiparticle while the plasmon mode emerges. Furthermore, Fig.~\ref{Fig_iso_Z_f_u-p_T01-mu-01-q-05}(b) clearly shows that $Z_f$ decreases as $X $ increases. This is due to the fact that the quasiparticles get destroyed and their weight is shifted to the plasmonic excitations as $X$ is cranked up.

\section{Summary and discussions}
\label{Sec_Summary}

To summarize, we have investigated the parameter regimes for the emergence of plasmons for isotropic, anisotropic, and band-mass symmetric and asymmetric Luttinger semimetals. A nonzero value of $T$ or $\mu$ is a necessity to get a plasmon, as otherwise the zero density of states at the QBCP can never lead to the appearance of this collective mode. The action of a nonzero temperature is to excite particle-hole pairs about the Fermi level due to thermal effects (even at zero doping), creating the possibility of the emergence of thermal plasmons~\cite{Mandal2019AP}. Needless to explain that a finite doping, on the other hand, sets the Fermi level away from the QBCP at any $T$, which naturally harbours a finite density of states.

We have also demonstrated the role of the material-dependent parameter $X =N_f\,\alpha_0 \,\tilde m$, which can favourably affect the emergence of plasmons. Although a nonzero $T$ or $\mu$ is necessary for plasmons to materialize, we have found that the specific values of $T$ and $\mu$ do not matter much, and the plasmons peaks are quite insensitive to a variation of these values. On the contrary, the parameter regimes for the existence of plasmon get broadened on increasing the value of $X$. Since $X$ is proportional to the number of fermion flavours, it gives us a powerful tuning parameter to influence the existence of the plasmons in LSMs.

There are two important properties that are elucidated through our numerical results and plots: (1) decay rate $\tau^{-1} $ of the quasiparticles, and (2) quasiparticle weight or residue $Z_f$. The values of both these factors determine whether conditions are favourable for the existence of well-defined and long-lived plasmons or quasiparticles. The favourable conditions for one are detrimental to the other. These questions were studied earlier exclusively for isotropic LSMs in  Refs.~\cite{Mandal2019AP,Tchoumakov-Krempa2019PRB,polini}.
While Ref.~\cite{Mandal2019AP} considered a finite $T$, Refs.~\cite{Tchoumakov-Krempa2019PRB,polini}
examined the effects of finite doping at $T=0$. Ref.~\cite{polini} also took band-mass asymmetries into account.
In this paper, we have considered the effects of nonzero values $T$ and $\mu$ simultaneously. But the most  important aspect of our computations is that we have considered generic LSM Hamiltonians, which include cubic anisotropies and band-mass asymmetries. These parameters are more likely to capture the features of realistic materials. Since it is not possible to get closed-form analytical approximations for the essential physical quantities in such a complicated system (not even for the retarded bare polarization bubble), we have obtained all our results by extensive numerical simulations. Our results will provide valuable information for future experiments engineered to measure transport or spectral properties of generic LSMs.

\section*{Acknowledgments}

J.W. is grateful to Zhao-Kun Yang for helpful discussions on the implementation parallel computing, and
has been partially supported by the National Natural Science Foundation of China under grant number
11504360.

\section*{Data availability statement}

No data was used for the research described in the article.

\section*{Author contribution statement}
I.M. conceived the original idea, developed the theory, and supervised the findings of this paper. J.W. performed the analytical calculations and the numerical simulations. All authors discussed the results and contributed to the final manuscript.


\appendix


\section{Appendix: Details for obtaining the one-loop polarization function}
\label{app}

We will outline here some of the important intermediate steps for obtaining the expressions for the bare polarization bubbles, shown in Sec.~\ref{Sec_bare-Pi}.

\subsection{Isotropic LSM}
\label{Subsec_A_Pi}

\begin{figure}
\centering
\subfigure[]{\includegraphics[width=0.45\textwidth]{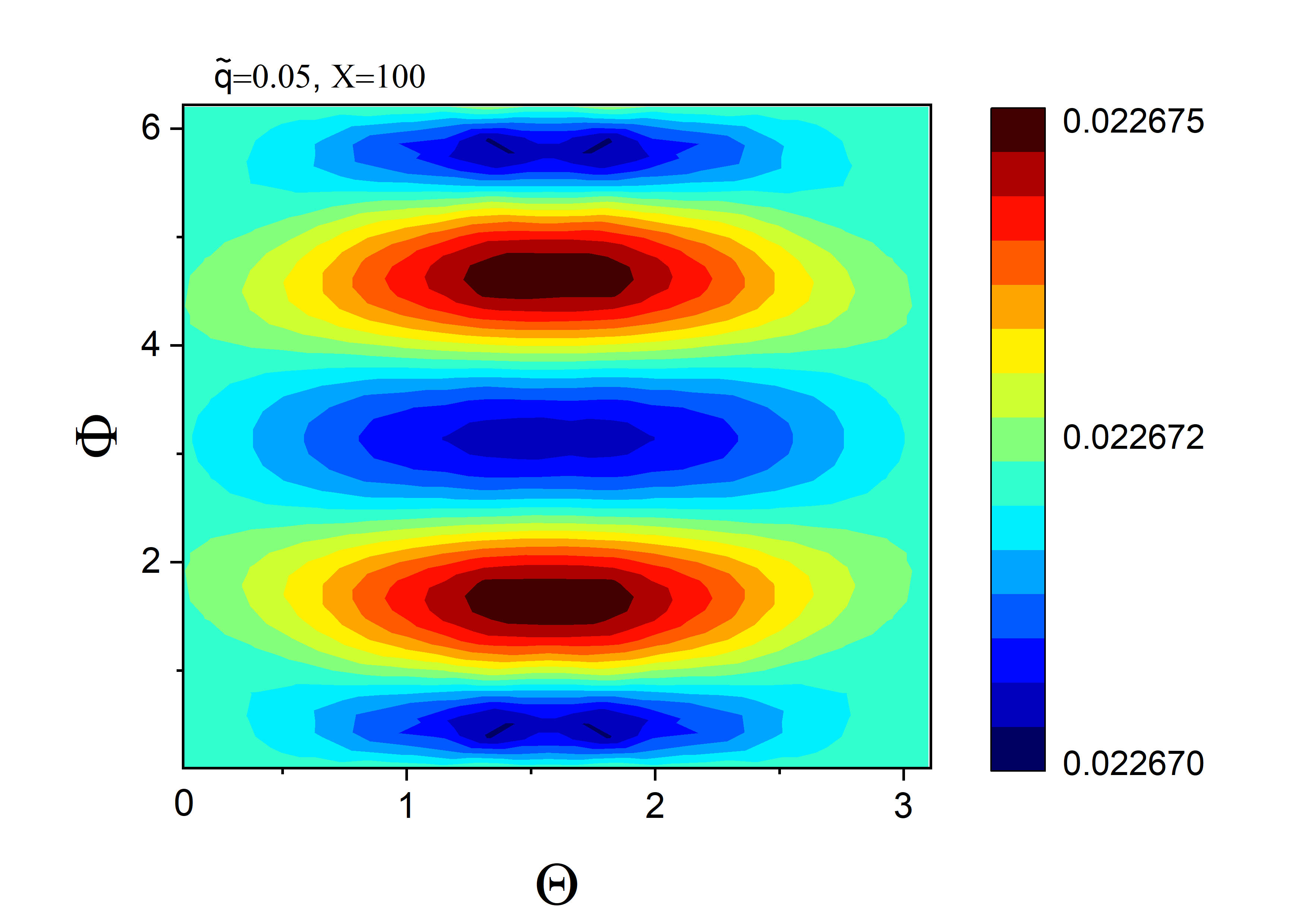}}\quad
\subfigure[]{\includegraphics[width=0.416\textwidth]{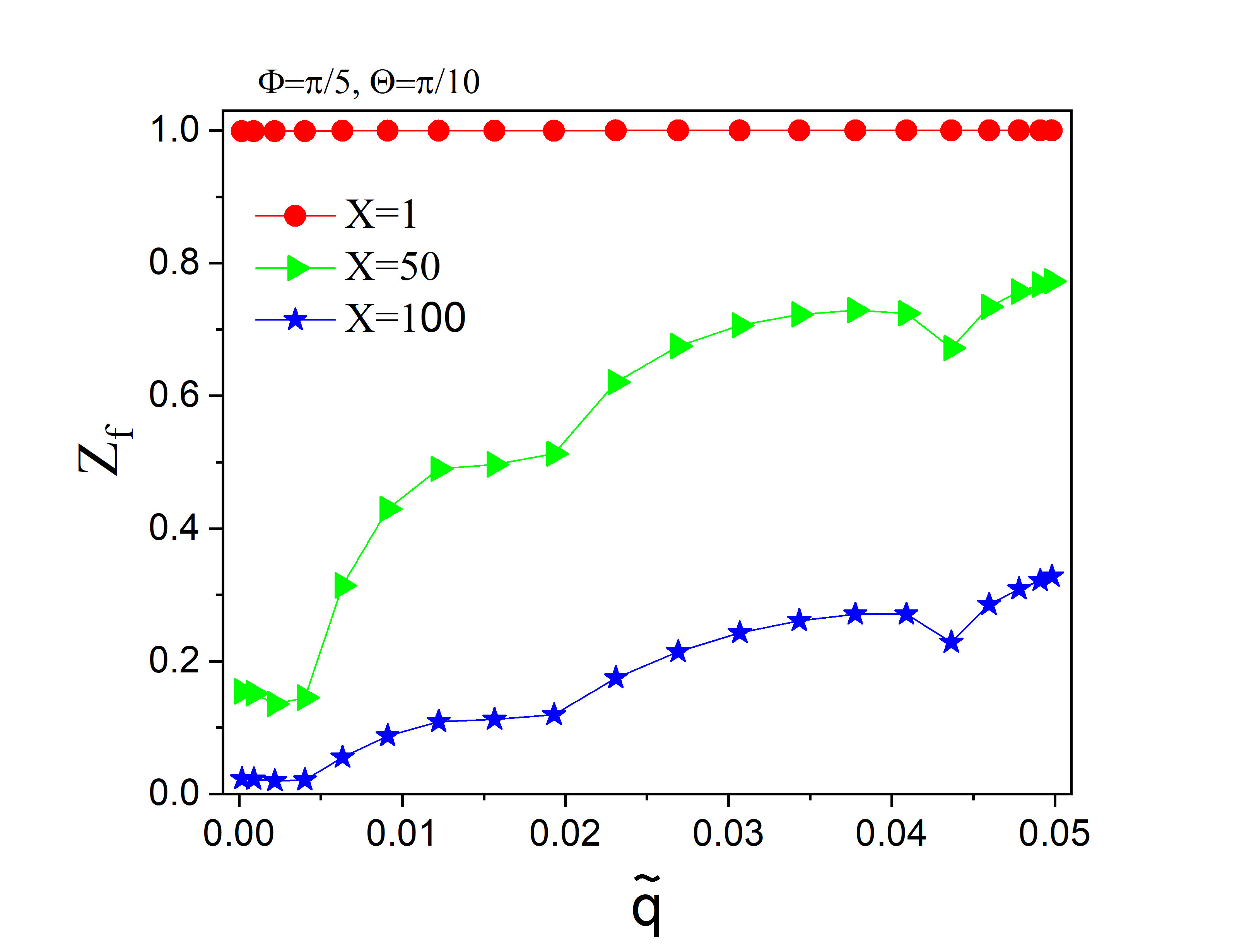}}\\
\caption{Behaviour of $Z_f$ for an isotropic LSM
with $\tilde T=0. 1$ and $\tilde \mu=0.1$:
Subfigure (a) captures the dependence on the angular variables
$\Theta$ and $\Phi$, for $ \tilde q = 0.05$ and $X = 100 $.
Subfigure (b) shows the dependence on $\tilde q$, with $\Phi= \pi/ 5 $ and
$\Theta= \pi / 10$, for $X= 1$ (red), $X= 50$ (green), and $X= 100$ (blue).
\label{Fig_iso_Z_f_u-p_T01-mu-01-q-05}
}
\end{figure}

Starting with Eq.~\eqref{eqmatpol}, and using Eq.~\eqref{Eq_G_0-mu}, we obtain
\begin{align}
 \Pi^R(\omega,\mathbf{q})
&=-\sum \limits_{\zeta_1, \,\zeta_2 =\pm1 }
\frac{\zeta_1 \,\zeta_2}  {8}
\int\frac{d^3\mathbf{p}}{(2 \, \pi)^3}
\frac{
\tanh \left [\frac{\beta \left \lbrace\zeta_1 \, |\mathbf{d}(\mathbf{p})|
-\mu \right \rbrace }{2}
\right ]-
\tanh\left[ \frac{\beta \left \lbrace \zeta_2 \, |\mathbf{d}(\mathbf{p+q})|
-\mu \right \rbrace}
{2}  \right ]
}
{\omega  + \zeta_1 \,|\mathbf{d}(\mathbf{p})|
- \zeta_2 \,|\mathbf{d}(\mathbf{p+q})|+i\,0^+}
\nonumber\\ &  \hspace{ 4.5 cm}\times
\frac{
\mathrm{Tr}\bigg[
\left \lbrace  \zeta_1 \, |\mathbf{d}(\mathbf{p})|+\mathbf{d}(\mathbf{p})\cdot {\mathbf \Gamma}
\right \rbrace
\left \lbrace \zeta_2\, |\mathbf{d}(\mathbf{p+q})|+\mathbf{d}(\mathbf{p+q})\cdot {\mathbf \Gamma}
\right \rbrace   \bigg]
}
{  |\mathbf{d}(\mathbf{p})|\, |\mathbf{d}(\mathbf{p+q})|}\,.
\end{align}

For computational convenience, we switch to the spherical coordinates, using the relations
\begin{align}
p_x= p \sin\theta\cos\phi \,,\quad
p_y=p  \sin\theta\sin\phi \,, \quad
p_z=p  \cos\theta \,,
\end{align}
and
\begin{align}
q_x=q\sin\Theta\cos\Phi \,,\quad
q_y=q\sin\Theta\sin\Phi\,, \quad
q_z=q\cos\Theta \,.
\end{align}
Using the identity
\begin{align}
\frac{1}{x-y+i\,0^+}
={\mathcal P} \left(\frac{1}{x-y}\right)-i\,\pi\,\delta(x-y)\,,
\end{align}
and defining $\Lambda_0$ as the inverse of the lattice spacing, we get
\begin{align}
\frac{\mathrm{Re} \, \Pi^R(\tilde \omega, \tilde q,\Theta,\Phi)}
{\tilde m  \, \Lambda_0^2 }  =\sum \limits_{\zeta_1, \,\zeta_2 =\pm1 }
\mathcal{K}_{\zeta_1\, \zeta_2 }  (\tilde \omega, \tilde q,\Theta,\Phi)
\,, \quad
\frac{\mathrm{Im} \,\Pi^R(\tilde \omega, \tilde q,\Theta,\Phi)}
{\tilde m  \, \Lambda_0^2}
& = \sum \limits_{\zeta_1, \,\zeta_2 =\pm1 }
\mathcal{F}_{\zeta_1\, \zeta_2 }  (\tilde \omega, \tilde q,\Theta,\Phi) \,.
\end{align}
The functions $\mathcal{K}_{\zeta_1\, \zeta_2 } $ and $ \mathcal{F}_{\zeta_1\, \zeta_2 } $
are given by
\begin{align}
\label{Eq_iso-KF}
& \mathcal{K}_{\zeta_1 \zeta_2} (\tilde \omega, \tilde q,\Theta,\Phi)
\nn &=-
\frac{\zeta_1\, \zeta_2 }  {16\,\pi^2}
\int_0^{1}  d {\tilde p} \,{\tilde p^2}
\int^{\pi}_0 d\theta \sin \theta
\int^{2 \, \pi}_0   d\phi \,
\left[\tanh\left ( \frac{ \zeta_1 \,| \boldsymbol{\eta}_{\mathbf p}|-\tilde \mu }
{2\,\tilde T}\right )
- \tanh\left ( \frac{ \zeta_2\,
 | \boldsymbol{\eta}_{\mathbf{ p+q}} |-\tilde \mu }
 {2\,\tilde T} \right ) \right]
\Bigl[
|{\boldsymbol \eta}_{\mathbf p}| \,| {\boldsymbol \eta}_{ \mathbf p +\mathbf q}|
+ {\boldsymbol \eta}_{\mathbf p} \cdot {\boldsymbol \eta}_{ \mathbf p +\mathbf q}
\Bigr]
\,
 \nn
& \hspace{ 6 cm} \times
{\mathcal P}
\bigg(\frac{1}
{ |{\boldsymbol \eta}_{\mathbf p}| \,| {\boldsymbol \eta}_{ \mathbf p +\mathbf q}|
\, (\tilde \omega-\zeta_1 \,|{\boldsymbol \eta}_{\mathbf p}|
-\zeta_2  |{\boldsymbol \eta}_{\mathbf{ p+q} }|)}
\bigg)\,,\nn
& \mathcal{F}_{\zeta_1\, \zeta_2 }  (\tilde \omega, \tilde q,\Theta,\Phi)
\nn &=
\frac{\zeta_1\, \zeta_2 }  {16\,\pi^2}
\int_0^{1}  d {\tilde p} \,{\tilde p^2}
\int^{\pi}_0 d\theta \sin \theta
\int^{2 \, \pi}_0   d\phi \,
\left[\tanh\left ( \frac{ \zeta_1 \,| \boldsymbol{\eta}_{\mathbf p}|-\tilde \mu }
{2\,\tilde T}\right )
- \tanh\left ( \frac{ \zeta_2\,
 | \boldsymbol{\eta}_{\mathbf{ p+q}} |-\tilde \mu }
 {2\,\tilde T} \right ) \right]
\nonumber\\
& \hspace{ 5.75 cm} \times
\left( 1 + \frac{
{\boldsymbol \eta}_{\mathbf p} \cdot {\boldsymbol \eta}_{ \mathbf p +\mathbf q}}
{|{\boldsymbol \eta}_{\mathbf p}| \, |{\boldsymbol \eta}_{\mathbf{ p+q}}|} \right)
\delta\Bigl( \tilde \omega +\zeta_1 \,|{\boldsymbol \eta}_{\mathbf p}|
-  \zeta_2\, |{\boldsymbol \eta}_{\mathbf{ p+q}}|
\Bigr) \,,
\end{align}
where
\begin{align}
\tilde m = m/\Lambda_0\,, \quad \tilde \omega = \omega /\Lambda_0\,,\quad
\tilde \mu = \mu /\Lambda_0\,, \quad \tilde T = T  /\Lambda_0\,,
\quad \tilde q = q/\Lambda_0\,, \quad
\tilde p =p/\Lambda_0\,, \quad
{\boldsymbol \eta}_{\mathbf p} =
\frac{2\, m\, {\mathbf d } ({\mathbf p})} {\Lambda_0^2}\,.
\end{align}

\subsection{Anisotropic and band-mass asymmetric LSMs}
\label{Subsec_B_Pi}

Following the same steps as outlined  for the isotropic case in the previous subsection, and using Eq.~\eqref{Eq_aniso-G_0}, the expression in Eq.~\eqref{eqmatpol} now leads to
\begin{align}
\frac{\mathrm{Re} \, \Pi^R(\tilde \omega, \tilde q,\Theta,\Phi)}
{\tilde m  \, \Lambda_0^2 }  =\sum \limits_{\zeta_1, \,\zeta_2 =\pm1 }
\mathcal{M}_{\zeta_1\, \zeta_2 }  (\tilde \omega, \tilde q,\Theta,\Phi)
\,, \quad
\frac{\mathrm{Im} \,\Pi^R(\tilde \omega, \tilde q,\Theta,\Phi)}
{\tilde m  \, \Lambda_0^2}
& = \sum \limits_{\zeta_1, \,\zeta_2 =\pm1 }
\mathcal{J}_{\zeta_1\, \zeta_2 }  (\tilde \omega, \tilde q,\Theta,\Phi) \,,
\end{align}
where
\begin{align}
\label{eq_MJ}
& \mathcal{M}_{\zeta_1 \zeta_2} (\tilde \omega, \tilde q,\Theta,\Phi)
\nn &=-
\frac{\zeta_1 \, \zeta_2 } {16 \,\pi^2}
\int_0^{1}  d \tilde p \,{\tilde p^2}
\int^{\pi}_0 d\theta \sin\theta
\int^{2 \, \pi}_0 d\phi
\left[
\tanh \bigg (\frac{ \mathcal{B}_{\zeta_1} (\mathbf p)- \mu } {2\, T} \bigg )
- \tanh \bigg (\frac{ \mathcal{B}_{\zeta_2} (\mathbf {p+q} )- \mu} {2 \, T} \bigg )\right]
\nonumber\\
&  \hspace{ 6 cm}\times
\mathcal{P} \bigg(
\frac{\Lambda_0^4}
{ m^2 \mathcal{A} (\mathbf p) \, \mathcal{A} (\mathbf {p+q})}
\times \frac{1}
{\tilde \omega+ m\,\frac{
\mathcal{B}_{\zeta_1}(\mathbf p) -\mathcal{B}_{\zeta_2} (\mathbf {p+q} )}
{ \Lambda_0^2} }
\bigg)
\nonumber\\
&  \hspace{ 6 cm}\times
\Bigg \{
\zeta_1 \,\zeta_2  + \frac{
(1+\eta^2)\,{\mathbf d}(\mathbf{p}) \cdot {\mathbf d} (\mathbf{p+q})
+ 2\, \eta \sum \limits^5_{a=1}s_a \, d_a (\mathbf{p}) \,d_a(\mathbf{p+q})
}
{
\mathcal{A} (\mathbf p) \, \mathcal{A} (\mathbf p +\mathbf q)
}
\Bigg \}\,, \nn
& \mathcal{J}_{\zeta_1 \zeta_2} (\tilde \omega, \tilde q,\Theta,\Phi)
\nn &=\frac{\zeta_1 \, \zeta_2 } {16 \,\pi^2}
\int_0^{1}  d \tilde p \,{\tilde p^2}
\int^{\pi}_0 d\theta \sin\theta
\int^{2 \, \pi}_0 d\phi
\left[
\tanh \bigg (\frac{ \mathcal{B}_{\zeta_1} (\mathbf p)- \mu } {2\, T} \bigg )
- \tanh \bigg (\frac{ \mathcal{B}_{\zeta_2} (\mathbf {p+q} )- \mu} {2 \, T} \bigg )\right]
\delta \bigg(
\tilde \omega+ m\,\frac{
 \mathcal{B}_{\zeta_1} (\mathbf p) - \mathcal{B}_{\zeta_2} (\mathbf {p+q} )}
{ \Lambda_0^2}
\bigg)
\nonumber\\
&  \hspace{ 5.75 cm}\times
\Bigg \{
\zeta_1 \,\zeta_2  + \frac{
(1+\eta^2)\,{\mathbf d}(\mathbf{p}) \cdot {\mathbf d} (\mathbf{p+q})
+ 2\, \eta \sum \limits^5_{a=1}s_a \, d_a (\mathbf{p}) \,d_a(\mathbf{p+q})
}
{
\mathcal{A} (\mathbf p) \, \mathcal{A} (\mathbf p +\mathbf q)
}
\Bigg \}\,,
\end{align}
where
\begin{align}
\mathcal{A} (\mathbf p) & =
\sqrt{(1+\eta)\, {\mathbf d}^2 (\mathbf{p})
+2 \,\eta  \sum \limits^5_{a=1} s_{a} \,d_{a}^2(\mathbf{p})
} \,,\quad
\mathcal{B}_{\zeta} (\mathbf p)
=\frac{\alpha\, {\mathbf p}^2}{2\,m}-{\zeta}\, \mathcal{A} (\mathbf p)\,.
\end{align}

\bibliography{biblio}

\begin{thebibliography}{37}%
\makeatletter
\providecommand \@ifxundefined [1]{%
 \@ifx{#1\undefined}
}%
\providecommand \@ifnum [1]{%
 \ifnum #1\expandafter \@firstoftwo
 \else \expandafter \@secondoftwo
 \fi
}%
\providecommand \@ifx [1]{%
 \ifx #1\expandafter \@firstoftwo
 \else \expandafter \@secondoftwo
 \fi
}%
\providecommand \natexlab [1]{#1}%
\providecommand \enquote  [1]{``#1''}%
\providecommand \bibnamefont  [1]{#1}%
\providecommand \bibfnamefont [1]{#1}%
\providecommand \citenamefont [1]{#1}%
\providecommand \href@noop [0]{\@secondoftwo}%
\providecommand \href [0]{\begingroup \@sanitize@url \@href}%
\providecommand \@href[1]{\@@startlink{#1}\@@href}%
\providecommand \@@href[1]{\endgroup#1\@@endlink}%
\providecommand \@sanitize@url [0]{\catcode `\\12\catcode `\$12\catcode
  `\&12\catcode `\#12\catcode `\^12\catcode `\_12\catcode `\%12\relax}%
\providecommand \@@startlink[1]{}%
\providecommand \@@endlink[0]{}%
\providecommand \url  [0]{\begingroup\@sanitize@url \@url }%
\providecommand \@url [1]{\endgroup\@href {#1}{\urlprefix }}%
\providecommand \urlprefix  [0]{URL }%
\providecommand \Eprint [0]{\href }%
\providecommand \doibase [0]{https://doi.org/}%
\providecommand \selectlanguage [0]{\@gobble}%
\providecommand \bibinfo  [0]{\@secondoftwo}%
\providecommand \bibfield  [0]{\@secondoftwo}%
\providecommand \translation [1]{[#1]}%
\providecommand \BibitemOpen [0]{}%
\providecommand \bibitemStop [0]{}%
\providecommand \bibitemNoStop [0]{.\EOS\space}%
\providecommand \EOS [0]{\spacefactor3000\relax}%
\providecommand \BibitemShut  [1]{\csname bibitem#1\endcsname}%
\let\auto@bib@innerbib\@empty
\bibitem [{\citenamefont {Moon}\ \emph {et~al.}(2013)\citenamefont {Moon},
  \citenamefont {Xu}, \citenamefont {Kim},\ and\ \citenamefont
  {Balents}}]{Moon2013PRL}%
  \BibitemOpen
  \bibfield  {author} {\bibinfo {author} {\bibfnamefont {E.-G.}\ \bibnamefont
  {Moon}}, \bibinfo {author} {\bibfnamefont {C.}~\bibnamefont {Xu}}, \bibinfo
  {author} {\bibfnamefont {Y.~B.}\ \bibnamefont {Kim}},\ and\ \bibinfo {author}
  {\bibfnamefont {L.}~\bibnamefont {Balents}},\ }\bibfield  {title} {\bibinfo
  {title} {Non-{F}ermi-liquid and topological states with strong spin-orbit
  coupling},\ }\href {https://doi.org/10.1103/PhysRevLett.111.206401}
  {\bibfield  {journal} {\bibinfo  {journal} {Phys. Rev. Lett.}\ }\textbf
  {\bibinfo {volume} {111}},\ \bibinfo {pages} {206401} (\bibinfo {year}
  {2013})}\BibitemShut {NoStop}%
\bibitem [{\citenamefont {{Kondo}}\ \emph {et~al.}(2015)\citenamefont
  {{Kondo}}, \citenamefont {{Nakayama}}, \citenamefont {{Chen}}, \citenamefont
  {{Ishikawa}}, \citenamefont {{Moon}}, \citenamefont {{Yamamoto}},
  \citenamefont {{Ota}}, \citenamefont {{Malaeb}}, \citenamefont {{Kanai}},
  \citenamefont {{Nakashima}}, \citenamefont {{Ishida}}, \citenamefont
  {{Yoshida}}, \citenamefont {{Yamamoto}}, \citenamefont {{Matsunami}},
  \citenamefont {{Kimura}}, \citenamefont {{Inami}}, \citenamefont {{Ono}},
  \citenamefont {{Kumigashira}}, \citenamefont {{Nakatsuji}}, \citenamefont
  {{Balents}},\ and\ \citenamefont {{Shin}}}]{kondo2015NC}%
  \BibitemOpen
  \bibfield  {author} {\bibinfo {author} {\bibfnamefont {T.}~\bibnamefont
  {{Kondo}}}, \bibinfo {author} {\bibfnamefont {M.}~\bibnamefont {{Nakayama}}},
  \bibinfo {author} {\bibfnamefont {R.}~\bibnamefont {{Chen}}}, \bibinfo
  {author} {\bibfnamefont {J.~J.}\ \bibnamefont {{Ishikawa}}}, \bibinfo
  {author} {\bibfnamefont {E.~G.}\ \bibnamefont {{Moon}}}, \bibinfo {author}
  {\bibfnamefont {T.}~\bibnamefont {{Yamamoto}}}, \bibinfo {author}
  {\bibfnamefont {Y.}~\bibnamefont {{Ota}}}, \bibinfo {author} {\bibfnamefont
  {W.}~\bibnamefont {{Malaeb}}}, \bibinfo {author} {\bibfnamefont
  {H.}~\bibnamefont {{Kanai}}}, \bibinfo {author} {\bibfnamefont
  {Y.}~\bibnamefont {{Nakashima}}}, \bibinfo {author} {\bibfnamefont
  {Y.}~\bibnamefont {{Ishida}}}, \bibinfo {author} {\bibfnamefont
  {R.}~\bibnamefont {{Yoshida}}}, \bibinfo {author} {\bibfnamefont
  {H.}~\bibnamefont {{Yamamoto}}}, \bibinfo {author} {\bibfnamefont
  {M.}~\bibnamefont {{Matsunami}}}, \bibinfo {author} {\bibfnamefont
  {S.}~\bibnamefont {{Kimura}}}, \bibinfo {author} {\bibfnamefont
  {N.}~\bibnamefont {{Inami}}}, \bibinfo {author} {\bibfnamefont
  {K.}~\bibnamefont {{Ono}}}, \bibinfo {author} {\bibfnamefont
  {H.}~\bibnamefont {{Kumigashira}}}, \bibinfo {author} {\bibfnamefont
  {S.}~\bibnamefont {{Nakatsuji}}}, \bibinfo {author} {\bibfnamefont
  {L.}~\bibnamefont {{Balents}}},\ and\ \bibinfo {author} {\bibfnamefont
  {S.}~\bibnamefont {{Shin}}},\ }\bibfield  {title} {\bibinfo {title}
  {Quadratic {F}ermi node in a 3d strongly correlated semimetal},\ }\href
  {https://doi.org/10.1038/ncomms10042} {\bibfield  {journal} {\bibinfo
  {journal} {Nature Communications}\ }\textbf {\bibinfo {volume} {6}},\
  \bibinfo {eid} {10042} (\bibinfo {year} {2015})}\BibitemShut {NoStop}%
\bibitem [{\citenamefont {D\'ora}\ and\ \citenamefont
  {Herbut}(2016)}]{Herbut2017PRB}%
  \BibitemOpen
  \bibfield  {author} {\bibinfo {author} {\bibfnamefont {B.}~\bibnamefont
  {D\'ora}}\ and\ \bibinfo {author} {\bibfnamefont {I.~F.}\ \bibnamefont
  {Herbut}},\ }\bibfield  {title} {\bibinfo {title} {Quadratic band touching
  with long-range interactions in and out of equilibrium},\ }\href
  {https://doi.org/10.1103/PhysRevB.94.155134} {\bibfield  {journal} {\bibinfo
  {journal} {Phys. Rev. B}\ }\textbf {\bibinfo {volume} {94}},\ \bibinfo
  {pages} {155134} (\bibinfo {year} {2016})}\BibitemShut {NoStop}%
\bibitem [{\citenamefont {Boettcher}\ and\ \citenamefont
  {Herbut}(2017)}]{Herbut2017PRB_2}%
  \BibitemOpen
  \bibfield  {author} {\bibinfo {author} {\bibfnamefont {I.}~\bibnamefont
  {Boettcher}}\ and\ \bibinfo {author} {\bibfnamefont {I.~F.}\ \bibnamefont
  {Herbut}},\ }\bibfield  {title} {\bibinfo {title} {Anisotropy induces
  non-{F}ermi-liquid behavior and nematic magnetic order in three-dimensional
  {L}uttinger semimetals},\ }\href {https://doi.org/10.1103/PhysRevB.95.075149}
  {\bibfield  {journal} {\bibinfo  {journal} {Phys. Rev. B}\ }\textbf {\bibinfo
  {volume} {95}},\ \bibinfo {pages} {075149} (\bibinfo {year}
  {2017})}\BibitemShut {NoStop}%
\bibitem [{\citenamefont {Nandkishore}\ and\ \citenamefont
  {Parameswaran}(2017)}]{Nandkishore2017PRB}%
  \BibitemOpen
  \bibfield  {author} {\bibinfo {author} {\bibfnamefont {R.~M.}\ \bibnamefont
  {Nandkishore}}\ and\ \bibinfo {author} {\bibfnamefont {S.~A.}\ \bibnamefont
  {Parameswaran}},\ }\bibfield  {title} {\bibinfo {title} {Disorder-driven
  destruction of a non-{F}ermi liquid semimetal studied by renormalization
  group analysis},\ }\href {https://doi.org/10.1103/PhysRevB.95.205106}
  {\bibfield  {journal} {\bibinfo  {journal} {Phys. Rev. B}\ }\textbf {\bibinfo
  {volume} {95}},\ \bibinfo {pages} {205106} (\bibinfo {year}
  {2017})}\BibitemShut {NoStop}%
\bibitem [{\citenamefont {Boettcher}\ and\ \citenamefont
  {Herbut}(2016)}]{Herbut2016PRB}%
  \BibitemOpen
  \bibfield  {author} {\bibinfo {author} {\bibfnamefont {I.}~\bibnamefont
  {Boettcher}}\ and\ \bibinfo {author} {\bibfnamefont {I.~F.}\ \bibnamefont
  {Herbut}},\ }\bibfield  {title} {\bibinfo {title} {Superconducting quantum
  criticality in three-dimensional {L}uttinger semimetals},\ }\href
  {https://doi.org/10.1103/PhysRevB.93.205138} {\bibfield  {journal} {\bibinfo
  {journal} {Phys. Rev. B}\ }\textbf {\bibinfo {volume} {93}},\ \bibinfo
  {pages} {205138} (\bibinfo {year} {2016})}\BibitemShut {NoStop}%
\bibitem [{\citenamefont {Mandal}(2018)}]{ips-qbt-sc}%
  \BibitemOpen
  \bibfield  {author} {\bibinfo {author} {\bibfnamefont {I.}~\bibnamefont
  {Mandal}},\ }\bibfield  {title} {\bibinfo {title} {Fate of superconductivity
  in three-dimensional disordered {L}uttinger semimetals},\ }\href
  {https://doi.org/https://doi.org/10.1016/j.aop.2018.03.004} {\bibfield
  {journal} {\bibinfo  {journal} {Annals of Physics}\ }\textbf {\bibinfo
  {volume} {392}},\ \bibinfo {pages} {179 } (\bibinfo {year}
  {2018})}\BibitemShut {NoStop}%
\bibitem [{\citenamefont {Mandal}\ and\ \citenamefont
  {Nandkishore}(2018)}]{Mandal2018PRB}%
  \BibitemOpen
  \bibfield  {author} {\bibinfo {author} {\bibfnamefont {I.}~\bibnamefont
  {Mandal}}\ and\ \bibinfo {author} {\bibfnamefont {R.~M.}\ \bibnamefont
  {Nandkishore}},\ }\bibfield  {title} {\bibinfo {title} {Interplay of
  {C}oulomb interactions and disorder in three-dimensional quadratic band
  crossings without time-reversal symmetry and with unequal masses for
  conduction and valence bands},\ }\href
  {https://doi.org/10.1103/PhysRevB.97.125121} {\bibfield  {journal} {\bibinfo
  {journal} {Phys. Rev. B}\ }\textbf {\bibinfo {volume} {97}},\ \bibinfo
  {pages} {125121} (\bibinfo {year} {2018})}\BibitemShut {NoStop}%
\bibitem [{\citenamefont {Zhai}\ and\ \citenamefont
  {Wang}(2020)}]{Wang2019EPJB}%
  \BibitemOpen
  \bibfield  {author} {\bibinfo {author} {\bibfnamefont {Y.-H.}\ \bibnamefont
  {Zhai}}\ and\ \bibinfo {author} {\bibfnamefont {J.}~\bibnamefont {Wang}},\
  }\bibfield  {title} {\bibinfo {title} {Effects of fermion-fermion
  interactions and impurity scatterings on fermion velocities in the line-nodal
  superconductors},\ }\href {https://doi.org/10.1140/epjb/e2020-10049-x}
  {\bibfield  {journal} {\bibinfo  {journal} {The European Physical Journal B}\
  }\textbf {\bibinfo {volume} {93}},\ \bibinfo {pages} {86} (\bibinfo {year}
  {2020})}\BibitemShut {NoStop}%
\bibitem [{\citenamefont {Mandal}(2019)}]{Mandal2019AP}%
  \BibitemOpen
  \bibfield  {author} {\bibinfo {author} {\bibfnamefont {I.}~\bibnamefont
  {Mandal}},\ }\bibfield  {title} {\bibinfo {title} {Search for plasmons in
  isotropic {L}uttinger semimetals},\ }\href
  {https://doi.org/https://doi.org/10.1016/j.aop.2019.04.002} {\bibfield
  {journal} {\bibinfo  {journal} {Annals of Physics}\ }\textbf {\bibinfo
  {volume} {406}},\ \bibinfo {pages} {173} (\bibinfo {year}
  {2019})}\BibitemShut {NoStop}%
\bibitem [{\citenamefont {Mandal}\ and\ \citenamefont
  {Freire}(2021)}]{ips-herm1}%
  \BibitemOpen
  \bibfield  {author} {\bibinfo {author} {\bibfnamefont {I.}~\bibnamefont
  {Mandal}}\ and\ \bibinfo {author} {\bibfnamefont {H.}~\bibnamefont
  {Freire}},\ }\bibfield  {title} {\bibinfo {title} {{Transport in the
  non-{F}ermi liquid phase of isotropic {L}uttinger semimetals}},\ }\href
  {https://doi.org/10.1103/PhysRevB.103.195116} {\bibfield  {journal} {\bibinfo
   {journal} {Phys. Rev. B}\ }\textbf {\bibinfo {volume} {103}},\ \bibinfo
  {pages} {195116} (\bibinfo {year} {2021})}\BibitemShut {NoStop}%
\bibitem [{\citenamefont {Freire}\ and\ \citenamefont
  {Mandal}(2021)}]{ips-herm2}%
  \BibitemOpen
  \bibfield  {author} {\bibinfo {author} {\bibfnamefont {H.}~\bibnamefont
  {Freire}}\ and\ \bibinfo {author} {\bibfnamefont {I.}~\bibnamefont
  {Mandal}},\ }\bibfield  {title} {\bibinfo {title} {{Thermoelectric and
  thermal properties of the weakly disordered non-{F}ermi liquid phase of
  {L}uttinger semimetals}},\ }\href
  {https://doi.org/10.1016/j.physleta.2021.127470} {\bibfield  {journal}
  {\bibinfo  {journal} {Physics Letters A}\ }\textbf {\bibinfo {volume}
  {407}},\ \bibinfo {pages} {127470} (\bibinfo {year} {2021})}\BibitemShut
  {NoStop}%
\bibitem [{\citenamefont {{Mandal}}\ and\ \citenamefont
  {{Freire}}(2022)}]{ips-herm3}%
  \BibitemOpen
  \bibfield  {author} {\bibinfo {author} {\bibfnamefont {I.}~\bibnamefont
  {{Mandal}}}\ and\ \bibinfo {author} {\bibfnamefont {H.}~\bibnamefont
  {{Freire}}},\ }\bibfield  {title} {\bibinfo {title} {{Raman response and
  shear viscosity in the non-{F}ermi liquid phase of {L}uttinger semimetals}},\
  }\href {https://doi.org/10.1088/1361-648X/ac6785} {\bibfield  {journal}
  {\bibinfo  {journal} {Journal of Physics Condensed Matter}\ }\textbf
  {\bibinfo {volume} {34}},\ \bibinfo {eid} {275604} (\bibinfo {year}
  {2022})}\BibitemShut {NoStop}%
\bibitem [{\citenamefont {Mandal}(2020)}]{ips-tunnel-qbcp}%
  \BibitemOpen
  \bibfield  {author} {\bibinfo {author} {\bibfnamefont {I.}~\bibnamefont
  {Mandal}},\ }\bibfield  {title} {\bibinfo {title} {Tunneling in {F}ermi
  systems with quadratic band crossing points},\ }\href
  {https://doi.org/https://doi.org/10.1016/j.aop.2020.168235} {\bibfield
  {journal} {\bibinfo  {journal} {Annals of Physics}\ }\textbf {\bibinfo
  {volume} {419}},\ \bibinfo {pages} {168235} (\bibinfo {year}
  {2020})}\BibitemShut {NoStop}%
\bibitem [{\citenamefont {Bera}\ and\ \citenamefont {Mandal}(2021)}]{Bera2021}%
  \BibitemOpen
  \bibfield  {author} {\bibinfo {author} {\bibfnamefont {S.}~\bibnamefont
  {Bera}}\ and\ \bibinfo {author} {\bibfnamefont {I.}~\bibnamefont {Mandal}},\
  }\bibfield  {title} {\bibinfo {title} {Floquet scattering of quadratic
  band-touching semimetals through a time-periodic potential well},\ }\href
  {https://doi.org/10.1088/1361-648x/ac020a} {\bibfield  {journal} {\bibinfo
  {journal} {Journal of Physics: Condensed Matter}\ }\textbf {\bibinfo {volume}
  {33}},\ \bibinfo {pages} {295502} (\bibinfo {year} {2021})}\BibitemShut
  {NoStop}%
\bibitem [{\citenamefont {Castro~Neto}\ \emph {et~al.}(2009)\citenamefont
  {Castro~Neto}, \citenamefont {Guinea}, \citenamefont {Peres}, \citenamefont
  {Novoselov},\ and\ \citenamefont {Geim}}]{Neto2009RMP}%
  \BibitemOpen
  \bibfield  {author} {\bibinfo {author} {\bibfnamefont {A.~H.}\ \bibnamefont
  {Castro~Neto}}, \bibinfo {author} {\bibfnamefont {F.}~\bibnamefont {Guinea}},
  \bibinfo {author} {\bibfnamefont {N.~M.~R.}\ \bibnamefont {Peres}}, \bibinfo
  {author} {\bibfnamefont {K.~S.}\ \bibnamefont {Novoselov}},\ and\ \bibinfo
  {author} {\bibfnamefont {A.~K.}\ \bibnamefont {Geim}},\ }\bibfield  {title}
  {\bibinfo {title} {The electronic properties of graphene},\ }\href
  {https://doi.org/10.1103/RevModPhys.81.109} {\bibfield  {journal} {\bibinfo
  {journal} {Rev. Mod. Phys.}\ }\textbf {\bibinfo {volume} {81}},\ \bibinfo
  {pages} {109} (\bibinfo {year} {2009})}\BibitemShut {NoStop}%
\bibitem [{\citenamefont {Yanagishima}\ and\ \citenamefont
  {Maeno}(2001)}]{Yanagishima2001}%
  \BibitemOpen
  \bibfield  {author} {\bibinfo {author} {\bibfnamefont {D.}~\bibnamefont
  {Yanagishima}}\ and\ \bibinfo {author} {\bibfnamefont {Y.}~\bibnamefont
  {Maeno}},\ }\bibfield  {title} {\bibinfo {title} {Metal-nonmetal changeover
  in pyrochlore iridates},\ }\href {https://doi.org/10.1143/JPSJ.70.2880}
  {\bibfield  {journal} {\bibinfo  {journal} {Journal of the Physical Society
  of Japan}\ }\textbf {\bibinfo {volume} {70}},\ \bibinfo {pages} {2880}
  (\bibinfo {year} {2001})}\BibitemShut {NoStop}%
\bibitem [{\citenamefont {Matsuhira}\ \emph {et~al.}(2007)\citenamefont
  {Matsuhira}, \citenamefont {Wakeshima}, \citenamefont {Nakanishi},
  \citenamefont {Yamada}, \citenamefont {Nakamura}, \citenamefont {Kawano},
  \citenamefont {Takagi},\ and\ \citenamefont {Hinatsu}}]{Matsuhira2007}%
  \BibitemOpen
  \bibfield  {author} {\bibinfo {author} {\bibfnamefont {K.}~\bibnamefont
  {Matsuhira}}, \bibinfo {author} {\bibfnamefont {M.}~\bibnamefont
  {Wakeshima}}, \bibinfo {author} {\bibfnamefont {R.}~\bibnamefont
  {Nakanishi}}, \bibinfo {author} {\bibfnamefont {T.}~\bibnamefont {Yamada}},
  \bibinfo {author} {\bibfnamefont {A.}~\bibnamefont {Nakamura}}, \bibinfo
  {author} {\bibfnamefont {W.}~\bibnamefont {Kawano}}, \bibinfo {author}
  {\bibfnamefont {S.}~\bibnamefont {Takagi}},\ and\ \bibinfo {author}
  {\bibfnamefont {Y.}~\bibnamefont {Hinatsu}},\ }\bibfield  {title} {\bibinfo
  {title} {{Metal–Insulator transition in pyrochlore iridates
  Ln$_2$Ir$_2$O$_7$ (Ln = Nd, Sm, and Eu)}},\ }\href
  {https://doi.org/10.1143/JPSJ.76.043706} {\bibfield  {journal} {\bibinfo
  {journal} {Journal of the Physical Society of Japan}\ }\textbf {\bibinfo
  {volume} {76}},\ \bibinfo {pages} {043706} (\bibinfo {year}
  {2007})}\BibitemShut {NoStop}%
\bibitem [{\citenamefont {Groves}\ and\ \citenamefont {Paul}(1963)}]{paul}%
  \BibitemOpen
  \bibfield  {author} {\bibinfo {author} {\bibfnamefont {S.}~\bibnamefont
  {Groves}}\ and\ \bibinfo {author} {\bibfnamefont {W.}~\bibnamefont {Paul}},\
  }\bibfield  {title} {\bibinfo {title} {Band structure of gray tin},\ }\href
  {https://doi.org/10.1103/PhysRevLett.11.194} {\bibfield  {journal} {\bibinfo
  {journal} {Phys. Rev. Lett.}\ }\textbf {\bibinfo {volume} {11}},\ \bibinfo
  {pages} {194} (\bibinfo {year} {1963})}\BibitemShut {NoStop}%
\bibitem [{\citenamefont {{Barbedienne}}\ \emph {et~al.}(2018)\citenamefont
  {{Barbedienne}}, \citenamefont {{Varignon}}, \citenamefont {{Reyren}},
  \citenamefont {{Marty}}, \citenamefont {{Vergnaud}}, \citenamefont {{Jamet}},
  \citenamefont {{Gomez-Carbonell}}, \citenamefont {{Lema{\^\i}tre}},
  \citenamefont {{Le F{\`e}vre}}, \citenamefont {{Bertran}}, \citenamefont
  {{Taleb-Ibrahimi}}, \citenamefont {{Jaffr{\`e}s}}, \citenamefont {{George}},\
  and\ \citenamefont {{Fert}}}]{gray-tin}%
  \BibitemOpen
  \bibfield  {author} {\bibinfo {author} {\bibfnamefont {Q.}~\bibnamefont
  {{Barbedienne}}}, \bibinfo {author} {\bibfnamefont {J.}~\bibnamefont
  {{Varignon}}}, \bibinfo {author} {\bibfnamefont {N.}~\bibnamefont
  {{Reyren}}}, \bibinfo {author} {\bibfnamefont {A.}~\bibnamefont {{Marty}}},
  \bibinfo {author} {\bibfnamefont {C.}~\bibnamefont {{Vergnaud}}}, \bibinfo
  {author} {\bibfnamefont {M.}~\bibnamefont {{Jamet}}}, \bibinfo {author}
  {\bibfnamefont {C.}~\bibnamefont {{Gomez-Carbonell}}}, \bibinfo {author}
  {\bibfnamefont {A.}~\bibnamefont {{Lema{\^\i}tre}}}, \bibinfo {author}
  {\bibfnamefont {P.}~\bibnamefont {{Le F{\`e}vre}}}, \bibinfo {author}
  {\bibfnamefont {F.}~\bibnamefont {{Bertran}}}, \bibinfo {author}
  {\bibfnamefont {A.}~\bibnamefont {{Taleb-Ibrahimi}}}, \bibinfo {author}
  {\bibfnamefont {H.}~\bibnamefont {{Jaffr{\`e}s}}}, \bibinfo {author}
  {\bibfnamefont {J.-M.}\ \bibnamefont {{George}}},\ and\ \bibinfo {author}
  {\bibfnamefont {A.}~\bibnamefont {{Fert}}},\ }\bibfield  {title} {\bibinfo
  {title} {Angular-resolved photoemission electron spectroscopy and transport
  studies of the elemental topological insulator $\alpha $-{S}n},\ }\href
  {https://doi.org/10.1103/PhysRevB.98.195445} {\bibfield  {journal} {\bibinfo
  {journal} {Phys. Rev. B}\ }\textbf {\bibinfo {volume} {98}},\ \bibinfo
  {pages} {195445} (\bibinfo {year} {2018})}\BibitemShut {NoStop}%
\bibitem [{\citenamefont {Tsidilkovski}(1997)}]{Tsidilkovski1997Book}%
  \BibitemOpen
  \bibfield  {author} {\bibinfo {author} {\bibfnamefont {I.~M.}\ \bibnamefont
  {Tsidilkovski}},\ }\bibfield  {title} {\bibinfo {title} {Band-structure
  calculation methods},\ }in\ \href@noop {} {\emph {\bibinfo {booktitle}
  {Electron Spectrum of Gapless Semiconductors}}}\ (\bibinfo  {publisher}
  {Springer},\ \bibinfo {year} {1997})\ pp.\ \bibinfo {pages}
  {3--52}\BibitemShut {NoStop}%
\bibitem [{\citenamefont {Luttinger}(1956)}]{Luttinger1956PR}%
  \BibitemOpen
  \bibfield  {author} {\bibinfo {author} {\bibfnamefont {J.~M.}\ \bibnamefont
  {Luttinger}},\ }\bibfield  {title} {\bibinfo {title} {Quantum theory of
  cyclotron resonance in semiconductors: General theory},\ }\href
  {https://doi.org/10.1103/PhysRev.102.1030} {\bibfield  {journal} {\bibinfo
  {journal} {Phys. Rev.}\ }\textbf {\bibinfo {volume} {102}},\ \bibinfo {pages}
  {1030} (\bibinfo {year} {1956})}\BibitemShut {NoStop}%
\bibitem [{\citenamefont {Abrikosov}\ and\ \citenamefont
  {Beneslavski\u{i}}()}]{abrikosov1996}%
  \BibitemOpen
  \bibfield  {author} {\bibinfo {author} {\bibfnamefont {A.~A.}\ \bibnamefont
  {Abrikosov}}\ and\ \bibinfo {author} {\bibfnamefont {S.~D.}\ \bibnamefont
  {Beneslavski\u{i}}},\ }\bibinfo {title} {Possible existence of substances
  intermediate between metals and dielectrics},\ in\ \href
  {https://doi.org/10.1142/9789814317344_0010} {\emph {\bibinfo {booktitle} {30
  Years of the Landau Institute — Selected Papers}}},\ pp.\ \bibinfo {pages}
  {64--73}\BibitemShut {NoStop}%
\bibitem [{\citenamefont {Murakami}\ \emph {et~al.}(2004)\citenamefont
  {Murakami}, \citenamefont {Nagosa},\ and\ \citenamefont
  {Zhang}}]{Shuichi2004PRB}%
  \BibitemOpen
  \bibfield  {author} {\bibinfo {author} {\bibfnamefont {S.}~\bibnamefont
  {Murakami}}, \bibinfo {author} {\bibfnamefont {N.}~\bibnamefont {Nagosa}},\
  and\ \bibinfo {author} {\bibfnamefont {S.-C.}\ \bibnamefont {Zhang}},\
  }\bibfield  {title} {\bibinfo {title} {$\text{SU}(2)$ non-{A}belian holonomy
  and dissipationless spin current in semiconductors},\ }\href
  {https://doi.org/10.1103/PhysRevB.69.235206} {\bibfield  {journal} {\bibinfo
  {journal} {Phys. Rev. B}\ }\textbf {\bibinfo {volume} {69}},\ \bibinfo
  {pages} {235206} (\bibinfo {year} {2004})}\BibitemShut {NoStop}%
\bibitem [{\citenamefont {Abrikosov}(1974)}]{Abrikosov}%
  \BibitemOpen
  \bibfield  {author} {\bibinfo {author} {\bibfnamefont {A.~A.}\ \bibnamefont
  {Abrikosov}},\ }\bibfield  {title} {\bibinfo {title} {Calculation of critical
  indices for zero-gap semiconductors},\ }\href@noop {} {\bibfield  {journal}
  {\bibinfo  {journal} {Sov. Phys.-JETP}\ }\textbf {\bibinfo {volume} {39}},\
  \bibinfo {pages} {709} (\bibinfo {year} {1974})}\BibitemShut {NoStop}%
\bibitem [{\citenamefont {Sun}\ \emph {et~al.}(2009)\citenamefont {Sun},
  \citenamefont {Yao}, \citenamefont {Fradkin},\ and\ \citenamefont
  {Kivelson}}]{Fradkin2009PRL}%
  \BibitemOpen
  \bibfield  {author} {\bibinfo {author} {\bibfnamefont {K.}~\bibnamefont
  {Sun}}, \bibinfo {author} {\bibfnamefont {H.}~\bibnamefont {Yao}}, \bibinfo
  {author} {\bibfnamefont {E.}~\bibnamefont {Fradkin}},\ and\ \bibinfo {author}
  {\bibfnamefont {S.~A.}\ \bibnamefont {Kivelson}},\ }\bibfield  {title}
  {\bibinfo {title} {Topological insulators and nematic phases from spontaneous
  symmetry breaking in 2{D} {F}ermi systems with a quadratic band crossing},\
  }\href {https://doi.org/10.1103/PhysRevLett.103.046811} {\bibfield  {journal}
  {\bibinfo  {journal} {Phys. Rev. Lett.}\ }\textbf {\bibinfo {volume} {103}},\
  \bibinfo {pages} {046811} (\bibinfo {year} {2009})}\BibitemShut {NoStop}%
\bibitem [{\citenamefont {Murray}\ and\ \citenamefont
  {Vafek}(2014)}]{Vafek2014PRB}%
  \BibitemOpen
  \bibfield  {author} {\bibinfo {author} {\bibfnamefont {J.~M.}\ \bibnamefont
  {Murray}}\ and\ \bibinfo {author} {\bibfnamefont {O.}~\bibnamefont {Vafek}},\
  }\bibfield  {title} {\bibinfo {title} {Renormalization group study of
  interaction-driven quantum anomalous {H}all and quantum spin {H}all phases in
  quadratic band crossing systems},\ }\href
  {https://doi.org/10.1103/PhysRevB.89.201110} {\bibfield  {journal} {\bibinfo
  {journal} {Phys. Rev. B}\ }\textbf {\bibinfo {volume} {89}},\ \bibinfo
  {pages} {201110} (\bibinfo {year} {2014})}\BibitemShut {NoStop}%
\bibitem [{\citenamefont {Venderbos}\ \emph {et~al.}(2016)\citenamefont
  {Venderbos}, \citenamefont {Manzardo}, \citenamefont {Efremov}, \citenamefont
  {van~den Brink},\ and\ \citenamefont {Ortix}}]{Venderbos2016}%
  \BibitemOpen
  \bibfield  {author} {\bibinfo {author} {\bibfnamefont {J.~W.~F.}\
  \bibnamefont {Venderbos}}, \bibinfo {author} {\bibfnamefont {M.}~\bibnamefont
  {Manzardo}}, \bibinfo {author} {\bibfnamefont {D.~V.}\ \bibnamefont
  {Efremov}}, \bibinfo {author} {\bibfnamefont {J.}~\bibnamefont {van~den
  Brink}},\ and\ \bibinfo {author} {\bibfnamefont {C.}~\bibnamefont {Ortix}},\
  }\bibfield  {title} {\bibinfo {title} {Engineering interaction-induced
  topological insulators in a
  $\sqrt{3}\ifmmode\times\else\texttimes\fi{}\sqrt{3}$ substrate-induced
  honeycomb superlattice},\ }\href {https://doi.org/10.1103/PhysRevB.93.045428}
  {\bibfield  {journal} {\bibinfo  {journal} {Phys. Rev. B}\ }\textbf {\bibinfo
  {volume} {93}},\ \bibinfo {pages} {045428} (\bibinfo {year}
  {2016})}\BibitemShut {NoStop}%
\bibitem [{\citenamefont {Wu}\ \emph {et~al.}(2016)\citenamefont {Wu},
  \citenamefont {He}, \citenamefont {Fang}, \citenamefont {Meng},\ and\
  \citenamefont {Lu}}]{Wu2016}%
  \BibitemOpen
  \bibfield  {author} {\bibinfo {author} {\bibfnamefont {H.-Q.}\ \bibnamefont
  {Wu}}, \bibinfo {author} {\bibfnamefont {Y.-Y.}\ \bibnamefont {He}}, \bibinfo
  {author} {\bibfnamefont {C.}~\bibnamefont {Fang}}, \bibinfo {author}
  {\bibfnamefont {Z.~Y.}\ \bibnamefont {Meng}},\ and\ \bibinfo {author}
  {\bibfnamefont {Z.-Y.}\ \bibnamefont {Lu}},\ }\bibfield  {title} {\bibinfo
  {title} {Diagnosis of interaction-driven topological phase via exact
  diagonalization},\ }\href {https://doi.org/10.1103/PhysRevLett.117.066403}
  {\bibfield  {journal} {\bibinfo  {journal} {Phys. Rev. Lett.}\ }\textbf
  {\bibinfo {volume} {117}},\ \bibinfo {pages} {066403} (\bibinfo {year}
  {2016})}\BibitemShut {NoStop}%
\bibitem [{\citenamefont {Wang}\ \emph {et~al.}(2017)\citenamefont {Wang},
  \citenamefont {Ortix}, \citenamefont {van~den Brink},\ and\ \citenamefont
  {Efremov}}]{Wang2017PRB}%
  \BibitemOpen
  \bibfield  {author} {\bibinfo {author} {\bibfnamefont {J.}~\bibnamefont
  {Wang}}, \bibinfo {author} {\bibfnamefont {C.}~\bibnamefont {Ortix}},
  \bibinfo {author} {\bibfnamefont {J.}~\bibnamefont {van~den Brink}},\ and\
  \bibinfo {author} {\bibfnamefont {D.~V.}\ \bibnamefont {Efremov}},\
  }\bibfield  {title} {\bibinfo {title} {Fate of interaction-driven topological
  insulators under disorder},\ }\href
  {https://doi.org/10.1103/PhysRevB.96.201104} {\bibfield  {journal} {\bibinfo
  {journal} {Phys. Rev. B}\ }\textbf {\bibinfo {volume} {96}},\ \bibinfo
  {pages} {201104} (\bibinfo {year} {2017})}\BibitemShut {NoStop}%
\bibitem [{\citenamefont {Dong}\ \emph {et~al.}(2020)\citenamefont {Dong},
  \citenamefont {Zhai}, \citenamefont {Zheng},\ and\ \citenamefont
  {Wang}}]{Wang2020PRB}%
  \BibitemOpen
  \bibfield  {author} {\bibinfo {author} {\bibfnamefont {Y.-M.}\ \bibnamefont
  {Dong}}, \bibinfo {author} {\bibfnamefont {Y.-H.}\ \bibnamefont {Zhai}},
  \bibinfo {author} {\bibfnamefont {D.-X.}\ \bibnamefont {Zheng}},\ and\
  \bibinfo {author} {\bibfnamefont {J.}~\bibnamefont {Wang}},\ }\bibfield
  {title} {\bibinfo {title} {Stability of two-dimensional asymmetric materials
  with a quadratic band crossing point under four-fermion interaction and
  impurity scattering},\ }\href {https://doi.org/10.1103/PhysRevB.102.134204}
  {\bibfield  {journal} {\bibinfo  {journal} {Phys. Rev. B}\ }\textbf {\bibinfo
  {volume} {102}},\ \bibinfo {pages} {134204} (\bibinfo {year}
  {2020})}\BibitemShut {NoStop}%
\bibitem [{\citenamefont {Tchoumakov}\ and\ \citenamefont
  {Witczak-Krempa}(2019)}]{Tchoumakov-Krempa2019PRB}%
  \BibitemOpen
  \bibfield  {author} {\bibinfo {author} {\bibfnamefont {S.}~\bibnamefont
  {Tchoumakov}}\ and\ \bibinfo {author} {\bibfnamefont {W.}~\bibnamefont
  {Witczak-Krempa}},\ }\bibfield  {title} {\bibinfo {title} {Dielectric and
  electronic properties of three-dimensional {L}uttinger semimetals with a
  quadratic band touching},\ }\href
  {https://doi.org/10.1103/PhysRevB.100.075104} {\bibfield  {journal} {\bibinfo
   {journal} {Phys. Rev. B}\ }\textbf {\bibinfo {volume} {100}},\ \bibinfo
  {pages} {075104} (\bibinfo {year} {2019})}\BibitemShut {NoStop}%
\bibitem [{\citenamefont {Mauri}\ and\ \citenamefont {Polini}(2019)}]{polini}%
  \BibitemOpen
  \bibfield  {author} {\bibinfo {author} {\bibfnamefont {A.}~\bibnamefont
  {Mauri}}\ and\ \bibinfo {author} {\bibfnamefont {M.}~\bibnamefont {Polini}},\
  }\bibfield  {title} {\bibinfo {title} {Dielectric function and plasmons of
  doped three-dimensional {L}uttinger semimetals},\ }\href
  {https://doi.org/10.1103/PhysRevB.100.165115} {\bibfield  {journal} {\bibinfo
   {journal} {Phys. Rev. B}\ }\textbf {\bibinfo {volume} {100}},\ \bibinfo
  {pages} {165115} (\bibinfo {year} {2019})}\BibitemShut {NoStop}%
\bibitem [{\citenamefont {Herbut}(2012)}]{Herbut2012PRB}%
  \BibitemOpen
  \bibfield  {author} {\bibinfo {author} {\bibfnamefont {I.~F.}\ \bibnamefont
  {Herbut}},\ }\bibfield  {title} {\bibinfo {title} {Isospin of topological
  defects in {D}irac systems},\ }\href
  {https://doi.org/10.1103/PhysRevB.85.085304} {\bibfield  {journal} {\bibinfo
  {journal} {Phys. Rev. B}\ }\textbf {\bibinfo {volume} {85}},\ \bibinfo
  {pages} {085304} (\bibinfo {year} {2012})}\BibitemShut {NoStop}%
\bibitem [{\citenamefont {Kozii}\ and\ \citenamefont
  {Fu}(2018)}]{Kozii-Fu2018PRB}%
  \BibitemOpen
  \bibfield  {author} {\bibinfo {author} {\bibfnamefont {V.}~\bibnamefont
  {Kozii}}\ and\ \bibinfo {author} {\bibfnamefont {L.}~\bibnamefont {Fu}},\
  }\bibfield  {title} {\bibinfo {title} {Thermal plasmon resonantly enhances
  electron scattering in {D}irac/{W}eyl semimetals},\ }\href
  {https://doi.org/10.1103/PhysRevB.98.041109} {\bibfield  {journal} {\bibinfo
  {journal} {Phys. Rev. B}\ }\textbf {\bibinfo {volume} {98}},\ \bibinfo
  {pages} {041109} (\bibinfo {year} {2018})}\BibitemShut {NoStop}%
\bibitem [{\citenamefont {Abrikosov}(1963)}]{abrikosov2012methods}%
  \BibitemOpen
  \bibfield  {author} {\bibinfo {author} {\bibfnamefont {A.}~\bibnamefont
  {Abrikosov}},\ }\href {https://books.google.co.in/books?id=1vxQAAAAMAAJ}
  {\emph {\bibinfo {title} {Methods of Quantum Field Theory in Statistical
  Physics}}},\ Dover books on advanced mathematics\ (\bibinfo  {publisher}
  {Prentice-Hall},\ \bibinfo {year} {1963})\BibitemShut {NoStop}%
\bibitem [{\citenamefont {Mahan}(1990)}]{Mahan2000Book}%
  \BibitemOpen
  \bibfield  {author} {\bibinfo {author} {\bibfnamefont {G.}~\bibnamefont
  {Mahan}},\ }\href {https://books.google.co.in/books?id=v8du6cp0vUAC} {\emph
  {\bibinfo {title} {Many-Particle Physics}}},\ Physics of Solids and Liquids\
  (\bibinfo  {publisher} {Springer US},\ \bibinfo {year} {1990})\BibitemShut
  {NoStop}%
\end{thebibliography}%

\end{document}